\documentclass[11pt,english]{article}
\usepackage[T1]{fontenc}
\usepackage[latin9]{inputenc}
\usepackage{geometry}
\usepackage{verbatim}
\usepackage{float}
\usepackage{rotfloat}
\usepackage{amsmath}
\usepackage{amssymb}
\usepackage{graphicx}
\usepackage{esint}

\makeatletter

\newcommand*\LyXZeroWidthSpace{\hspace{0pt}}
\providecommand{\tabularnewline}{\\}

\usepackage{times}

\newtheorem{proposition}{}

\newtheorem{lemma}{}

\newtheorem{assumption}{Assumption}
\newtheorem{Aassumption}{Assumption}
\newtheorem{BDassumption}{Assumption}

\DeclareMathOperator*{\plim}{plim}

\DeclareRobustCommand{\stirling}{\genfrac\{\}{0pt}{}}

\newcommand{\ackname}{Acknowledgements:}\newenvironment{acknowledgements}
{\chapter*{\ackname}\addcontentsline{toc}{chapter}{\ackname}}{\clearpage}

\usepackage[round,authoryear]{natbib}

\usepackage{babel}

\makeatother

\usepackage{babel}
\begin{document}

\title{A DIAGNOSTIC CRITERION \\
 FOR APPROXIMATE FACTOR STRUCTURE}

\author{Patrick Gagliardini$^{a}$, Elisa Ossola$^{b}$ and Olivier Scaillet$^{c}${*}}

\date{First draft: February 2014. This version: July 2017.}
\maketitle
\begin{abstract}
{\footnotesize{}{}We build a simple diagnostic criterion for approximate
factor structure in large cross-sectional equity datasets. Given a
model for asset returns with observable factors, the criterion checks
whether the error terms are weakly cross-sectionally correlated or
share at least one unobservable common factor. It only requires computing
the largest eigenvalue of the empirical cross-sectional covariance
matrix of the residuals of a large unbalanced panel. A general version
of this criterion allows us to determine the number of omitted common
factors. The panel data model accommodates both time-invariant and
time-varying factor structures. The theory applies to random coefficient
panel models with interactive fixed effects under large cross-section
and time-series dimensions. The empirical analysis runs on monthly
and quarterly returns for about ten thousand US stocks from January
1968 to December 2011 for several time-invariant and time-varying
specifications. For monthly returns, we can choose either among time-invariant
specifications with at least four financial factors, or a scaled three-factor
specification. For quarterly returns, we cannot select macroeconomic
models without the market factor.}{\footnotesize \par}

\noindent \textit{\footnotesize{}{}JEL Classification:}{\footnotesize{}{}
C12, C13, C23, C51, C52, C58, G12.}{\footnotesize \par}

\noindent \textit{\footnotesize{}{}Keywords:}{\footnotesize{}{}
large panel, approximate factor model, asset pricing, model selection,
interactive fixed effects.}{\footnotesize \par}

\noindent {\scriptsize{}{}$^{a}$Università della Svizzera Italiana
(USI Lugano) and Swiss Finance Institute, $^{b}$ European Commission,
Joint Research Centre, $^{c}$University of Geneva and Swiss Finance
Institute.}{\scriptsize \par}

\noindent {\tiny{}{}\begin{acknowledgements} The second author gratefully
acknowledges the Swiss National Science Foundation (SNSF) for generously
funding her research with a Marie Heim-Voegtlin fellowship. We are
grateful to Lu Zhang and Alex Horenstein for sharing their data with
us. We thank L.\ Mancini, E.\ Sentana, M.\ Weidmer, participants
at ESEM 2014, COMPSTAT 2014, EWM 2015, IMS-China conference 2015,
Bristol Econometric Study Group 2015, ESWC 2015, Workshop ``Pietro
Balestra'' on panel data, INFER workshop, DagStat 2016, CORE@50 Conference,
SoFiE Annual Meeting 2016, ESEM 2016, FMA 2016 meeting Las Vegas,
ICEEE 2017, CIB conference on stochastic models in Lausanne, \textquotedbl{}Meeting
in Econometrics\textquotedbl{} conference in Toulouse, and participants
at seminars at the University of Chicago, University of Orléans, University
of St Gallen, Groningen, Tilburg, EPFL, ULB, University of Venice,
University College of London, CEMFI, Pittsburgh University, PennState
University, Federal Reserve Board of Governors, University of Surrey,
University of Cergy Pontoise, University of Illinois Urbana-Champaign,
and University of Aarhus. {*}Disclaimer: The content of this article
does not reflect the official opinion of the European Commission.
Responsibility for the information and views expressed therein lies
entirely with the authors.\end{acknowledgements}}{\tiny \par}
\end{abstract}
\pagebreak{}

\section{Introduction}

Empirical work in asset pricing vastly relies on linear multi-factor
models with either time-invariant coefficients (unconditional models)
or time-varying coefficients (conditional models). The factor structure
is often based on observable variables (empirical factors) and supposed
to be rich enough to extract systematic risks while idiosyncratic
risk is left over to the error term. Linear factor models are rooted
in the Arbitrage Pricing Theory (APT, \cite{Ross_1976}, \cite{Chamberlain_Rothschild_1983})
or come from a loglinearization of nonlinear consumption-based models
(\cite{Campbell_1996}). Conditional linear factor models aim at capturing
the time-varying influence of financial and macroeconomic variables
in a simple setting (see e.g. \cite{Shanken_1990}, \cite{Cochrane_1996},
\cite{Ferson_Schadt_1996}, \cite{Ferson_Harvey_1991,Ferson_Harvey_1999},
\cite{Lettau_Ludvigson_2001}, \cite{Petkova_Zhang_2005}). Time variation
in risk biases time-invariant estimates of alphas and betas, and therefore
asset pricing test conclusions (\cite{Jagannathan_Wang_1996}, \cite{Lewellen_Nagel_2006},
\cite{Boguth_Carlson_Fisher_Simutin_2011}). \cite{Ghysels_1998}
discusses the pros and cons of modeling time-varying betas.

A central and practical issue is to determine whether there are one
or more factors omitted in the chosen specification. Approximate factor
structures with nondiagonal error covariance matrices (\cite{Chamberlain_Rothschild_1983})
answer the potential empirical mismatch of exact factor structures
with diagonal error covariance matrices underlying the original APT
of \cite{Ross_1976}. If the set of observable factors is correctly
specified, the errors are weakly cross-sectionally correlated, namely
the covariance matrix of the error terms in the factor model has a
fastly vanishing largest eigenvalue. %
If the set of observable factors is not correctly specified, the no-arbitrage
restrictions derived from APT will not hold, and the risk premia estimated
by the two-pass regression approach will be meaningless (see Appendix
H in Gagliardini, Ossola, Scaillet (2016, GOS)\nocite{GOS_2016} for
a discussion of misspecification in the two-pass methodology). Even
if the omitted factors are not priced, i.e., their associated risk
premia are nil, direct computations of the limits of first pass and
second pass estimators under misspecification show that second pass
estimates will not converge to the risk premia of the priced factors,
and that biases on betas and risk premia will not compensate each
other. Hence detecting an omitted factor is also important in that
case to produce correct expected excess returns from the no arbitrage
restrictions. \cite{Giglio_Xiu_2016} have proposed a three-pass methodology
allowing consistent estimation by exploiting an invariance property
in time invariant models with omitted factors in balanced panels.
Given the large menu of factors available in the literature (the factor
zoo of \cite{Cochrane_2011}, see also \cite{Harvey_Liu_Zhu_2015},
\cite{Harvey_Liu_2016}), we need a simple diagnostic criterion to
decide whether we can feel comfortable with the chosen set of observable
factors before proceeding further in the empirical analysis of large
cross sectional equity data sets under the APT setting. For example,
if the factor model passes the diagnostic, and we reject that alphas
are zero using a GRS-type statistic (\cite{Gibbons_Ross_Shanken_1989}),
it will not be because of an omitted factor.

For models with unobservable (latent) factors only, \cite{Connor_Korajczyk_1993}
are the first to develop a test for the number of factors for large
balanced panels of individual stock returns in time-invariant models
under covariance stationarity and homoskedasticity. Unobservable factors
are estimated by the method of asymptotic principal components developed
by \cite{Connor_Korajczyk_1986} (see also \cite{Stock_Watson_2002b}).
For heteroskedastic settings, the recent literature on large panels
with static factors (see \cite{Hallin_Liska_2007} and \cite{Jan_Otter_2008}
for a selection procedure in the generalized dynamic factor model
of \cite{Forni_al_2000}) has extended the toolkit available to researchers.
A first strand of that literature focuses on consistent estimation
procedures for the number of factors. \cite{Bai_Ng_2002} introduce
a penalized least-squares strategy to estimate the number of factors,
at least one (see \cite{Amengual_Watson_2007} to include dynamic
factors). \cite{Onatski_2010} looks at the behavior of the adjacent
eigenvalues to determine the number of factors when the cross-sectional
dimension ($n$) and the time-series dimension ($T$) are comparable.
\cite{Ahn_Horenstein_2013} opt for the same strategy and cover the
possibility of zero factors. Caner and Han (2014)\nocite{Caner_Han_2014}
propose an estimator with a group bridge penalization to determine
the number of unobservable factors. A second strand of that literature
develops inference procedures for hypotheses on the number of latent
factors. \cite{Onatski_2009} deploys a characterization of the largest
eigenvalues of a Wishart-distributed covariance matrix with large
dimensions in terms of the Tracy-Widom Law. To get a Wishart distribution,
\cite{Onatski_2009} assumes either Gaussian errors or $T$ much larger
than $n$. \cite{Kapetanios_2010} uses subsampling to estimate the
limit distribution of the adjacent eigenvalues. \cite{Harding_2013}
uses free probability theory to derive analytic expressions for the
limiting moments of the spectral distribution.

Our paper expands the first strand of the literature by developing
a consistent estimation procedure for the number of latent factors
in the error terms in a model with observable factors when the cross-section
dimension can be much larger than the time series dimension. Concluding
for zero omitted factors means weakly cross-sectionally correlated
errors. We require $n=O(T^{\bar{\gamma}})$, $\bar{\gamma}>0$, and
$T=O(n^{\gamma})$, $\gamma\in(0,1]$, which is equivalent to $C_{1}n^{1/\bar{\gamma}}\leq T\leq C_{2}n^{\gamma}$
for some positive constants $C_{1}$, $C_{2}$. The case $\gamma<1$
implies $T/n=o(1)$, namely $n$ is much larger than $T$, and the
case $\bar{\gamma}=\gamma=1$ implies that $n$ and $T$ are comparable.
In our empirical application, we have monthly and quarterly returns
for about ten thousand US stocks from January 1968 to December 2011,
and this explains why we also investigate the setting $T/n=o(1)$.
The asymptotic distribution of the eigenvalues is degenerate under
the usual standardisation of the $T\times T$ covariance matrix by
$n^{-1}$ when the ratio $T/n$ goes to zero as $T,n\rightarrow\infty$.
In such a setting, we cannot exploit well-defined limiting characterizations
(Marchenko-Pastur distribution, Tracy-Widom distribution) obtained
when $T/n$ converges to a strictly positive constant. Without such
distributional characterizations, we do not see hope for testing procedures
as developed by \cite{Onatski_2009}. However, a key theoretical result
of our paper is that we can still have an asymptotically valid selection
procedure for the number of latent factors even in the presence of
a degenerate distribution of the eigenvalues of the sample covariance
matrix of the errors. We show that this extends to sample covariance
matrices of residuals of an estimated linear model with observable
factors in unbalanced panels.\LyXZeroWidthSpace{} An extension to
residuals instead of true errors is not trivial since we need to cope
with a projection matrix in the estimated errors, and there are little
results about the analysis of the spectrum of matrix products (as
opposed to the many results for matrix sums). The unbalanced nature
makes things worse since we also have to take care of the matrix of
observability indicators in the product. This further explains why
we shy away from putting an additional structure on the errors (\cite{Onatski_2010},
\cite{Ahn_Horenstein_2013}) or estimated errors in unbalanced panels
in our assumptions. Most of our assumptions are weaker, and the arguments
developed in the proofs of the theorems supporting our extension are
new to the literature as further commented below. 

For applications of factor models in empirical finance, \cite{Bai_Ng_2006}
analyze statistics to test whether the observable factors in time-invariant
models span the space of unobservable factors (see also \cite{Lehmann_Modest_1988}
and \cite{Connor_Korajczyk_1988}). They find that the three-factor
model of \cite[FF]{Fama_French_1993} is the most satisfactory proxy
for the unobservable factors estimated from balanced panels of portfolio
and individual stock returns. \cite{Ahn_Horenstein_Wang_2015} study
a rank estimation method to also check whether time-invariant factor
models are compatible with a number of unobservable factors. For portfolio
returns, they find that the FF model exhibits a full rank beta (factor
loading) matrix. \cite{Goncalves_Perron_Djogbenou_2015} consider
bootstrap prediction intervals for factor models. Factor analysis
for large cross-sectional datasets also find applications in studying
bond risk premia (\cite{Ludvigson_Ng_2007,Ludvigson_Ng_2009}) and
measuring time-varying macroeconomic uncertainty (\cite{Jurako_Ludvigson_Ng_2015}).
\cite{Connor_Hagmann_Linton_2012} showed that large cross sections
exploit data more efficiently in a semiparametric characteristic-based
factor model of stock returns. Recent papers (\cite{Fan_Furger_Xiu_2016},
\cite{Pelger_2015}, \cite{AitSahalia_Xiu_2016}) have also investigated
large-dimensional factor modeling with in-fill asymptotics for high-frequency
data.

In this paper, we build a simple diagnostic criterion for approximate
factor structure in large cross-sectional datasets. The criterion
checks whether the error terms in a given model with observable factors
are weakly cross-sectionally correlated or share at least one common
factor. It only requires computing the largest eigenvalue of the empirical
cross-sectional covariance matrix of the residuals of a large unbalanced
panel and subtracting a penalization term vanishing to zero for large
$n$ and $T$. The steps of the diagnostic are easy: 1) compute the
largest eigenvalue, 2) subtract a penalty, 3) conclude to validity
of the proposed approximate factor structure if the difference is
negative, or conclude to at least one omitted factor if the difference
is positive. Our main theoretical contribution shows that step 3)
yields asymptotically the correct model selection. The mechanics of
the selection are easy to grasp. If we have an approximate factor
structure, we expect a vanishing largest eigenvalue because of a lack
of a common signal in the error terms. So, if we take a penalizing
term with a slower rate towards zero, a negative criterion points
to weak cross-sectional correlation. On the contrary, the largest
eigenvalue remains bounded from below away from zero if we face omitted
factors. We have at least one non vanishing eigenvalue because of
a common signal due to omitted factors. The positive largest eigenvalue
dominates the vanishing penalizing term, and this explains why we
conclude against weak cross sectional correlation when the criterion
is positive. We also propose a general version of the diagnostic criterion
that determines the number of omitted common factors. As shown below,
the criterion coincides with the penalized least-squares approach
of \cite{Bai_Ng_2002} applied on residuals of an unbalanced panel.
We derive all properties for unbalanced panels in the setting of \cite{Connor_Korajczyk_1987}
to avoid the survivorship bias inherent to studies restricted to balanced
subsets of available stock return databases (\cite{Brown_Goetzmann_Ross_1995}).
The panel data model is sufficiently general to accommodate both time-invariant
and time-varying factor structures. Allowing for time-varying factor
loadings presents challenges for finance theory and econometric modelling,
but GOS explain how to solve these issues and give empirical evidence
of time-varying\LyXZeroWidthSpace{} risk premia. We recast the factor
models as generic random coefficient panel models and develop the
theory for large cross-section and time-series dimensions with $n=O(T^{\bar{\gamma}})$,
$\bar{\gamma}>0$, and $T=O(n^{\gamma})$, $\gamma\in(0,1]$. Omitted
latent factors are also called interactive fixed effects in the panel
literature (\cite{Pesaran_2006}, \cite{Bai_2009}, \cite{Moon_Weidner_2015},
\cite{Magnac_Gobillon_2014}). \LyXZeroWidthSpace \cite{King_Sentana_Wadhwani_1994}
use them to capture the correlation between the unanticipated innovations
in observable descriptors of economic performance (e.g. industrial
production, inflation, etc.) and stock returns.

For our empirical contribution, we consider the Center for Research
in Security Prices (CRSP) database and take the Compustat database
to match firm characteristics. The merged dataset comprises about
ten thousands stocks with returns from January 1968 to December 2011.
We look at a variety of empirical factors and we build factor models
popular in the empirical literature to explain monthly and quarterly
equity returns. They differ by the choice of the observable factors.
We analyze monthly returns using recent financial specifications such
as the five factors of \cite{Fama_French_2015}, the profitability
and investment factors of \cite{Hou_Xue_Zhang_2015}, the quality
minus junk and bet against beta factors of \cite{Asness_Frazzini_Pedersen_2014}
and \cite{Frazzini_Pedersen_2014}, as well as other specifications
described below. We analyze quarterly returns using macroeconomic
specifications including consumption growth (CCAPM), market returns
and consumption growth (\cite{Epstein_Zin_1989}), the three factors
in \cite{Yogo_2006}, the three factors in \cite{Li_Vassalou_Xing_2006},
and the five factors of \cite{Chen_Roll_Ross_1986}. We study time-invariant
and time-varying versions of the financial factor models (\cite{Shanken_1990},
\cite{Cochrane_1996}, \cite{Ferson_Schadt_1996}, \cite{Ferson_Harvey_1999}).
For the latter, we use both macrovariables and firm characteristics
as instruments (\cite{Avramov_Chordia_2006}). For monthly returns,
our diagnostic criterion is met by time-invariant specifications with
at least four financial factors, and a scaled three-factor FF time-varying
specification. For quarterly returns, we cannot select macroeconomic
models without the market factor.

The outline of the paper is as follows. In Section \ref{sec:Conditional-factor-model},
we consider a general framework of conditional linear factor model
for asset returns. In Section \ref{sec:Diagnostic-criterion}, we
present our diagnostic criterion for approximate factor structure
in random coefficient panel models. In Section \ref{sec:DeterminingNumberFactors},
we provide the diagnostic criterion to determine the number of omitted
factors. Section \ref{sec:Implementation} explains how to implement
the criterion in practice and how to design a simple graphical diagnostic
tool related to the well-known scree plot in principal component analysis.
Section \ref{sec:Empirical-results} contains the empirical results.
In Appendices 1 and 2, we gather the theoretical assumptions and some
proofs. We use high-level assumptions on cross-sectional and serial
dependence of error terms, and show in \ref{sec:Appendix_CheckBlockDep}
that we meet them under a block cross-sectional dependence structure
in a serially i.i.d. framework. We place all omitted proofs in the
online supplementary materials. There we link our approach to the
expectation-maximization (EM) algorithm proposed by \cite{Stock_Watson_2002b}
for unbalanced panels. We also include some Monte-Carlo simulation
results under a design mimicking our empirical application to show
the practical relevance of our selection procedure in finite samples.
The additional empirical results, discussed but not reported in the
paper, are available on request.

\section{Conditional factor model of asset returns \label{sec:Conditional-factor-model}}

In this section, we consider a conditional linear factor model with
time-varying coefficients. We work in a multi-period economy (\cite{Hansen_Richard_1987})
under an approximate factor structure (\cite{Chamberlain_Rothschild_1983})
with a continuum of assets as in GOS. Such a construction is close
to the setting advocated by \cite{AlNajjar_1995,AlNajjar_1998,AlNajjar_1999a}
in a static framework with an exact factor structure. He discusses
several key advantages of using a continuum economy in arbitrage pricing
and risk decomposition. A key advantage is robustness of factor structures
to asset repackaging (\cite{AlNajjar_1999b}; see GOS for a proof).

Let ${\cal F}_{t}$, with $t=1,2,...$, be the information available
to investors. Without loss of generality, the continuum of assets
is represented by the interval $\left[0,1\right]$. The excess returns
$R_{t}\left(\gamma\right)$ of asset $\gamma\in\left[0,1\right]$
at dates $t=1,2,...$ satisfy the conditional linear factor model:
\begin{equation}
R_{t}(\gamma)=a_{t}(\gamma)+b_{t}(\gamma)^{^{\prime}}f_{t}+\varepsilon_{t}(\gamma),\label{condLinearFactMod}
\end{equation}
where vector $f_{t}$ gathers the values of $K$ observable factors
at date $t$. The intercept $a_{t}(\gamma)$ and factor sensitivities
$b_{t}(\gamma)$ are $\mathcal{F}_{t-1}$-measurable. The error terms
$\varepsilon_{t}\left(\gamma\right)$ have mean zero and are uncorrelated
with the factors conditionally on information $\mathcal{F}_{t-1}$.
Moreover, we exclude asymptotic arbitrage opportunities in the economy:
there are no portfolios that approximate arbitrage opportunities when
the number of assets increases. In this setting, GOS show that the
following asset pricing restriction holds: 
\begin{equation}
a_{t}(\gamma)=b_{t}(\gamma)^{\prime}\nu_{t},\ \text{for almost all }\gamma\in[0,1],\label{condNoArbRest}
\end{equation}
almost surely in probability, where random vector $\nu_{t}\in\mathbb{R}^{K}$
is unique and is $\mathcal{F}_{t-1}$-measurable. The asset pricing
restriction (\ref{condNoArbRest}) is equivalent to $E\left[R_{t}(\gamma)\vert\mathcal{F}_{t-1}\right]=b_{t}(\gamma)^{\prime}\lambda_{t},$
where $\lambda_{t}=\nu_{t}+E\left[f_{t}\vert\mathcal{F}_{t-1}\right]$
is the vector of the conditional risk premia.

To have a workable version of Equations (\ref{condLinearFactMod})
and (\ref{condNoArbRest}), we define how the conditioning information
is generated and how the model coefficients depend on it via simple
functional specifications. The conditioning information $\mathcal{F}_{t-1}$
contains $Z_{\underline{t-1}}$ and $Z_{\underline{t-1}}(\gamma)$,
for all $\gamma\in[0,1]$, where the vector of lagged instruments
$Z_{t-1}\in\mathbb{R}^{p}$ is common to all stocks, the vector of
lagged instruments $Z_{t-1}(\gamma)\in\mathbb{R}^{q}$ is specific
to stock $\gamma$, and $Z_{\underline{t}}=\{Z_{t},Z_{t-1},...\}$.
Vector $Z_{t-1}$ may include the constant and past observations of
the factors and some additional variables such as macroeconomic variables.
Vector $Z_{t-1}(\gamma)$ may include past observations of firm characteristics
and stock returns. To end up with a linear regression model, we assume
that: (i) the vector of factor loadings $b_{t}\left(\gamma\right)$
is a linear function of lagged instruments $Z_{t-1}$ (\cite{Shanken_1990},
\cite{Ferson_Harvey_1991}) and $Z_{t-1}\left(\gamma\right)$ (\cite{Avramov_Chordia_2006});
(ii) the vector of risk premia $\lambda_{t}$ is a linear function
of lagged instruments $Z_{t-1}$ (\cite{Cochrane_1996}, \cite{Jagannathan_Wang_1996});
(iii) the conditional expectation of $f_{t}$ given the information
${\cal F}_{t-1}$ depends on $Z_{t-1}$ only and is linear (as e.g.
if $Z_{t}$ follows a Vector Autoregressive (VAR) model of order $1$).

To ensure that cross-sectional limits exist and are invariant to reordering
of the assets, we introduce a sampling scheme as in GOS. We formalize
it so that observable assets are random draws from an underlying population
(\cite{Andrews_2005}). In particular, we rely on a sample of $n$
assets by randomly drawing i.i.d. indices $\gamma_{i}$ from the population
according to a probability distribution $G$ on $[0,1]$. For any
$n,T\in\mathbb{N}$, the excess returns are $R_{i,t}=R_{t}(\gamma_{i})$.
Similarly, let $a_{i,t}=a_{t}(\gamma_{i})$ and $b_{i,t}=b_{t}\left(\gamma_{i}\right)$
be the coefficients, $\varepsilon_{i,t}=\varepsilon_{t}(\gamma_{i})$
be the error terms, and $Z_{i,t}=Z_{t}(\gamma_{i})$ be the stock
specific instruments. By random sampling, we get a random coefficient
panel model (e.g.\ \cite{Hsiao_2003}, Chapter 6). Such a formalisation
is key to reconcile finance theory and econometric modelling. Without
drawings, cross-sectional averages such as ${\displaystyle \frac{1}{n}\sum_{i}b_{i}}$
correspond to determinist sequences since the $b_{i}$s are then parameters.
Working with the standard arbitrage pricing theory with approximate
factor models has three issues as discussed in GOS. First, cross-sectional
limits depend in general on the ordering of the financial assets,
and there is no natural ordering between assets (firms). Second, we
cannot exploit either a law of large numbers to guarantee existence
of those limits, nor a central limit theorem to get distributional
results. Third, the asset pricing restrictions derived under no arbitrage
are not testable, the so-called Shanken critique (\cite{Shanken_1982}).
In available datasets, we do not observe asset returns for all firms
at all dates. Thus, we account for the unbalanced nature of the panel
through a collection of indicator variables $I_{i,t}$, for any asset
$i$ at time $t$. We define $I_{i,t}=1$ if the return of asset $i$
is observable at date $t$, and 0 otherwise (\cite{Connor_Korajczyk_1987}).

Through appropriate redefinitions of the regressors and coefficients,
GOS show that we can rewrite the model for Equations (\ref{condLinearFactMod})
and (\ref{condNoArbRest}) as a generic random coefficient panel model:
\begin{equation}
R_{i,t}=x_{i,t}^{\prime}\beta_{i}+\varepsilon_{i,t},\label{LinearFactorRegression}
\end{equation}
where the regressor $x_{i,t}=\left(x_{1,i,t}^{\prime},x_{2,i,t}^{\prime}\right)^{\prime}$
has dimension $d=d_{1}+d_{2}$ and includes vectors $x_{1,i,t}=\left(vech\left[X_{t}\right]^{\prime},Z_{t-1}^{\prime}\otimes Z_{i,t-1}^{\prime}\right)^{\prime}\in\mathbb{R}^{d_{1}}$
and $x_{2,i,t}=\left(f_{t}^{\prime}\otimes Z_{t-1}^{\prime},f_{t}^{\prime}\otimes Z_{i,t-1}^{\prime}\right)^{\prime}\in\mathbb{R}^{d_{2}}$
with $d_{1}=p(p+1)/2+pq$ and $d_{2}=K(p+q)$. In vector $x_{2,i,t}$,
the first components with common instruments take the interpretation
of scaled factors (\cite{Cochrane_2005_book}), while the second components
do not since they depend on $i$. The symmetric matrix $X_{t}=[X_{t,k,l}]\in\mathbb{R}^{p\times p}$
is such that $X_{t,k,l}=Z_{t-1,k}^{2}$, if $k=l$, and $X_{t,k,l}=2Z_{t-1,k}Z_{t-1,l}$,
otherwise, $k,l=1,\ldots,p$, where $Z_{t,k}$ denotes the $k$th
component of the vector $Z_{t}$. The vector-half operator $vech\left[\cdot\right]$
stacks the elements of the lower triangular part of a $p\times p$
matrix as a $p\left(p+1\right)/2\times1$ vector (see Chapter 2 in
\cite{Magnus_Neudecker_2007} for properties of this matrix tool).
The vector of coefficients $\beta_{i}$ is a function of asset specific
parameters defining the dynamics of $a_{i,t}$ and $b_{i,t}$ detailed
in GOS. In matrix notation, for any asset $i$, we have 
\begin{equation}
R_{i}=X_{i}\beta_{i}+\varepsilon_{i},\label{LinearFactorRegression_MatrixNotat}
\end{equation}
where $R_{i}$ and $\varepsilon_{i}$ are $T\times1$ vectors. Regression
(\ref{LinearFactorRegression}) contains both explanatory variables
that are common across assets (scaled factors) and asset-specific
regressors. It includes models with time-invariant coefficients as
a particular case. In such a case, the regressor reduces to $x_{t}=\left(1,f_{t}^{\prime}\right)^{\prime}$
and is common across assets, and the regression coefficient vector
is $\beta_{i}=\left(a_{i},b_{i}^{\prime}\right)^{\prime}$ of dimension
$d=K+1$.

In order to build the diagnostic criterion for the set of observable
factors, we consider the following rival models: 
\[
\mathcal{M}_{1}:\quad\text{the linear regression model (\ref{LinearFactorRegression}), where the errors \ensuremath{(\varepsilon_{i,t})\ }are weakly cross-sectionally dependent,}
\]
and 
\[
\mathcal{M}_{2}:\quad\text{the linear regression model (\ref{LinearFactorRegression}), where the errors \ensuremath{(\varepsilon_{i,t})\ }satisfy a factor structure.}
\]
Under model ${\cal M}_{1}$, the observable factors fully capture
the systematic risk, and the error terms do not feature pervasive
forms of cross-sectional dependence (see Assumption \ref{A_Ass_restricSerialCrossDep}
in \ref{sec:Appendix_RegularityConditions}). Under model ${\cal M}_{2}$,
the following error factor structure holds 
\begin{equation}
\varepsilon_{i,t}=\theta_{i}^{\prime}h_{t}+u_{i,t},\label{FactorStructure_M2}
\end{equation}
where the $m\times1$ vector $h_{t}$ includes unobservable (i.e.,
latent or hidden) factors, and the $u_{i,t}$ are weakly cross-sectionally
correlated. The latent factors may include scaled factors to cover
latent time-varying factor loadings with common instruments. We cannot
allow for latent time-varying factor loadings with stock-specific
instruments because of identification issues. In (\ref{FactorStructure_M2}),
the $\theta_{i}$'s are also called interactive fixed effects in the
panel literature. The $m\times1$ vector $\theta_{i}$ corresponds
to the factor loadings, and the number $m$ of common factors is assumed
unknown. In vector notation, we have: 
\begin{equation}
\varepsilon_{i}=H\theta_{i}+u_{i},\label{FactirStructure_M2_MatrixNotat}
\end{equation}
where $H$ is the $T\times m$ matrix of unobservable factor values,
and $u_{i}$ is a $T\times1$ vector.

\begin{assumption} \label{ass_ftheta} Under model $\mathcal{M}_{2}$:
(i) Matrix ${\displaystyle \frac{1}{T}\sum_{t}h_{t}h_{t}'}$ converges
in probability to a positive definite matrix $\Sigma_{h}$, as $T\rightarrow\infty$.
(ii) ${\displaystyle \mu_{1}\left(\frac{1}{n}\sum_{i}\theta_{i}\theta_{i}'\right)\geq C}$,
w.p.a. $1$ as $n\rightarrow\infty$, for a constant $C>0$, where
$\mu_{1}\left(.\right)$ denotes the largest eigenvalue of a symmetric
matrix. \end{assumption}

\noindent Assumption \ref{ass_ftheta} (i) is a standard condition
in linear latent factor models (see Assumption A in \cite{Bai_Ng_2002})
and we can normalize matrix $\Sigma_{h}$ to be the identity matrix
$I_{m}$ for identification. Assumption \ref{ass_ftheta} (ii) requires
that at least one factor in the error terms is strong. It is satisfied
if the second-order matrix of the loadings ${\displaystyle \frac{1}{n}\sum_{i}\theta_{i}\theta_{i}'}$
converges in probability to a positive definite matrix (see Assumption
B in \cite{Bai_Ng_2002}).

We work with the condition: 
\begin{equation}
E[x_{i,t}h_{t}']=0,\quad\forall i,\label{orth}
\end{equation}
that is, orthogonality between latent factors and observable regressors
for all stocks. This condition allows us to follow a two-step approach:
we first regress stock returns on observable regressors to compute
residuals, and then search for latent common factors in the panel
of residuals (see next section). We can interpret condition (\ref{orth})
via an analogy with the partitioned regression: ${\displaystyle Y=X_{1}\beta_{1}+X_{2}\beta_{2}+\varepsilon}$.
The Frisch-Waugh-Lovell Theorem (\cite{Frish_Waugh_Frederick_1933},
\cite{Lovell_1963})\textbf{ }states that the ordinary least squares
(OLS) estimate of $\beta_{2}$ is identical to the OLS estimate of
$\beta_{2}$ in the regression $M_{X_{1}}Y=M_{X_{1}}X_{2}\beta_{2}+\eta$,
where $M_{X_{1}}=I_{n}-X_{1}\left(X_{1}^{\prime}X_{1}\right)^{-1}X_{1}^{\prime}$.
Condition (\ref{orth}) is tantamount to the orthogonality condition
$X_{1}^{\prime}X_{2}=0$ ensuring that we can estimate $\beta_{2}$
from regressing the residuals $M_{X_{1}}Y$ on $X_{2}$ only, instead
of the residuals $M_{X_{1}}X_{2}$ coming from the regression of $X_{2}$
on $X_{1}$. When condition (\ref{orth}) is not satisfied, joint
estimation of regression coefficients, latent factor betas and factor
values is required (see e.g. \cite{Bai_2009}, \cite{Moon_Weidner_2015}
in a model with homogeneous regression coefficients $\beta_{i}=\beta$
for all $i$, and \cite{Ando_Bai_2015} for heterogeneous $\beta_{i}$
in balanced panels). If the regressors are common across stocks, i.e.,
$x_{i,t}=x_{t}$, we can obtain condition (\ref{orth}) by transformation
of the latent factors. It simply corresponds to an identification
restriction on the latent factors, and is then not an assumption.
If the regressors are stock-specific, ensuring orthogonality between
the latent factors $h_{t}$ and the observable regressors $x_{i,t}$
for all $i$ is more than an identification restriction. It requires
an additional assumption where we decompose common and stock-specific
components in the regressors vector by writing $x_{i,t}=(x_{t}',\tilde{x}_{i,t}')'$,
where $x_{t}:=(vech[X_{t}]',f_{t}'\otimes Z_{t-1}')'$ and $\tilde{x}_{i,t}:=(Z_{t-1}'\otimes Z_{i,t-1}',f_{t}'\otimes Z_{i,t-1}')'$.

\begin{assumption} \label{A1-1} The best linear prediction of the
unobservable factor $EL(h_{t}\vert\{x_{i,t},\ i=1,2,...\})$ is independent
of $\{\tilde{x}_{i,t}$, $i=1,2,...\}$. \end{assumption}

\noindent Assumption \ref{A1-1} amounts to contemporaneous Granger
non-causality from the stock-specific regressors to the latent factors,
conditionally on the common regressors. Assumption \ref{A1-1} is
verified e.g. if the latent factors are independent of the lagged
stock-specific instruments, conditional on the observable factors
and the lagged common instruments (see the supplementary materials
for a derivation). We keep Assumption \ref{A1-1} as a maintained
assumption on the factor structure under $\mathcal{M}_{2}$. Under
Assumption \ref{A1-1}, $EL(h_{t}\vert\{x_{i,t},\ i=1,2,...\})=:\Psi x_{t}$
is a linear function of $x_{t}$. Therefore, by transformation of
the latent factor $h_{t}\rightarrow h_{t}-\Psi x_{t},$ we can assume
that $EL(h_{t}\vert\{x_{i,t},\ i=1,2,...\})=0$, without loss of generality.
This condition implies (\ref{orth}).

\section{Diagnostic criterion \label{sec:Diagnostic-criterion}}

In this section, we provide the diagnostic criterion that checks whether
the error terms are weakly cross-sectionally correlated or share at
least one common factor. To compute the criterion, we estimate the
generic panel model (\ref{LinearFactorRegression}) by OLS applied
asset by asset, and we get estimators ${\displaystyle \hat{\beta}_{i}=\hat{Q}_{x,i}^{-1}\frac{1}{T_{i}}\sum_{t}I_{i,t}x_{i,t}R_{i,t},}$
for $i=1,...,n$, where ${\displaystyle \hat{Q}_{x,i}=\frac{1}{T_{i}}\sum_{t}I_{i,t}x_{i,t}x_{i,t}^{\prime}}$.
We get the residuals $\hat{\varepsilon}_{i,t}=R_{i,t}-x_{i,t}^{\prime}\hat{\beta}_{i}$,
where ${\displaystyle \hat{\varepsilon}}_{i,t}$ is observable only
if ${\displaystyle I_{i,t}=1}$. In available panels, the random sample
size ${\displaystyle T_{i}}$ for asset $i$ can be small, and the
inversion of matrix $\hat{Q}_{x,i}$ can be numerically unstable.
To avoid unreliable estimates of $\beta_{i}$, we apply a trimming
approach as in GOS. We define $\boldsymbol{1}_{i}^{\chi}=\boldsymbol{1}\left\{ CN\left(\hat{Q}_{x,i}\right)\leq\chi_{1,T},\tau_{i,T}\leq\chi_{2,T}\right\} $,
where ${\displaystyle CN\left(\hat{Q}_{x,i}\right)=\sqrt{\mu_{1}\left(\hat{Q}_{x,i}\right)/\mu_{d}\left(\hat{Q}_{x,i}\right)}}$
is the condition number of the $d\times d$ matrix $\hat{Q}_{x,i}$,
$\mu_{d}\left(\hat{Q}_{x,i}\right)$ is its smallest eigenvalue and
$\tau_{i,T}=T/T_{i}$. The two sequences $\chi_{1,T}>0$ and $\chi_{2,T}>0$
diverge asymptotically (Assumption \ref{A_Ass_CHI}). The first trimming
condition $\{CN\left(\hat{Q}_{x,i}\right)\leq\chi_{1,T}\}$ keeps
in the cross-section only assets for which the time-series regression
is not too badly conditioned. A too large value of $CN\left(\hat{Q}_{x,i}\right)$
indicates multicollinearity problems and ill-conditioning (\cite{Belsley_Kuh_Welsch_2004},
\cite{Greene_2008}). The second trimming condition $\{\tau_{i,T}\leq\chi_{2,T}\}$
keeps in the cross-section only assets for which the time series is
not too short. We also use both trimming conditions in the proofs
of the asymptotic results.

We consider the following diagnostic criterion: 
\begin{equation}
\xi=\mu_{1}\left(\frac{1}{nT}\sum_{i}\boldsymbol{1}_{i}^{\chi}\bar{\varepsilon}_{i}\bar{\varepsilon}_{i}^{\prime}\right)-g(n,T),\label{Criterion}
\end{equation}
where the vector $\bar{\varepsilon}_{i}$ of dimension $T$ gathers
the values $\bar{\varepsilon}_{i,t}=I_{i,t}\hat{\varepsilon}_{i,t}$,
the penalty $g(n,T)$ is such that $g(n,T)\rightarrow0$ and $C_{n,T}^{2}g(n,T)\rightarrow\infty$,
when $n,T\rightarrow\infty$, for $C_{n,T}^{2}=\min\{n,T\}$. \cite{Bai_Ng_2002}
consider several simple potential candidates for the penalty $g(n,T)$.
We discuss them in Section \ref{sec:Implementation}. In vector $\bar{\varepsilon}_{i}$,
the unavailable residuals are replaced by zeros. We use the following
assumption on $n$ and $T$.

\begin{assumption} \label{A1nT} The cross-sectional dimension $n$
and time series dimension $T$ are such that $n=O(T^{\bar{\gamma}})$,
$\bar{\gamma}>0$, and $T=O(n^{\gamma})$, $\gamma\in(0,1]$. \end{assumption}

The following model selection rule explains our choice of the diagnostic
criterion (\ref{Criterion}) for approximate factor structure in large
unbalanced cross-sectional datasets.

\begin{proposition}\label{Prop_ModelSelectionRule}Model selection
rule: We select ${\cal M}_{1}$ if $\xi<0$, and we select $\mathcal{M}_{2}$
if $\xi>0$, since under Assumptions \ref{ass_ftheta}-\ref{A1nT}
and Assumptions \ref{A_Ass_ErrTerm_u_M2}-\ref{A_Ass_CHI}, (a) ${\displaystyle Pr}\left(\xi<0\mid{\cal M}_{1}\right)\rightarrow1$,
and (b) ${\displaystyle Pr}\left(\xi>0\mid{\cal M}_{2}\right)\rightarrow1$,
when $n,T\rightarrow\infty$.\end{proposition}

\ref{Prop_ModelSelectionRule} characterizes an asymptotically valid
model selection rule, which treats both models symmetrically. The
model selection rule is valid since parts (a) and (b) of \ref{Prop_ModelSelectionRule}
imply ${\displaystyle Pr\left({\cal M}_{1}|\xi<0\right)=}$\linebreak{}
 ${\displaystyle Pr\left(\xi<0|{\cal M}_{1}\right)Pr\left({\cal M}_{1}\right)\left[Pr\left(\xi<0|{\cal M}_{1}\right)Pr\left({\cal M}_{1}\right)+Pr\left(\xi<0|{\cal M}_{2}\right)Pr\left({\cal M}_{2}\right)\right]^{-1}\rightarrow1,}$
as $n,T\rightarrow\infty$, by Bayes Theorem. Similarly, we have ${\displaystyle Pr\left({\cal M}_{2}|\xi>0\right)\rightarrow1}$.
The diagnostic criterion in \ref{Prop_ModelSelectionRule} is not
a testing procedure since we do not use a critical region based on
an asymptotic distribution and a chosen significance level. The zero
threshold corresponds to an implicit critical value yielding a test
size asymptotically equal to zero since $Pr(\xi<0|{\cal M}_{1})\rightarrow1$.
The selection procedure is conservative in diagnosing zero factor
by construction. We do not allow type I error under ${\cal M}_{1}$
asymptotically, and really want to ensure that there is no omitted
factor as required in the APT setting. This also means that we will
not suffer from false discoveries related to a multiple testing problem
(see e.g. \cite{Barras_Scaillet_Wermers_2010}, \cite{Harvey_Liu_Zhu_2015})
in our empirical application where we consider a large variety of
factor models on monthly and quarterly data. However, a possibility
to achieve $p$-values is to use a randomisation procedure as in Trapani
(2017)\nocite{Trapani_2017} (see \cite{Bandi_Corradi_2014} and \cite{Corradi_Swanson_2006}
for recent applications in econometrics). This type of procedure controls
for an error of the first type, conditional on the information provided
by the sample and under a randomness induced by auxiliary experiments.

The proof of \ref{Prop_ModelSelectionRule} shows that the largest
eigenvalue in (\ref{Criterion}) vanishes at a faster rate (\ref{Lemma_L1}
in Appendix \ref{Appendix_ProofProp1}) than the penalization term
under ${\cal M}_{1}$ when $n$ and $T$ go to infinity. Under ${\cal M}_{1}$,
we expect a vanishing largest eigenvalue because of a lack of a common
signal in the error terms. The negative penalizing term $-g(n,T)$
dominates in (\ref{Criterion}), and this explains why we select the
first model when $\xi$ is negative. On the contrary, the largest
eigenvalue remains bounded from below away from zero (\ref{Lemma_L4}
in Appendix \ref{Appendix_ProofProp1}) under ${\cal M}_{2}$ when
$n$ and $T$ go to infinity. Under ${\cal M}_{2}$, we have at least
one non vanishing eigenvalue because of a common signal due to omitted
factors. The largest eigenvalue dominates in (\ref{Criterion}), and
this explains why we select the second model when $\xi$ is positive.
We can interpret the criterion (\ref{Criterion}) as the adjusted
gain in fit including a single additional (unobservable) factor in
model ${\cal M}_{1}$. We can rewrite (\ref{Criterion}) as ${\displaystyle \xi=SS_{0}-SS_{1}-g\left(n,T\right)},$
where ${\displaystyle SS_{0}=\frac{1}{nT}\sum_{i}\sum_{t}\boldsymbol{1}_{i}^{\chi}\bar{\varepsilon}_{i,t}^{2}}$
is the sum of squared errors and ${\displaystyle SS_{1}=\min\frac{1}{nT}\sum_{i}\sum_{t}\boldsymbol{1}_{i}^{\chi}\left(\bar{\varepsilon}_{i,t}-\theta_{i}h_{t}\right)^{2},}$
where the minimization is w.r.t. the vectors $H\in\mathbb{R}^{T}$
of factor values and $\Theta=(\theta_{1},...,\theta_{n})'\in\mathbb{R}^{n}$
of factor loadings in a one-factor model, subject to the normalization
constraint ${\displaystyle \frac{H^{\prime}H}{T}=1}.$ Indeed, the
largest eigenvalue ${\displaystyle \mu_{1}\left(\frac{1}{nT}\sum_{i}\boldsymbol{1}_{i}^{\chi}\bar{\varepsilon}_{i}\bar{\varepsilon}_{i}^{\prime}\right)}$
corresponds to the difference between $SS_{0}$ and $SS_{1}$. Furthermore,
the criterion $\xi$ is equal to the difference of the penalized criteria
for zero- and one-factor models defined in \cite{Bai_Ng_2002} applied
on the residuals. Indeed, ${\displaystyle \xi=PC\left(0\right)-PC\left(1\right),}$
where ${\displaystyle PC\left(0\right)=SS_{0}},$ and ${\displaystyle PC\left(1\right)=SS_{1}+g\left(n,T\right).}$

\ref{Lemma_L1} in Appendix \ref{Appendix_ProofProp1} gives an asymptotic
upper bound on the largest eigenvalue of a symmetric matrix based
on similar arguments as in \cite{Geman_1980}, \cite{Yin_Bai_Krishnaiah_1988},
and \cite{Bai_Yin_1993} without exploiting distributional results
from random matrix theory valid when $n$ is comparable with $T$.
This exemplifies a key difference with the proportional asymptotics
used in \cite{Onatski_2010} or \cite{Ahn_Horenstein_2013} for balanced
panel without observable factors. In \ref{Prop_ModelSelectionRule},
when $\gamma<1$, the condition $T/n=o\left(1\right)$ agrees with
the ``large $n$, small $T$'' case that we face in the empirical
application (ten thousand individual stocks monitored over forty-five
years of either monthly, or quarterly, returns). Another key difference
w.r.t. the available literature is the handling of unbalanced panels.
We need to address explicitly the presence of the observability indicators
$I_{i,t}$ and the trimming devices $\boldsymbol{1}_{i}^{\chi}$ in
the proofs of the asymptotic results.

The recent literature on the properties of the two-pass regressions
for fixed $n$ and large $T$ shows that the presence of useless factors
(Kan and Zhang (1999a,b)\nocite{Kan_Zhang_1999a}\nocite{Kan_Zhang_1999b},
\cite{Gospodinov_Kan_Robotti_2014}) or weak factor loadings (\cite{Kleibergen_2009})
does not affect the asymptotic distributional properties of factor
loading estimates, but alters the ones of the risk premia estimates.
Useless factors have zero loadings, and weak loadings drift to zero
at rate $1/\sqrt{T}$. The vanishing rate of the largest eigenvalue
of the empirical cross-sectional covariance matrix of the residuals
does not change if we face useless factors or weak factor loadings
in the observable factors under ${\cal M}_{1}$. The same remark applies
under ${\cal M}_{2}$. Hence the selection rule remains the same since
the probability of taking the right decision still approaches 1. If
we have a number of useless factors or weak factor loadings strictly
smaller than the number $m$ of the omitted factors under ${\cal M}_{2}$,
this does not impact the asymptotic rate of the diagnostic criterion
if Assumption \ref{ass_ftheta} holds. If we only have useless factors
in the omitted factors under ${\cal M}_{2}$, we face an identification
issue. Assumption \ref{ass_ftheta} (ii) is not satisfied. We cannot
distinguish such a specification from ${\cal M}_{1}$ since it corresponds
to a particular approximate factor structure. Again the selection
rule remains the same since the probability of taking the right decision
still approaches 1. Finally, let us study the case of only weak factor
loadings under ${\cal M}_{2}$. We consider a simplified setting:
\[
R_{i,t}=x_{i,t}^{\prime}\beta_{i}+\varepsilon_{i,t}
\]
where $\varepsilon_{i,t}=\theta_{i}h_{t}+u_{i,t}$ has only one factor
with a weak factor loading, namely $m=1$ and $\theta_{i}=\bar{\theta}_{i}/T^{c}$
with $c>0$. Let us assume that ${\displaystyle \frac{1}{n}\sum_{i}\bar{\theta}_{i}^{2}}$
is bounded from below away from zero (see Assumption \ref{ass_ftheta}
(ii)) and bounded from above. By the properties of the eigenvalues
of a scalar multiple of a matrix, we deduce that ${\displaystyle c_{1}/T^{2c}\leq\mu_{1}\left(\frac{1}{nT}\sum_{i}\theta_{i}^{2}HH^{\prime}\right)\leq c_{2}/T^{2c}}$,
$w.p.a.\ 1$, for some constants $c_{1},c_{2}$ such that $c_{2}\geq c_{1}>0$.
Hence, by similar arguments as in the proof of \ref{Prop_ModelSelectionRule},
we get: 
\[
c_{1}T^{-2c}-g(n,T)+O_{p}\left(C_{nT}^{-2}+\bar{\chi}_{T}T^{-1}\right)\leq\xi\leq c_{2}T^{-2c}-g(n,T)+O_{p}\left(C_{nT}^{-2}+\bar{\chi}_{T}T^{-1}\right),
\]
where we define $\bar{\chi}_{T}=\chi_{1,T}^{4}\chi_{2,T}^{2}$. To
conclude ${\cal M}_{2}$, we need that $C_{nT}^{-2}+\bar{\chi}_{T}T^{-1}$
and the penalty $g(n,T)$ vanish at a faster rate than $T^{-2c}$,
namely $C_{nT}^{-2}+\bar{\chi}_{T}T^{-1}=o\left(T^{-2c}\right)$ and
$g(n,T)=o\left(T^{-2c}\right)$. To conclude ${\cal M}_{1}$, we need
that $g(n,T)$ is the dominant term, namely $T^{-2c}=o\left(g(n,T)\right)$
and $C_{nT}^{-2}+\bar{\chi}_{T}T^{-1}=o\left(g(n,T)\right)$. As an
example, let us take $g(n,T)=T^{-1}\log T$ and $n=T^{\bar{\gamma}}$
with $\bar{\gamma}>1$, and assume that the trimming is such that
$\bar{\chi}_{T}=o(\log T)$. Then, we conclude ${\cal M}_{2}$ if
$c<1/2$ and ${\cal M}_{1}$ if $c>1/2$. This means that detecting
a weak factor loading structure is difficult if $c$ is not sufficiently
small. The factor loadings should drift to zero not too fast to conclude
${\cal M}_{2}$. Otherwise, we cannot distinguish it asymptotically
from weak cross-sectional correlation.

\section{Determining the number of factors\label{sec:DeterminingNumberFactors}}

In the previous section, we have studied a diagnostic criterion to
check whether the error terms are weakly cross-sectionally correlated
or share at least one unobservable common factor. This section aims
at answering: do we have one, two, or more omitted factors? The design
of the diagnostic criterion to check whether the error terms share
exactly $k$ unobservable common factors or share at least $k+1$
unobservable common factors follows the same mechanics. We consider
the following rival models: 
\begin{align*}
\mathcal{M}_{1}\left(k\right): & \quad\text{\mbox{the linear regression model (\ref{LinearFactorRegression}), where the errors }(\ensuremath{\varepsilon_{i,t}})\mbox{ satisfy a factor structure} }\\
 & \mbox{\quad with exactly }\ensuremath{k}\mbox{ unobservable factors,}
\end{align*}
and 
\begin{align*}
\mathcal{M}_{2}(k): & \quad\text{\mbox{the linear regression model (\ref{LinearFactorRegression}), where the errors }(\ensuremath{\varepsilon_{i,t}})\mbox{ satisfy a factor structure}}\\
 & \quad\mbox{with at least }k+1\mbox{ unobservable factors.}
\end{align*}
The above definitions yield ${\cal M}_{1}={\cal M}_{1}\left(0\right)$
and ${\cal M}_{2}={\cal M}_{2}\left(0\right)$.%

\begin{assumption} \label{ass3} Under model $\mathcal{M}_{2}(k)$,
we have ${\displaystyle \mu_{k+1}\left(\frac{1}{n}\sum_{i}\theta_{i}\theta_{i}'\right)\geq C}$,
w.p.a. $1$ as $n\rightarrow\infty$, for a constant $C>0$, where
$\mu_{k+1}\left(.\right)$ denotes the $\left(k+1\right)$-th largest
eigenvalue of a symmetric matrix. \end{assumption}

\noindent Models $\mathcal{M}_{1}(k)$ and $\mathcal{M}_{2}(k)$ with
$k\geq1$ are subsets of model $\mathcal{M}_{2}$. Hence, Assumption
\ref{ass_ftheta} (i) guarantees the convergence of matrix ${\displaystyle \frac{1}{T}\sum_{t}h_{t}h_{t}'}$
to a positive definite $k\times k$ matrix under $\mathcal{M}_{1}(k)$,
and to a positive definite $m\times m$ matrix under $\mathcal{M}_{2}(k)$,
with $m\geq k+1$. Assumption \ref{ass3} requires that there are
at least $k+1$ strong factors under $\mathcal{M}_{2}(k)$.

The diagnostic criterion exploits the $\left(k+1\right)$th largest
eigenvalue of the empirical cross-sectional covariance matrix of the
residuals:

\begin{equation}
\xi(k)=\mu_{k+1}\left(\frac{1}{nT}\sum_{i}\boldsymbol{1}_{i}^{\chi}\bar{\varepsilon}_{i}\bar{\varepsilon}_{i}^{\prime}\right)-g(n,T).\label{xik}
\end{equation}

As discussed in \cite{Ahn_Horenstein_2013} (see also \cite{Onatski_2013})
for balanced panels, we can rewrite (\ref{xik}) as $\xi(k)=SS_{k}-SS_{k+1}-g(n,T)$
where ${\displaystyle SS_{k}=\min\frac{1}{nT}\sum_{i}\sum_{t}\boldsymbol{1}_{i}^{\chi}\left(\bar{\varepsilon}_{i,t}-\theta_{i}'h_{t}\right)^{2}}$
and the minimization is w.r.t. $H\in\mathbb{R}^{T\times k}$ and $\Theta=(\theta_{1},...,\theta_{n})'\in\mathbb{R}^{n\times k}$.
The criterion $\xi(k)$ is equal to the difference of the penalized
criteria for $k$ and $(k+1)$-factor models defined in \cite{Bai_Ng_2002}
applied on the residuals. Indeed, $\xi(k)=PC(k)-PC(k+1)$, where $PC(k)=SS_{k}+kg(n,T)$
and $PC(k+1)=SS_{k+1}+(k+1)g(n,T)$.

The following model selection rule extends \ref{Prop_ModelSelectionRule}.

\begin{proposition} \label{Prop_propgeneral} Model selection rule:
We select ${\cal M}_{1}(k)$ if $\xi(k)<0$, and we select ${\cal M}_{2}(k)$
if $\xi(k)>0$, since under Assumptions \ref{ass_ftheta}(i), \ref{A1-1}-\ref{ass3},
and Assumptions \ref{A_Ass_ErrTerm_u_M2}-\ref{A_Ass_boundII}, (a)
$Pr[\xi(k)<0\vert{\cal M}_{1}(k)]\rightarrow1$ and (b) $Pr[\xi(k)>0\vert{\cal M}_{2}(k)]\rightarrow1$,
when $n,T\rightarrow\infty$. \end{proposition}

The proof of \ref{Prop_propgeneral} is more complicated than the
proof of \ref{Prop_ModelSelectionRule}. We need additional arguments
to derive an asymptotic upper bound when we look at the ($k+1$)th
eigenvalue of a symmetric matrix (\ref{lemma_L4} in Appendix \ref{Appendix_ProofProp3}).
We rely on the Courant-Fischer min-max theorem and Courant-Fischer
formula (see beginning of \ref{Appendix_Proof}) which represent eigenvalues
as solutions of constrained quadratic optimization problems. We know
that the largest eigenvalue $\mu_{1}(A)$ of a symmetric positive
semi-definite matrix $A$ is equal to its operator norm. There is
no such useful norm interpretation for the smaller eigenvalues $\mu_{k}(A)$,
$k\geq2$. We cannot directly exploit standard inequalities or bounds
associated to a norm when we investigate the asymptotic behavior of
the spectrum beyond its largest element. We cannot either exploit
distributional results from random matrix theory since we also allow
for $T/n=o(1)$. The slow convergence rate $\sqrt{T}$ for the individual
estimates $\hat{\beta_{i}}$ also complicates the proof. In the presence
of homogeneous regression coefficients $\beta_{i}=\beta$ for all
$i$, the estimate $\hat{\beta}$ in \cite{Bai_2009} and \cite{Moon_Weidner_2015}
has a fast convergence rate $\sqrt{nT}$. In that case, controlling
for the estimation error in $\hat{\varepsilon}_{i,t}=\varepsilon_{i,t}+x_{i,t}^{\prime}(\beta-\hat{\beta})$
is straightforward due to the small asymptotic contribution of $(\beta-\hat{\beta})$.
The approach of \cite{Onatski_2010} requires the convergence of the
upper edge of the spectrum (i.e., the first $k$ largest eigenvalues
of the covariance matrix, with $k/T=o(1)$) to a constant, while the
approach of \cite{Ahn_Horenstein_2013} requires an asymptotic lower
bound on the eigenvalues. Extending these approaches for residuals
of an unbalanced panel when $T/n=o(1)$ looks challenging.

We can use the results of \ref{Prop_propgeneral} in order to estimate
the number of unobservable factors. It suffices to choose the minimum
$k$ such that $\xi(k)<0$. The next proposition states the consistency
of that estimate even in the presence of a degenerate distribution
of the eigenvalues.

\begin{proposition} \label{Prop_choice} Let $\hat{k}=\min\left\{ k=0,1,...,T-1\ :\ \xi(k)<0\right\} $,
where $\hat{k}=T$ if $\xi(k)\geq0$ for all $k\leq T-1$. Then, under
Assumptions \ref{ass_ftheta}(i), \ref{A1-1}-\ref{ass3}, and Assumptions
\ref{A_Ass_ErrTerm_u_M2}-\ref{A_Ass_boundII}, and under $\mathcal{M}_{1}(k_{0})$,
we have $P[\hat{k}=k_{0}]\rightarrow1$, as $n,T\rightarrow\infty$.
\end{proposition}

In \ref{Prop_choice}, we do not need to give conditions on the growth
rate of the maximum possible number $kmax$ of factors as in \cite{Onatski_2010}
and \cite{Ahn_Horenstein_2013}. We believe that this is a strong
advantage since there are many possible choices for $kmax$ and the
estimated number of factors is sometimes sensitive to the choice of
$kmax$ (see the simulation results in those papers). In the online
supplementary materials, we show that our procedure selects the right
number of factors with an observed $100$ percent probability in most
Monte Carlo experiments when $n$ is comparable or much larger than
$T$.

\section{Implementation and graphical diagnostic tool \label{sec:Implementation}}

In this section, we discuss how we can implement the model selection
rule in practice and design simple graphical diagnostic tools to determine
the number of unobservable factors (see Figures \ref{fig_numOmitFactCAPM}-\ref{fig_numOmitFactCCAPM}
in the next section). Let us first recognize that 
\[
{\displaystyle \hat{\sigma}^{2}=\frac{1}{nT}\sum_{i}\sum_{t}\boldsymbol{1}_{i}^{\chi}\bar{\varepsilon}_{i,t}^{2}=tr\left(\frac{1}{nT}\sum_{i}\boldsymbol{1}_{i}^{\chi}\bar{\varepsilon}_{i}\bar{\varepsilon}_{i}^{\prime}\right)=\sum_{j=1}^{T}\mu_{j}\left(\frac{1}{nT}\sum_{i}\boldsymbol{1}_{i}^{\chi}\bar{\varepsilon}_{i}\bar{\varepsilon}_{i}^{\prime}\right)}.
\]
The ratio ${\displaystyle \mu_{j}\left(\frac{1}{nT}\sum_{i}\boldsymbol{1}_{i}^{\chi}\bar{\varepsilon}_{i}\bar{\varepsilon}_{i}^{\prime}\right)/\hat{\sigma}^{2}}$
gauges the contribution of the $j$th eigenvalue in percentage of
the variance $\hat{\sigma}^{2}$ of the residuals. Similarly, the
sum ${\displaystyle \sum_{j=1}^{k}\mu_{j}\left(\frac{1}{nT}\sum_{i}\boldsymbol{1}_{i}^{\chi}\bar{\varepsilon}_{i}\bar{\varepsilon}_{i}^{\prime}\right)/\hat{\sigma}^{2}}$
gauges the cumulated contribution of the $k$ largest eigenvalues
in percentage of $\hat{\sigma}^{2}$. From \ref{Prop_propgeneral},
when all eigenvalues in that sum are larger than $g(n,T)$, this is
equal to the percentage of $\hat{\sigma}^{2}$ explained by the $k$
unobservable factors. Therefore, we suggest to work in practice with
rescaled eigenvalues which are more informative. We can easily build
a scree plot where we display the rescaled eigenvalues ${\displaystyle \mu_{j}\left(\frac{1}{nT}\sum_{i}\boldsymbol{1}_{i}^{\chi}\bar{\varepsilon}_{i}\bar{\varepsilon}_{i}^{\prime}\right)/\hat{\sigma}^{2}}$
in descending order versus the number of omitted factors $k$, and
use the horizontal line set at $g(n,T)/\hat{\sigma}^{2}$ as the cut-off
point to determine the number of omitted factors. This yields exactly
the same choice as the one in \ref{Prop_choice}. The asymptotic validity
of the selection rule in unaffected since $\hat{\sigma}^{2}$ converges
to a strictly positive constant when $n,T\rightarrow\infty$. Such
a scree plot helps to visually assess which unobservable factors,
if needed, explain most of the variability in the residuals. We can
set ${\displaystyle g(n,T)/\hat{\sigma}^{2}=\left(\frac{n+T}{nT}\right)\ln\left(\frac{nT}{n+T}\right)}$
following a suggestion in \cite{Bai_Ng_2002}. Those authors propose
two other potential choices ${\displaystyle \left(\frac{n+T}{nT}\right)\ln C_{nT}^{2}}$
and ${\displaystyle \left(\frac{\ln C_{nT}^{2}}{C_{nT}^{2}}\right)}$.
In our empirical application, $n$ is much larger than $T$, and they
yield identical results.

From Section \ref{sec:Diagnostic-criterion}, we know that $\xi=SS_{0}-SS_{1}-g(n,T)$.
Given such an interpretation in terms of sums of squared errors, we
can think about another diagnostic criterion based on a logarithmic
version $\check{\xi}$ as in Corollary 2 of \cite{Bai_Ng_2002}. The
second diagnostic criterion is 
\begin{equation}
{\displaystyle \check{\xi}=\ln\left(\hat{\sigma}^{2}\right)-\ln\left(\hat{\sigma}^{2}-\mu_{1}\left(\frac{1}{nT}\sum_{i}\boldsymbol{1}_{i}^{\chi}\bar{\varepsilon}_{i}\bar{\varepsilon}_{i}^{\prime}\right)\right)-g(n,T).}\label{LogCriteria}
\end{equation}
We get $\hat{\sigma}^{2}=SS_{0}$, and ${\displaystyle \check{\xi}=\ln(SS_{0}/SS_{1})-g(n,T)}$
is equal to the difference of $IC\left(0\right)$ and $IC\left(1\right)$
criteria in \cite{Bai_Ng_2002}. Then, the model selection rule is
the same as in \ref{Prop_ModelSelectionRule} with $\check{\xi}$
substituted for $\xi$. For the logarithmic version, \cite{Bai_Ng_2002}
suggest to use the penalty \break ${\displaystyle g(n,T)=\left(\frac{n+T}{nT}\right)\ln\left(\frac{nT}{n+T}\right)}$
since the scaling by $\hat{\sigma}^{2}$ is implicitly performed by
the logarithmic transformation of $SS_{0}$ and $SS_{1}$. Since,
from Equation (\ref{LogCriteria}), ${\displaystyle \check{\xi}=\ln\left(1/\left(1-\mu_{1}\left(\frac{1}{nT}\sum_{i}\boldsymbol{1}_{i}^{\chi}\bar{\varepsilon}_{i}\bar{\varepsilon}_{i}^{\prime}\right)/\hat{\sigma}^{2}\right)\right)}$\linebreak{}
${\displaystyle -g(n,T)}$ and $x$ is close to $\ln(1/(1-x))$ for
a small $x$, we see that a rule based on the rescaled criterion $\xi/\hat{\sigma}^{2}$
is closely related to the logarithmic version when the rescaled eigenvalue
is small. This further explains why we are in favour of working in
practice with rescaled eigenvalues.

Prior to computation of the eigenvalues, \cite{Bai_Ng_2002} advocate
each series to be demeaned and standardized to have unit variance
(see also Section 4 in \cite{King_Sentana_Wadhwani_1994}). In our
setting, each time series of residuals $\bar{\varepsilon}_{i,t}$
have zero mean by construction, and we also standardize them to have
unit variance over the sample of $T$ observations before computing
the eigenvalues. Working with ${\displaystyle {\bar{\bar{\varepsilon}}}_{i,t}=\bar{\varepsilon}_{i,t}/\sqrt{\frac{1}{T}\sum_{t}\bar{\varepsilon}_{i,t}^{2}}}$
ensures that all series of residuals have a common scale of measurement
and improves the stability of the information extracted from the multivariate
time series (see e.g.\ \cite{Pena_Poncela_2006}). Since ${\displaystyle tr\left(\sum_{i}\boldsymbol{1}_{i}^{\chi}\bar{\bar{\varepsilon}}_{i}\bar{\bar{\varepsilon}}_{i}^{\prime}\right)=n^{\chi}T}$
with ${\displaystyle n^{\chi}=\sum_{i}\boldsymbol{1}_{i}^{\chi}}$,
we suggest to work with the normalised matrix ${\displaystyle \frac{1}{n^{\chi}T}\sum_{i}\boldsymbol{1}_{i}^{\chi}\bar{\bar{\varepsilon}}_{i}\bar{\bar{\varepsilon}}_{i}^{\prime}}$,
so that the variance ${\displaystyle \frac{1}{n^{\chi}T}\sum_{i}\sum_{t}\boldsymbol{1}_{i}^{\chi}\bar{\bar{\varepsilon}}_{i,t}^{2}}$
of the scaled residuals is 1 by construction, and we can interpret
${\displaystyle \mu_{j}\left(\frac{1}{n^{\chi}T}\sum_{i}\boldsymbol{1}_{i}^{\chi}\bar{\bar{\varepsilon}}_{i}\bar{\bar{\varepsilon}}_{i}^{\prime}\right)}$
directly as percentage of the variance of the normalised residuals.

From \cite{Johnston_2001}, we know that for a matrix of residuals,
all of whose entries are independent standard Gaussian variates in
a balanced panel, the distribution of the largest eigenvalue of the
corresponding Wishart variable suitably normalized approaches the
Tracy-Widom law of order 1 under proportional asymptotics. That result
implies that, for such standard Gaussian residuals, the largest eigenvalue
that we compute should be approximately $1/T$ if $T$ is smaller
than $n$ (see also \cite{Geman_1980}) without the need to rely on
a scaling by an estimated variance $\hat{\sigma}^{2}$. This further
explains why we are in favor of working with standardised residuals,
so that we are as close as possible to a standardized Gaussian reference
model. This is akin to use the standard rule of thumb based on a Gaussian
reference model in nonparametric density estimation (\cite{Silverman_1986}).
We know the rate of convergence of the kernel density estimate but
need an idea of the constant to use that information for practical
bandwidth choice. In our setting, we can set the constant to one,
when we face independent standard Gaussian residuals. The Gaussian
reference model also suggests to use the penalisation ${\displaystyle g(n,T)=\frac{\left(\sqrt{n}+\sqrt{T}\right)^{2}}{nT}\ln\left(\frac{nT}{\left(\sqrt{n}+\sqrt{T}\right)^{2}}\right)}$.
This is our choice in the empirical section with $n^{\chi}$ substituted
for $n$, and a data-driven constant substituted for the known constant
$1$ of the Gaussian reference model (see the Monte Carlo section
for a detailed explanation of the selection method based on the proposal
of \citet{Alessi_Barigozzi_Capasso_2010}; see also \cite{Hallin_Liska_2007}
in the general dynamic factor model). We show the good performance
of such a rule in the Monte Carlo results for unbalanced panels.

Finally, we can also investigate the ratio ${\displaystyle \mu_{j}^{2}\left(\frac{1}{nT}\sum_{i}\boldsymbol{1}_{i}^{\chi}\bar{\bar{\varepsilon}}_{i}\bar{\bar{\varepsilon}}_{i}^{\prime}\right)/\sum_{l=1}^{T}\mu_{l}^{2}\left(\frac{1}{nT}\sum_{i}\boldsymbol{1}_{i}^{\chi}\bar{\bar{\varepsilon}}_{i}\bar{\bar{\varepsilon}}_{i}^{\prime}\right)}$
and the cumulated contribution ${\displaystyle \sum_{j=1}^{k}\mu_{j}^{2}\left(\frac{1}{nT}\sum_{i}\boldsymbol{1}_{i}^{\chi}\bar{\bar{\varepsilon}}_{i}\bar{\bar{\varepsilon}}_{i}^{\prime}\right)/\sum_{l=1}^{T}\mu_{l}^{2}\left(\frac{1}{nT}\sum_{i}\boldsymbol{1}_{i}^{\chi}\bar{\bar{\varepsilon}}_{i}\bar{\bar{\varepsilon}}_{i}^{\prime}\right)}$
. The denominator corresponds to the square of the Frobenius (or Hilbert-Schmidt)
norm of the matrix ${\displaystyle \frac{1}{nT}\sum_{i}\boldsymbol{1}_{i}^{\chi}\bar{\bar{\varepsilon}}_{i}\bar{\bar{\varepsilon}}_{i}^{\prime}}$
since the sum of the squared eigenvalues of a positive semidefinite
symmetric matrix $A=(a_{ij})$ corresponds to ${\displaystyle tr(A'A)=\sum_{i,j}a_{ij}^{2}}$.
Those quantities measure the contributions of the omitted factors
in terms of the off-diagonal terms (correlation part) in addition
to the diagonal terms (residual variance). Here we follow Fiorentini
and Sentana (2015, pages 158-159)\nocite{Fiorentini_Sentana_2015}
who prefer to look at the fraction of the Frobenius norm instead of
the usual fraction of the trace of the sample covariance matrix to
judge the representativeness of principal components. \LyXZeroWidthSpace \cite{King_Sentana_Wadhwani_1994}\LyXZeroWidthSpace{}
use the Frobenius norm to decompose the sample covariance of residuals
of a Vector AutoRegressive (VAR) model and obtain starting values
for maximum likelihood estimation of the parameters of a factor model
for the error terms. A selection rule based on the squared eigenvalues
being above or below the squared penalty delivers the same diagnostic,
but helps to gauge the impact on correlation explanation by the omitted
factors.

\section{Empirical results \label{sec:Empirical-results}}

In this section, we compute the diagnostic criteria and the number
of omitted factors using a large variety of combinations of financial
and macroeconomic factors. We estimate linear factor models using
monthly and quarterly data from January 1968 to December 2011.

\subsection{Factor models and data description\label{sec:FactorMod_DataDescr}}

We consider several linear factor models that involve financial and
macroeconomic variables. Let us start with the financial specifications
listed in Table \ref{Tab_PureFinancial}. We estimate these linear
specifications using monthly data. We proxy the risk free rate with
the monthly 30-day T-bill beginning-of-month yield. The three factors
of \cite{Fama_French_1993} are the monthly excess return $r_{m,t}$
on CRSP NYSE/AMEX/Nasdaq value-weighted market portfolio over the
risk free rate, and the monthly returns on zero-investment factor-mimicking
portfolios for size and book-to-market, denoted by $r_{smb,t}$ and
$r_{hml,t}$. The monthly returns on portfolio for momentum is denoted
by $r_{mom,t}$ (\cite{Carhart_1997}). The two operative profitability
factors of \cite{Fama_French_2015} are the difference between monthly
returns on diversified portfolios with robust and weak profitability
and investments, and with low and high investment stocks, denoted
by $r_{rmw,t}$ and $r_{cma,t}$. We have downloaded the time series
of these factors from the website of Kenneth French. We denote the
monthly returns of size, investment, and profitability portfolios
introduced by \cite{Hou_Xue_Zhang_2015} by $r_{me,t}$, $r_{I/A,t}$
and $r_{ROE,t}$ (see also \cite{Hou_Xue_Zhang_2014}). Furthermore,
we include quality minus junk ($qmj_{t}$) and bet against beta (\textbf{
$bab_{t}$}) factors as described in \cite{Asness_Frazzini_Pedersen_2014}
and \cite{Frazzini_Pedersen_2014}. The factor return $qmj_{t}$ is
the average return on the two high quality portfolios minus the average
return on the two low quality (junk) portfolios. The bet against beta
factor is a portfolio that is long low-beta securities and short high-beta
securities. We have downloaded these data from the website of AQR.

As additional specifications, we consider the two reversal factors
which are monthly returns on portfolios for short-term and long-term
reversals from the website of Kenneth French. Besides, the monthly
returns of industry-adjusted value, momentum and profitability factors
are available from the website of Robert Novy-Marx (see \cite{NovyMarx_2013}).
We also include the three liquidity-related factors of \cite{Pastor_Stambaugh_2002}
that consist of monthly liquidity level, traded liquidity, and the
innovation in aggregate liquidity. We have downloaded them from the
website of Lubos Pastor.




In Table \ref{Tab_PureMacroeco}, we list the linear factor specifications
that involve financial and macroeconomic variables. We estimate these
specifications using quarterly data. We consider the aggregate consumption
growth $cg_{t}$ for the CCAPM (\cite{Lucas_1978}, \cite{Breeden_79})
and the \cite{Epstein_Zin_1989} model (see also \cite{Epstein_Zin_1991}),
the durable and nondurable-consumption growth rate introduced by \cite{Yogo_2006}
and denoted by $dcg_{t}$ and $ndcg_{t}$. The investment factors
used in \cite{Li_Vassalou_Xing_2006} track the changes in the gross
private investment for households, for non-financial corporate and
for non-financial non-corporate firms, and are denoted by $dhh_{t}$,
$dcorp_{t}$, and $dncorp_{t}$. Finally, we consider the five factors
of \cite{Chen_Roll_Ross_1986} available from the website of Laura
Xiaolei Liu. Those factors are the growth rate of industrial production
$mp_{t}$, the unexpected inflation $ui_{t}$, the change in the expected
inflation $dei_{t}$, the term spread $uts_{t}$, proxied by the difference
between yields on 10-year Treasury and 3-month T-bill, and the default
premia $upr_{t}$, proxied by the yield difference between Moody's
Baa-rated and Aaa-rated corporate bonds.

To account for time-varying coefficients, we consider two conditional
specifications: \break (i) $Z_{t-1}=\left(1,divY_{t-1}\right)'$
and (ii) $Z_{t-1}=\left(1,divY_{t-1}\right)'$, $Z_{i,t-1}=bm_{i,t-1}$,
where $divY_{t-1}$ is the lagged dividend yield and the asset specific
instrument $bm_{i,t-1}$ corresponds to the lagged book-to-market
equity of firm $i$. We compute the firm characteristic from Compustat
as in the appendix of \cite{Fama_French_2008}. We refer to \cite{Avramov_Chordia_2006}
for convincing theoretical and empirical arguments in favor of the
chosen conditional specifications. The parsimony and the empirical
results below explain why we have not included an additional firm
characteristic such as the size of firm $i$.

As additional specifications, we consider the lagged default spread,
term spread, monthly 30-day T-bill, aggregate consumption-to-wealth
ratio (\cite{Lettau_Ludvigson_2001}), and labour-to-consumption ratio
(\cite{Santos_Veronesi_2006}) as common instruments.

The CRSP database provides the monthly stock returns data and we exclude
financial firms (Standard Industrial Classification Codes between
6000 and 6999) as in \cite{Fama_French_2008}. The dataset after matching
CRSP and Compustat contents comprises $n=10,442$ stocks, and covers
the period from January 1968 to December 2011 with $T=546$ months.
We constructed the quarterly stock returns from the monthly data and
$T=176$. In order to account for the unbalanced characteristic, if
the monthly observability indicators $I_{i,t},I_{i,t+1}$ and $I_{i,t+2}$
are observed, we built the returns of the quarter $s=1,2,3,4$ as
the average of the three monthly returns at time $t,t+1$ and $t+2$.
Otherwise, the observability indicator of the quarter $s$ takes value
zero.

\subsection{Results for financial models\label{sec:EmpRes_PureFin}}

In this section, we compute the diagnostic criteria for the linear
factor models listed in Table \ref{Tab_PureFinancial}. We fix $\chi_{1,T}=15$
as advocated by \cite{Greene_2008} and $\chi_{2,T}=546/60$, i.e.,
at least 60 months of return observations as in \cite{Bai_Ng_2002}.
In Table \ref{Tab_nchi_PureFin_monthly}, we report the trimmed cross-sectional
dimension $n^{\chi}$. In some time-varying specifications, we face
severe multicollinearity problems due to the correlations within the
vector of regressors $x_{i,t}$, that involves cross-products of factors
$f_{t}$ and instruments $Z_{t-1}$. These problems explain why we
shrink from $n^{\chi}=6,775$ for time-invariant models to around
three thousand assets for time-varying models.

Table \ref{Tab_Results_timeInv_pureFin} reports the contribution
in percentage of the first eigenvalue $\mu_{1}$ with respect to the
variance of normalized residuals ${\displaystyle \frac{1}{n^{\chi}T}\sum_{i}\boldsymbol{1}_{i}^{\chi}\bar{\bar{\varepsilon}}_{i}\bar{\bar{\varepsilon}}_{i}^{\prime}}$,
that is equal to one by construction under our variance scaling to
one for each time series of residuals. We also report the number of
omitted factors $k$, the contribution of the first $k$ eigenvalues,
i.e., ${\displaystyle \sum_{j=1}^{k}\mu_{j}}$, and the incremental
contribution of the $\left(k+1\right)$-th eigenvalue $\mu_{k+1}$.
For each model, we also specify the numerical value of the penalisation
function $g\left(n^{\chi},T\right)$, as defined in Section \ref{sec:Implementation}.

\begin{figure}[H]
\textbf{\protect\caption{\textbf{\label{fig_numOmitFactCAPM}Number of omitted factors and
cumulated eigenvalues for the CAPM}}
}

\begin{centering}
{\small{}{}{}}%
\begin{tabular}{c}
{\small{}{}{}Time-invariant CAPM} \tabularnewline
\hspace{-2cm} {\small{}{}{}\includegraphics[scale=0.35]{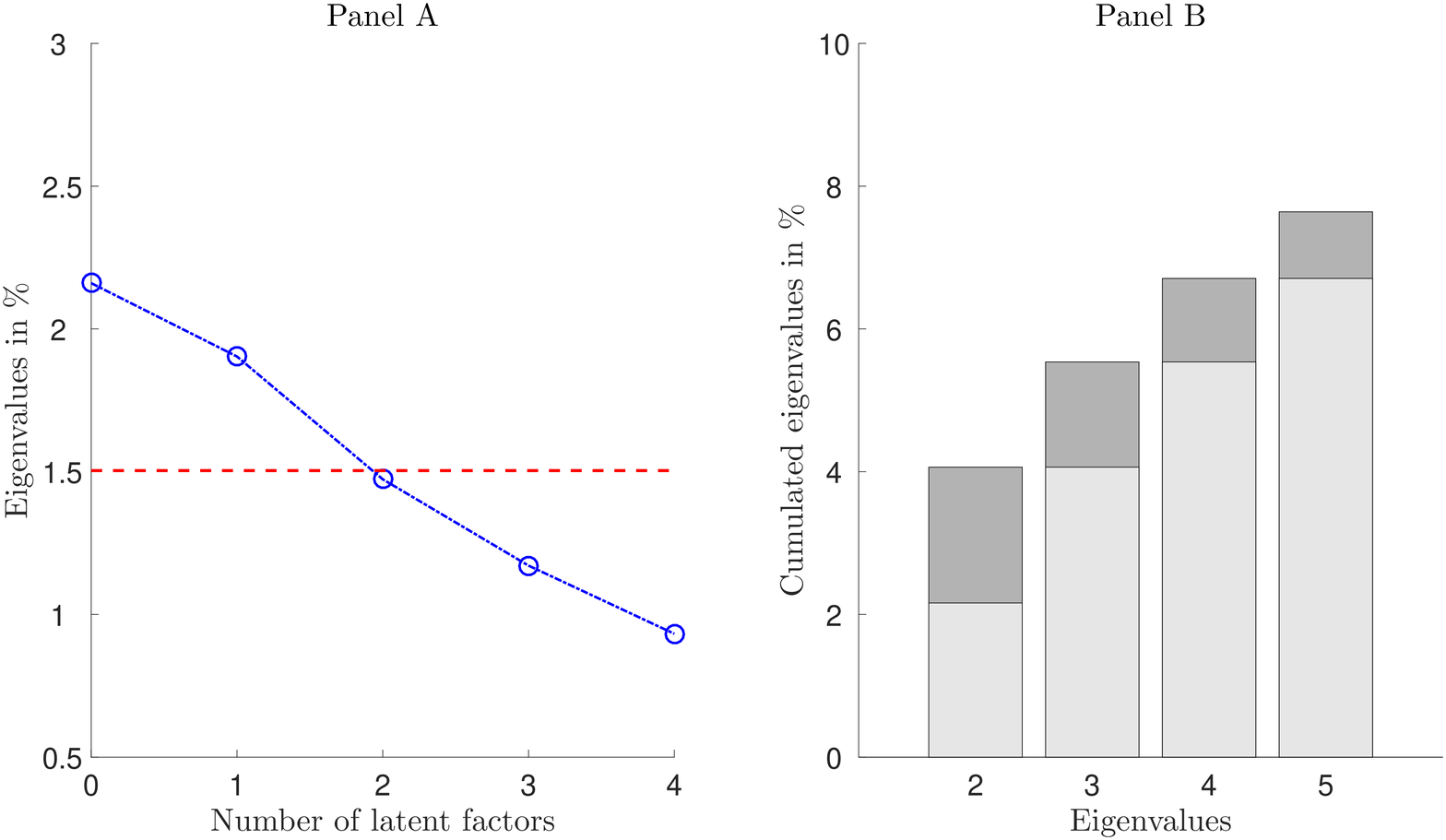}} \tabularnewline
{\small{}{}{}Time-varying CAPM} \tabularnewline
\tabularnewline
\hspace{-2cm} {\small{}{}{}\includegraphics[scale=0.35]{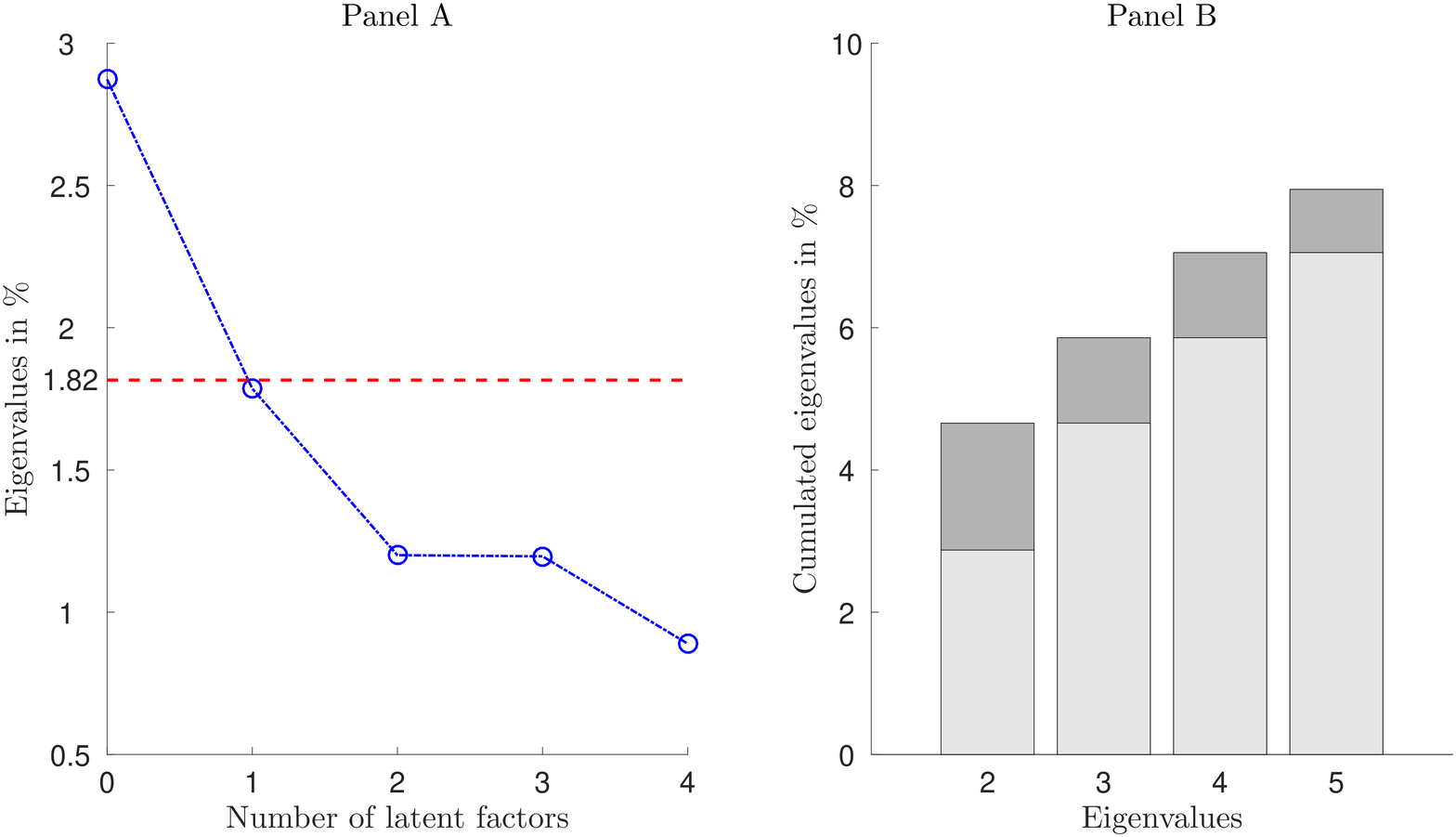}}\tabularnewline
\end{tabular}
\par\end{centering}
Panel A plots the scree-plot of the values of the first five eigenvalues
in percentage, i.e., ${\displaystyle \mu_{j}\left(\frac{1}{n^{\chi}T}\sum_{i}\boldsymbol{1}_{i}^{\chi}\bar{\bar{\varepsilon}}_{i}\bar{\bar{\varepsilon}}_{i}^{\prime}\right)}$
with $j=1,...,5$. The horizonal line corresponds to the penalty function.
Panel B plots the cumulated eigenvalues in percentage. The light grey
area corresponds to ${\displaystyle \sum_{l=1}^{j-1}\mu_{l}\left(\frac{1}{n^{\chi}T}\sum_{i}\boldsymbol{1}_{i}^{\chi}\bar{\bar{\varepsilon}}_{i}\bar{\bar{\varepsilon}}_{i}^{\prime}\right)}$,
the dark grey is the contribution of the $j$th eigenvalue in percentage.
The figure reports results for the CAPM for the time-invariant and
time-varying specifications with $Z_{t-1}=\left(1,divY_{t-1}\right)'$. 
\end{figure}

\begin{figure}[H]
\textbf{\protect\caption{\textbf{\label{fig_numOmitFactCAPM-1}Number of omitted factors and
cumulated squared eigenvalues for the CAPM}}
}

\begin{centering}
{\small{}{}{}}%
\begin{tabular}{c}
{\small{}{}{}Time-invariant CAPM} \tabularnewline
\hspace{-2cm} {\small{}{}{}\includegraphics[scale=0.35]{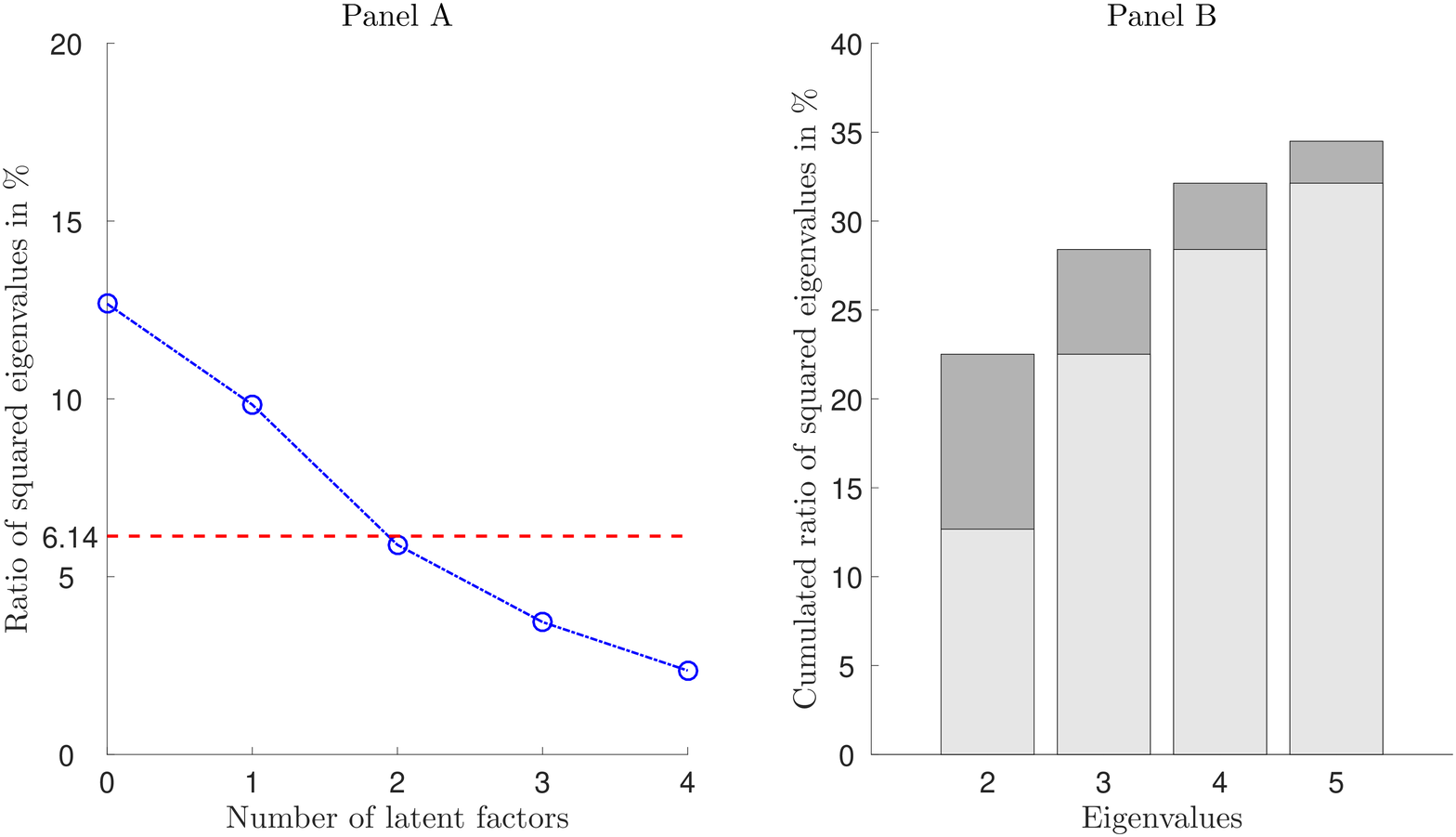}} \tabularnewline
{\small{}{}{}Time-varying CAPM} \tabularnewline
\hspace{-2cm} {\small{}{}{}\includegraphics[scale=0.35]{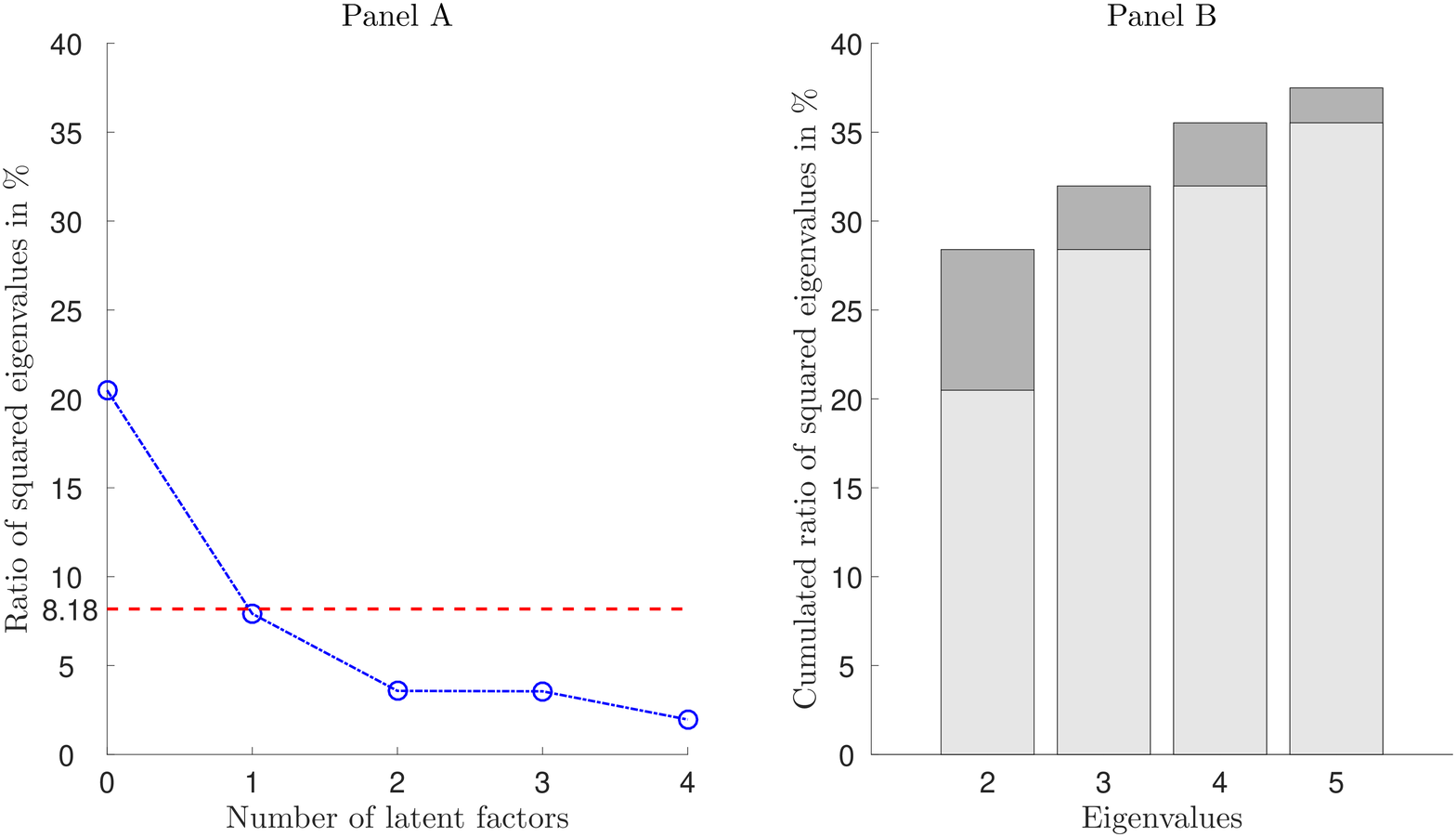}}\tabularnewline
\end{tabular}
\par\end{centering}
Panel A plots the scree-plot of the values of the first five squared
eigenvalues in percentage, i.e., ${\displaystyle \mu_{j}^{2}\left(\frac{1}{n^{\chi}T}\sum_{i}\boldsymbol{1}_{i}^{\chi}\bar{\bar{\varepsilon}}_{i}\bar{\bar{\varepsilon}}_{i}^{\prime}\right)/\sum_{l=1}^{T}\mu_{l}^{2}\left(\frac{1}{nT}\sum_{i}\boldsymbol{1}_{i}^{\chi}\bar{\bar{\varepsilon}}_{i}\bar{\bar{\varepsilon}}_{i}^{\prime}\right)}$
with $j=1,...,5$. The horizonal line corresponds to the penalty function
${\displaystyle g\left(n^{\chi},T\right)^{2}/\sum_{l=1}^{T}\mu_{l}^{2}\left(\frac{1}{nT}\sum_{i}\boldsymbol{1}_{i}^{\chi}\bar{\bar{\varepsilon}}_{i}\bar{\bar{\varepsilon}}_{i}^{\prime}\right)}$.
Panel B plots the cumulated squared eigenvalues in percentage. The
light grey area corresponds to ${\displaystyle \sum_{l=1}^{j-1}\mu_{l}^{2}\left(\frac{1}{n^{\chi}T}\sum_{i}\boldsymbol{1}_{i}^{\chi}\bar{\bar{\varepsilon}}_{i}\bar{\bar{\varepsilon}}_{i}^{\prime}\right)/\sum_{l=1}^{T}\mu_{l}^{2}\left(\frac{1}{nT}\sum_{i}\boldsymbol{1}_{i}^{\chi}\bar{\bar{\varepsilon}}_{i}\bar{\bar{\varepsilon}}_{i}^{\prime}\right)}$,
the dark grey is the contribution of the $j$th squared eigenvalue
in percentage. The figure reports results for the CAPM for the time-invariant
and time-varying specifications with $Z_{t-1}=\left(1,divY_{t-1}\right)'$. 
\end{figure}

\begin{figure}[H]
\textbf{\protect\caption{\textbf{\label{fig_numOmitFactFF}Number of omitted factors and cumulated
eigenvalues for the FF model}}
}

\begin{centering}
{\small{}{}{}}%
\begin{tabular}{c}
{\small{}{}{}Time-invariant FF} \tabularnewline
\hspace{-2cm} {\small{}{}{}\includegraphics[scale=0.35]{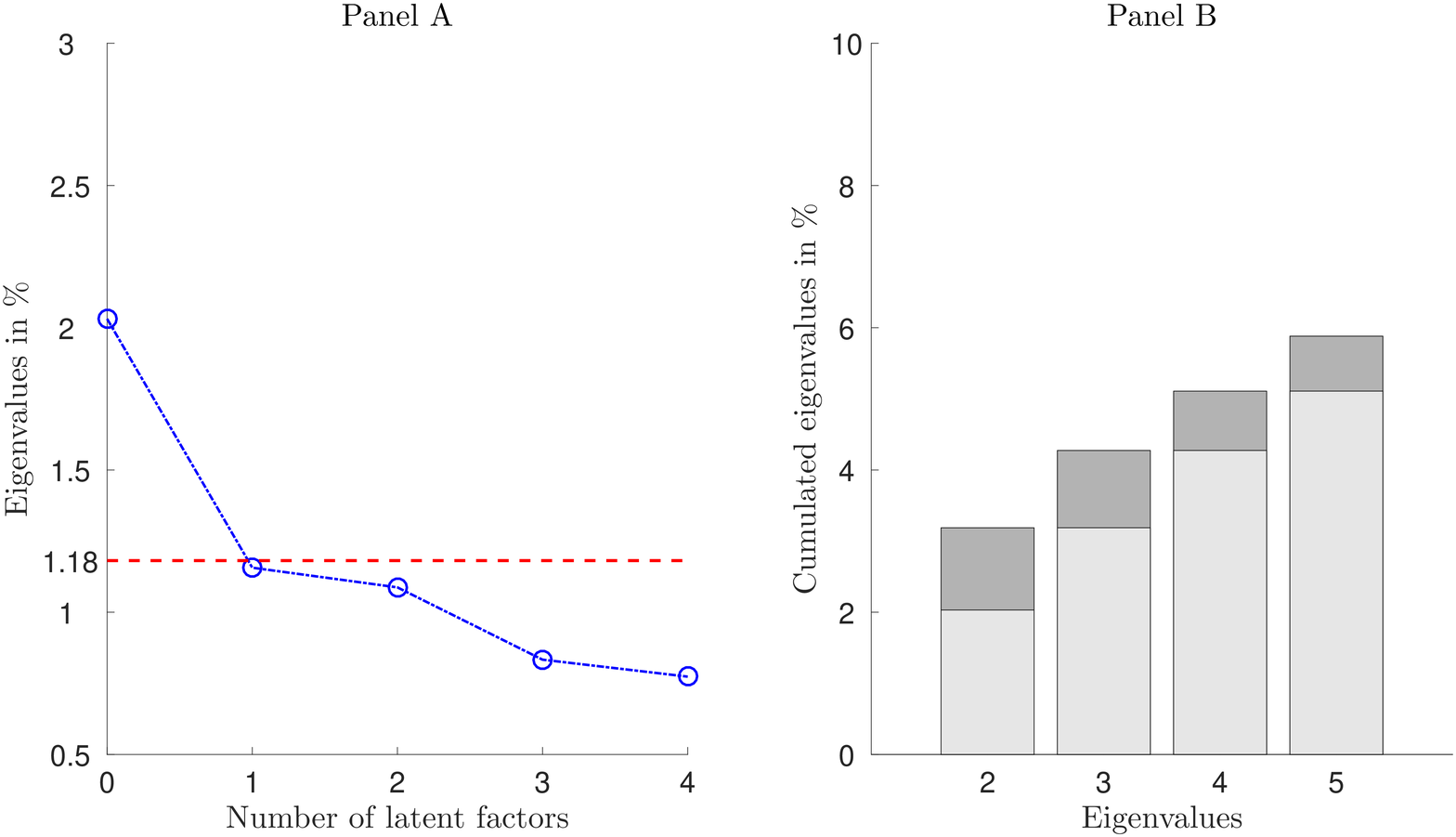}} \tabularnewline
{\small{}{}{}Time-varying FF} \tabularnewline
\tabularnewline
\hspace{-2cm} {\small{}{}{}\includegraphics[scale=0.35]{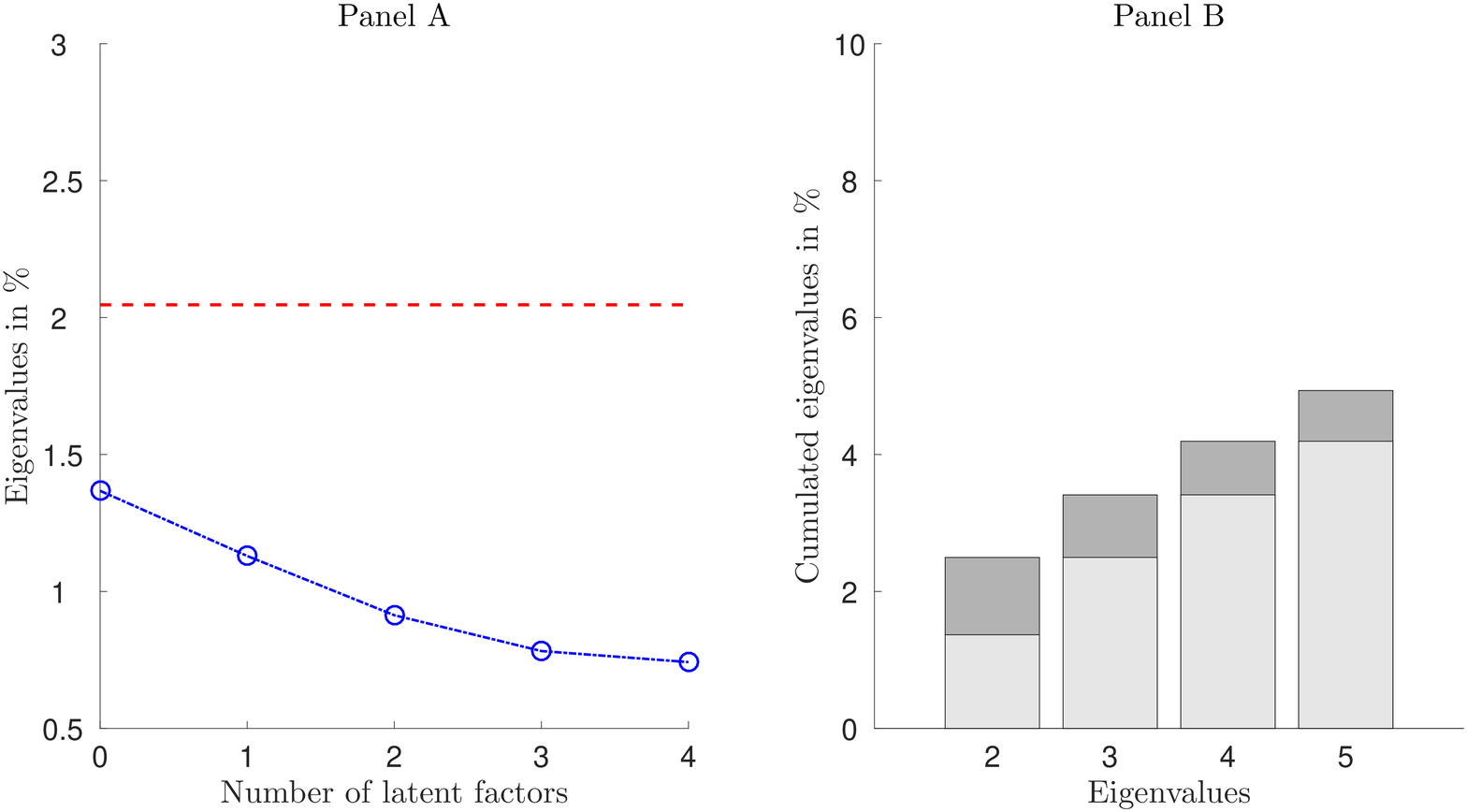}}\tabularnewline
\end{tabular}
\par\end{centering}
Panel A plots the scree-plot of the values of the first five eigenvalues
in percentage, i.e., ${\displaystyle \mu_{j}\left(\frac{1}{n^{\chi}T}\sum_{i}\boldsymbol{1}_{i}^{\chi}\bar{\bar{\varepsilon}}_{i}\bar{\bar{\varepsilon}}_{i}^{\prime}\right)}$
with $j=1,...,5$. The horizonal line corresponds to the penalty function.
Panel B plots the cumulated eigenvalues in percentage. The light grey
area corresponds to ${\displaystyle \sum_{l=1}^{j-1}\mu_{l}\left(\frac{1}{n^{\chi}T}\sum_{i}\boldsymbol{1}_{i}^{\chi}\bar{\bar{\varepsilon}}_{i}\bar{\bar{\varepsilon}}_{i}^{\prime}\right)}$,
the dark grey is the contribution of the $j$th eigenvalue in percentage.
The figure reports results for the FF model for the time-invariant
and time-varying specifications with $Z_{t-1}=\left(1,divY_{t-1}\right)'$. 
\end{figure}

\begin{figure}[H]
\textbf{\protect\caption{\textbf{\label{fig_numOmitFactFF-1}Number of omitted factors and
cumulated squared eigenvalues for the FF model}}
}

\begin{centering}
{\small{}{}{}}%
\begin{tabular}{c}
{\small{}{}{}Time-invariant FF} \tabularnewline
\hspace{-2cm} {\small{}{}{}\includegraphics[scale=0.35]{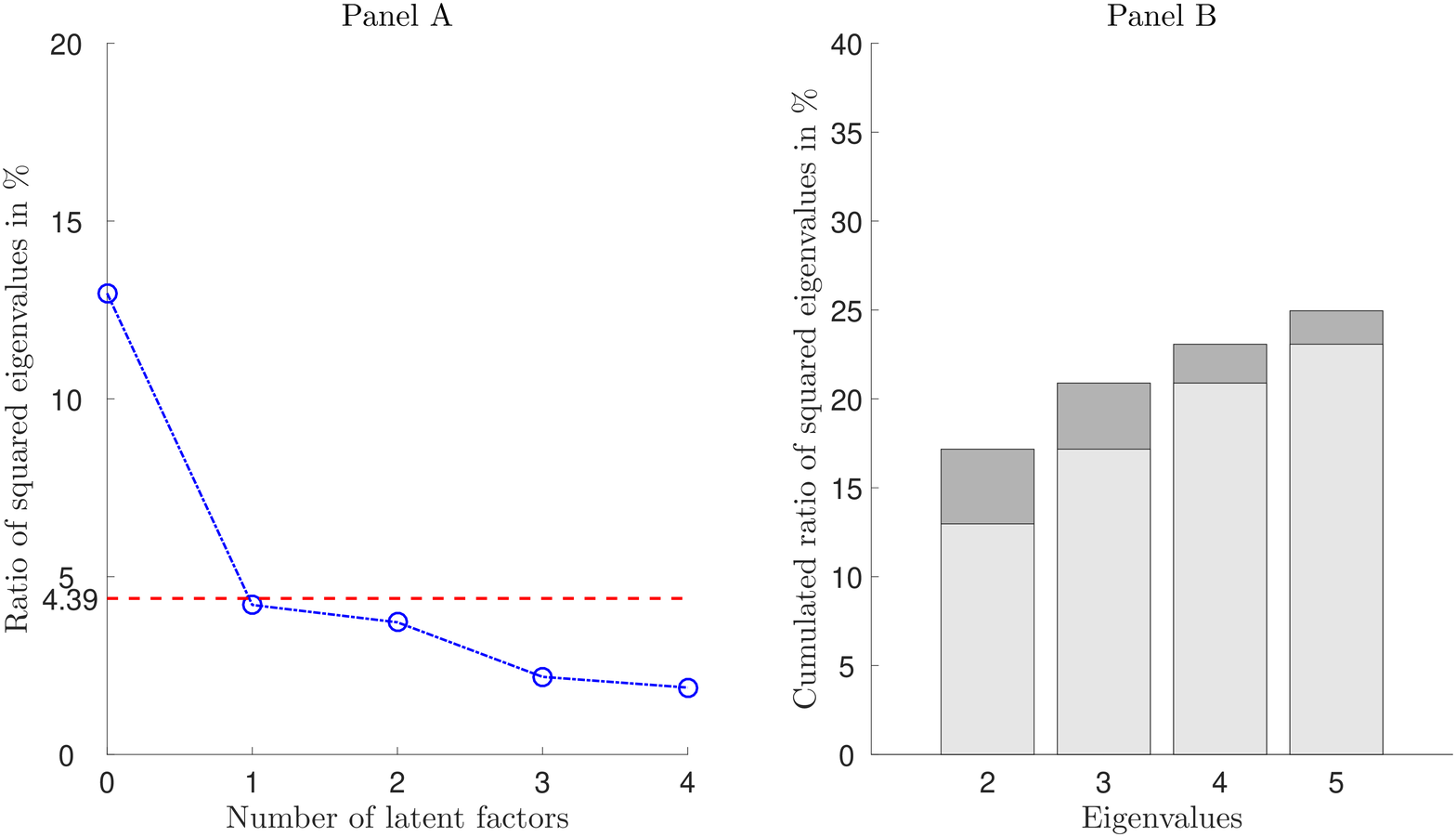}} \tabularnewline
{\small{}{}{}Time-varying FF} \tabularnewline
\hspace{-2cm} {\small{}{}{}\includegraphics[scale=0.35]{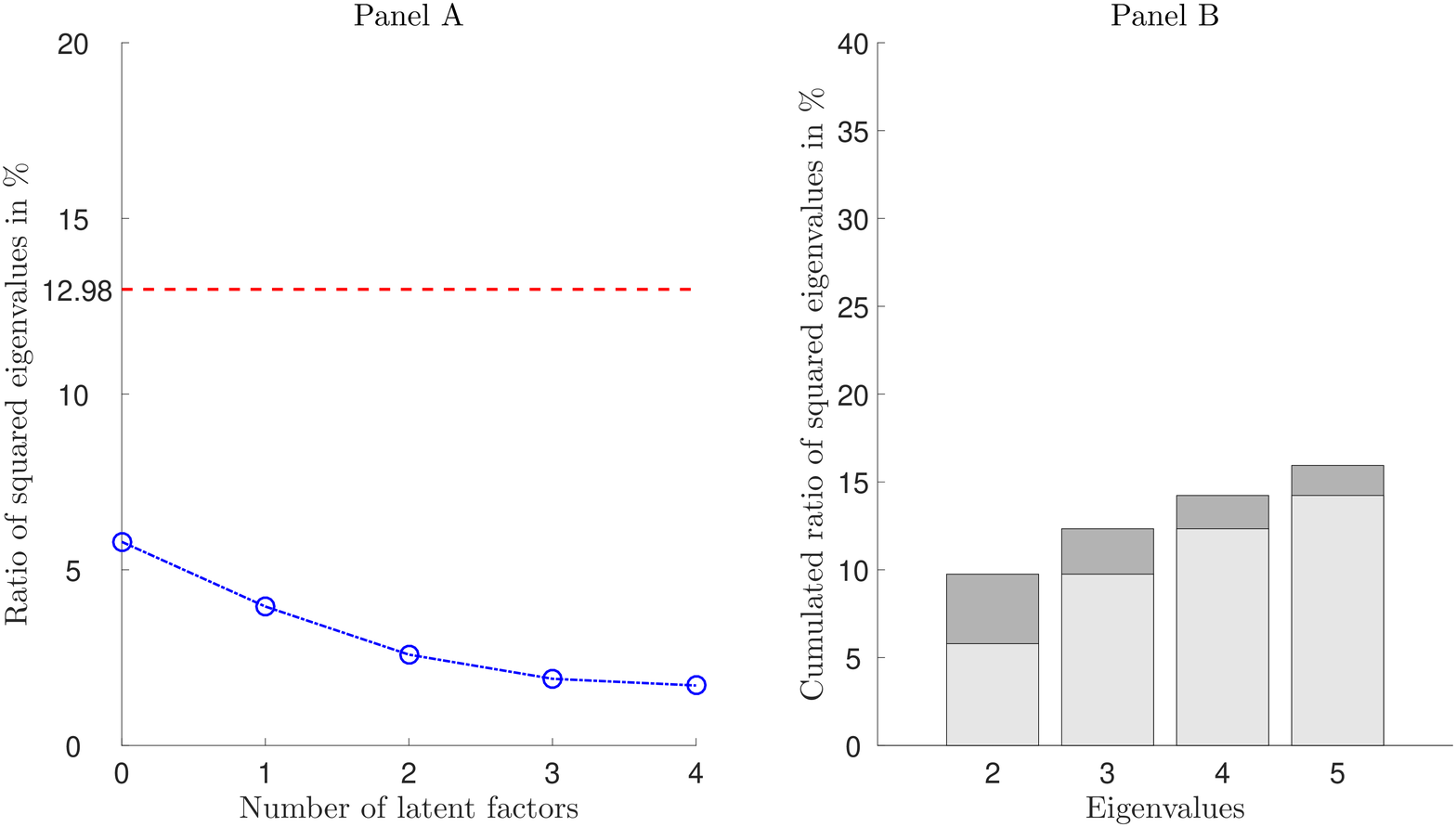}}\tabularnewline
\end{tabular}
\par\end{centering}
Panel A plots the scree-plot of the values of the first five squared
eigenvalues in percentage, i.e., ${\displaystyle \mu_{j}^{2}\left(\frac{1}{n^{\chi}T}\sum_{i}\boldsymbol{1}_{i}^{\chi}\bar{\bar{\varepsilon}}_{i}\bar{\bar{\varepsilon}}_{i}^{\prime}\right)/\sum_{l=1}^{T}\mu_{l}^{2}\left(\frac{1}{nT}\sum_{i}\boldsymbol{1}_{i}^{\chi}\bar{\bar{\varepsilon}}_{i}\bar{\bar{\varepsilon}}_{i}^{\prime}\right)}$
with $j=1,...,5$. The horizonal line corresponds to the penalty function
${\displaystyle g\left(n^{\chi},T\right)^{2}/\sum_{l=1}^{T}\mu_{l}^{2}\left(\frac{1}{nT}\sum_{i}\boldsymbol{1}_{i}^{\chi}\bar{\bar{\varepsilon}}_{i}\bar{\bar{\varepsilon}}_{i}^{\prime}\right)}$.
Panel B plots the cumulated squared eigenvalues in percentage. The
light grey area corresponds to ${\displaystyle \sum_{l=1}^{j-1}\mu_{l}^{2}\left(\frac{1}{n^{\chi}T}\sum_{i}\boldsymbol{1}_{i}^{\chi}\bar{\bar{\varepsilon}}_{i}\bar{\bar{\varepsilon}}_{i}^{\prime}\right)/\sum_{l=1}^{T}\mu_{l}^{2}\left(\frac{1}{nT}\sum_{i}\boldsymbol{1}_{i}^{\chi}\bar{\bar{\varepsilon}}_{i}\bar{\bar{\varepsilon}}_{i}^{\prime}\right)}$,
the dark grey is the contribution of the $j$th squared eigenvalue
in percentage. The figure reports results for the FF model for the
time-invariant and time-varying specifications with $Z_{t-1}=\left(1,divY_{t-1}\right)'$. 
\end{figure}

Let us start with the results for the time-invariant specifications.
The number $k$ of omitted factors is larger than one for the most
popular financial models, e.g., the CAPM (\cite{Sharpe_1964}), the
three-factor Fama-French model (FF) and the four-factor \cite{Carhart_1997}
model (CAR). On the contrary, for the recent proposals based on profitability
and investment (5FF, HXZ), quality minus junk (QMJ), and bet against
beta (BAB) factors, we find no omitted latent factor. We observe that
adding observable factors helps to reduce the contribution of the
first eigenvalue $\mu_{1}$ to the variance of residuals. However,
when we face latent factors, the omitted systematic contribution ${\displaystyle \sum_{j=1}^{k}\mu_{j}}$
only accounts for a small proportion of the residual variance. For
instance, we find $k=2$ omitted factors in the CAPM. Those two latent
factors only contribute to $\mu_{1}+\mu_{2}=4.06\%$ of the residual
variance. Figure \ref{fig_numOmitFactCAPM} summarizes this information
graphically by displaying the penalized scree plots and the plots
of cumulated eigenvalues for the CAPM. For instance, $\mu_{3}=1.47\%$
lies below the horizontal line $g(n^{\chi},T)=1.50\%$ in Panel A
for the time-invariant CAPM, so that $k=2$. In Panel B for the time-invariant
CAPM, the vertical bar $\mu_{1}+\mu_{2}=4.06\%$ is divided into the
contribution of $\mu_{1}=2.16\%$ (light grey area) and that of $\mu_{2}=1.90\%$
(dark grey area). Figure \ref{fig_numOmitFactCAPM-1} Panel A displays
the scree plots of squared eigenvalues for the CAPM and the square
$g^{2}\left(n^{\chi},T\right)$ of the penalisation function relative
to the squared Frobenius norm ${\displaystyle \sum_{l=1}^{T}\mu_{l}^{2}\left(\frac{1}{nT}\sum_{i}\boldsymbol{1}_{i}^{\chi}\bar{\bar{\varepsilon}}_{i}\bar{\bar{\varepsilon}}_{i}^{\prime}\right)}$.
By construction, the conclusion of the number of omitted factor is
the same as for the scree plot shown in Figure \ref{fig_numOmitFactCAPM}.
From the plot of cumulated squared eigenvalues in Figure \ref{fig_numOmitFactCAPM-1}
Panel B, we conclude that the two omitted factors contribute more
to the relative explanation of the correlation part than of the residual
variance. For example, we get that the sum of the square of the two
first eigenvalues accounts for $22.51\%$ of the square of the Frobenius
norm for the time-invariant CAPM. Thus, the two latent factors are
much more representative of the off-diagonal components. We conclude
similarly for the time-invariant FF model, even if the correlation
explanation provided by the single omitted factor is lower.

For the time-varying specifications (i) and (ii) of Table \ref{Tab_Results_timeInv_pureFin},
we still find one omitted factor for the CAPM. We see that the scaled
three-factor FF model with $Z_{t-1}=(1,divY_{t-1})'$ passes the diagnostic
criterion. The largest eigenvalue $\mu_{1}=1.37\%$ lies below the
horizontal line $g(n^{\chi},T)=2.05\%$ in Figure \ref{fig_numOmitFactFF}
Panel A, and its square $\mu_{1}^{2}$ only contributes to $5.80\%$
of the square of the Frobenius norm in Figure \ref{fig_numOmitFactFF-1}
Panel B for the scaled three-factor FF model, so that $k=0$. The
additional stock specific instrument $Z_{i,t-1}=bm_{i,t-1}$ is not
necessary to exhaust the cross-sectional dependence. Hence, the empirical
message of Table \ref{Tab_Results_timeInv_pureFin} is that we can
choose either among time-invariant specifications with at least four
financial factors, or a scaled FF model. The latter is more parsimonious
for the factor space in the conditional sense ($K=3$ versus $K=4$),
but less parsimonious for the parameter space ($d=9$ versus $d=5$).
From an econometric point of view, it is not clear which parsimony
we should favor to decide between the time-invariant specification
(more factors, less parameters) and the time-varying specification
(less factors, more parameters). From a finance point of view, the
first one is better suited for static (unconditional) investment decisions
while the second one is better suited for dynamic (conditional) investment
decisions. The choice between the two models should meet the investor
needs or answer the empirical research question at hand. For a balanced
panel of monthly returns for $4,883$ stocks on the period January
1994 to December 1998 ($T=60$), \cite{Bai_Ng_2002} find only two
latent factors.

As observed in GOS, measures of limits-to-arbitrage and missing factor
impact (not reported here) like those in \cite{Pontiff_2006}, \cite{Ang_Hodrick_Xing_Zhang_2009},
\cite{Lam_Wei_2011}, \cite{Stambaugh_Yu_Yuan_2015} decrease with
the number of observable factors.

Concerning the additional factors and instruments mentioned in Section
\ref{sec:FactorMod_DataDescr}, none of them allows to reach a more
parsimonious factor structure in a time-invariant or time-varying
setting. Moreover, neither the time-invariant CAPM, FF and CAR models,
nor their time-varying specifications with term spread, default spread,
and book-to-market equity used in GOS, pass the diagnostic criterion.
As conjectured in GOS, this might be one reason for the rejection
of the asset pricing restrictions. 

\subsection{Results for macroeconomic models}

\begin{figure}[H]
\textbf{\protect\caption{\textbf{\label{fig_numOmitFactCCAPM}Number of omitted factors and
cumulated squared eigenvalues for the CCAPM model}}
}

\begin{centering}
{\small{}{}{}}%
\begin{tabular}{c}
\hspace{-2cm} {\small{}{}{}\includegraphics[scale=0.35]{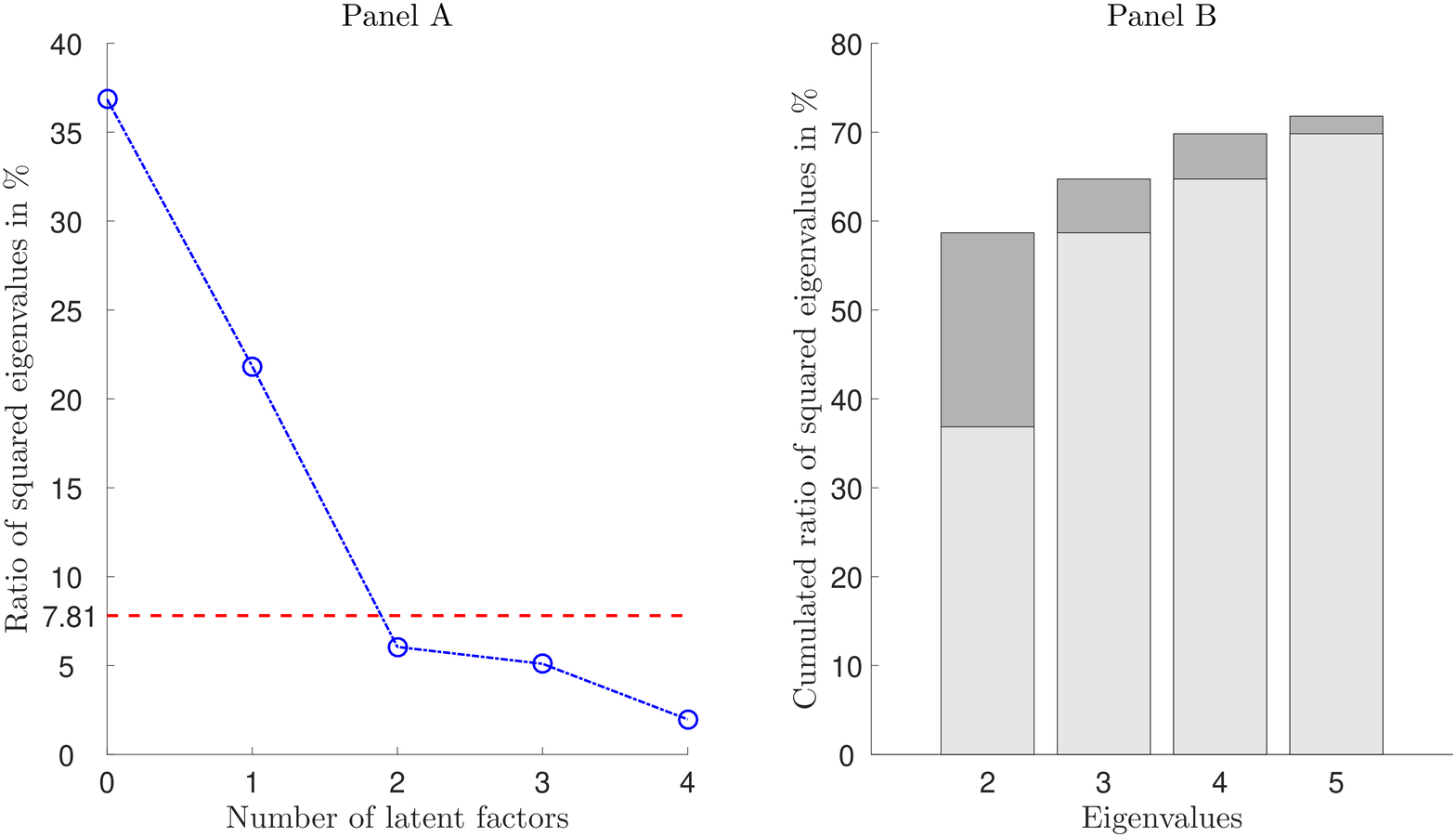}}\tabularnewline
\end{tabular}
\par\end{centering}
Panel A plots the scree-plot of the values of the first five squared
eigenvalues in percentage, i.e., ${\displaystyle \mu_{j}^{2}\left(\frac{1}{n^{\chi}T}\sum_{i}\boldsymbol{1}_{i}^{\chi}\bar{\bar{\varepsilon}}_{i}\bar{\bar{\varepsilon}}_{i}^{\prime}\right)/\sum_{l=1}^{T}\mu_{l}^{2}\left(\frac{1}{nT}\sum_{i}\boldsymbol{1}_{i}^{\chi}\bar{\bar{\varepsilon}}_{i}\bar{\bar{\varepsilon}}_{i}^{\prime}\right)}$
with $j=1,...,5$. The horizonal line corresponds to the penalty function
${\displaystyle g\left(n^{\chi},T\right)^{2}/\sum_{l=1}^{T}\mu_{l}^{2}\left(\frac{1}{nT}\sum_{i}\boldsymbol{1}_{i}^{\chi}\bar{\bar{\varepsilon}}_{i}\bar{\bar{\varepsilon}}_{i}^{\prime}\right)}$.
Panel B plots the cumulated squared eigenvalues in percentage. The
light grey area corresponds to ${\displaystyle \sum_{l=1}^{j-1}\mu_{l}^{2}\left(\frac{1}{n^{\chi}T}\sum_{i}\boldsymbol{1}_{i}^{\chi}\bar{\bar{\varepsilon}}_{i}\bar{\bar{\varepsilon}}_{i}^{\prime}\right)/\sum_{l=1}^{T}\mu_{l}^{2}\left(\frac{1}{nT}\sum_{i}\boldsymbol{1}_{i}^{\chi}\bar{\bar{\varepsilon}}_{i}\bar{\bar{\varepsilon}}_{i}^{\prime}\right)}$,
the dark grey is the contribution of the $j$th sqaured eigenvalue
in percentage. 
\end{figure}

In this section, we perform the empirical exercises on the macroeconomic
linear factor models listed in Table \ref{Tab_PureMacroeco}. We fix
$\chi_{1,T}=15$ and $\chi_{2,T}=176/20$, i.e., at least 20 quarterly
return observations. In Table \ref{Tab_Results_pureMacro}, we report
the trimmed cross-sectional dimension $n^{\chi}$. The quarterly dataset
has $6,707$ stocks with more than twenty quarterly observations.
The trimming is driven by the multicollinearity between regressors,
when $n^{\chi}<6,707$. Table \ref{Tab_Results_pureMacro} further
reports the empirical results for the macroeconomic models. The time-invariant
specifications which include only macroeconomic variables (CCAPM,
NDC and DC, LVX, and CRR) and exclude the market, do not pass the
diagnostic criterion. We find $k=1$ omitted factors. Moreover, $\mu_{1}$
is about $8\%$ of the residual variance in Table \ref{Tab_Results_pureMacro}
and $\mu_{1}^{2}$ accounts for $37\%$ of the square of the Frobenius
norm in Figure \ref{fig_numOmitFactCCAPM}, in contrast to the $4.06\%$
and $22.51\%$ found for the time-invariant CAPM with monthly returns.
The latent factors in the macro economic models are both representative
of the residual variance (diagonal values) and the correlation part
(off-diagonal values). When we incorporate the market (EZ and YO),
we find no omitted latent factors. This is not surprising since, for
quarterly data, the CAPM fully captures the systematic risk of individual
stocks, with $\mu_{1}=3.15\%$, $g(n^{\chi},T)=3.74\%$, and $n^{\chi}=6,707$.
We do not report results for time-varying specifications. We have
a limited sample size $T=176$. Because of multicollinearity problems
and the parameter dimension being up to $d=14$, the estimation yields
imprecise results. The trimmed sample size $n^{\chi}$ is often lower
than $T$, which casts doubt about empirical results obtained under
a large panel assumption.

\pagebreak{}

\bibliographystyle{plainnat}
\bibliography{myrefs_TestStat}

\pagebreak{}

\begin{table}[H]
\textbf{\small{}{}{}\protect\caption{\textbf{\small{}{}{}\label{Tab_PureFinancial}Financial linear factor
models }}
}{\small \par}

\smallskip{}
\begin{centering}
{\small{}{}{}}%
\begin{tabular}{llc}
\hline 
{\small{}{}{}Model }  & {\small{}{}{}Factors }  & {\small{}{}{}$K$ }\tabularnewline
\hline 
{\small{}{}{} CAPM }  & {\small{}{}{}$r_{m,t}$ }  & {\small{}{}{}1 }\tabularnewline
{\small{}{}{} FF }  & {\small{}{}{}$r_{m,t},r_{smb,t},r_{hml,t}$ }  & {\small{}{}{}3 }\tabularnewline
{\small{}{}{} CAR }  & {\small{}{}{}$r_{m,t},r_{smb,t},r_{hml,t},r_{mom,t}$ }  & {\small{}{}{}4 }\tabularnewline
{\small{}{}{} 5FF }  & {\small{}{}{}$r_{m,t},r_{smb,t},r_{hml,t},r_{rmw,t},r_{cma,t}$
}  & {\small{}{}{}5}\tabularnewline
{\small{}{}{} HXZ }  & {\small{}{}{}$r_{m,t},r_{me,t},r_{I/A,t},r_{ROE,t}$ }  & {\small{}{}{}4}\tabularnewline
{\small{}{}{} FF and QMJ }  & {\small{}{}{}$r_{m,t},r_{smb,t},r_{hml,t},qmj_{t}$ }  & {\small{}{}{}4}\tabularnewline
{\small{}{}{} FF and BAB }  & {\small{}{}{}$r_{m,t},r_{smb,t},r_{hml,t},bab_{t}$ }  & {\small{}{}{}4}\tabularnewline
\hline 
\end{tabular}
\par\end{centering}
\smallskip{}

The table lists the linear factor models based on financial variables.
We estimate these specifications by using monthly data. For each model,
we report the factors labeling and their number $K$. FF, CAR, 5FF,
HXZ, QMJ and BAB refer to the three Fama-French factors, the four
Carhart factors, the five Fama-French factors, the four Hou-Xue-Zhang
factors, quality minus junk factor, and bet against beta factor. 
\end{table}

\vspace{-1cm}

\begin{table}[H]
\textbf{\small{}{}{}\protect\caption{\textbf{\small{}{}{}\label{Tab_PureMacroeco}Macroeconomic linear
factor models}}
}{\small \par}

\smallskip{}
\begin{centering}
{\small{}{}{}}%
\begin{tabular}{llc}
\hline 
{\small{}{}{}Model }  & {\small{}{}{}Factors }  & {\small{}{}{}$K$ }\tabularnewline
\hline 
{\small{}{}{}CCAPM }  & {\small{}{}{}$cg_{t}$ }  & {\small{}{}{}1 }\tabularnewline
{\small{}{}{}EZ }  & {\small{}{}{}$r_{m,t},cg_{t}$ }  & {\small{}{}{}2 }\tabularnewline
{\small{}{}{}NDC and DC}  & {\small{}{}{}$ndcg_{t},dcg_{t}$ }  & {\small{}{}{}2 }\tabularnewline
{\small{}{}{}YO}  & {\small{}{}{}$r_{m,t},ndcg_{t},dcg_{t}$ }  & {\small{}{}{}3 }\tabularnewline
{\small{}{}{} LVX }  & {\small{}{}{}$dhh_{t},dcorp_{t},dncorp_{t}$ }  & {\small{}{}{}3}\tabularnewline
{\small{}{}{} CRR }  & {\small{}{}{}$mp_{t},ui_{t},dei_{t},uts_{t},upr_{t}$ }  & {\small{}{}{}5 }\tabularnewline
\hline 
\end{tabular}
\par\end{centering}
\smallskip{}

The table lists the linear factor models based on macroeconomic variables
and the market. We estimate these specifications by using quarterly
data. For each model, we report the factors labeling and their number
$K$. EZ, NDC and DC, YO, LVX and CRR refer to the two Epstein-Zin
factors, the two nondurable and durable consumption growth factors,
the three Yogo factors, the three Li-Vassalou-Xing factors, and the
five Chen-Roll-Ross factors. 
\end{table}


\newpage{}

\vspace{-4cm}

\begin{table}[H]
\textbf{\small{}{}{}\protect\caption{\textbf{\small{}{}{}\label{Tab_nchi_PureFin_monthly}Trimmed cross-sectional
dimensions $n^{\chi}$ and number $d$ of parameters to estimate for
financial models}}
}{\small \par}

\begin{centering}
\smallskip{}
\begin{tabular}{lc|cccccc}
\hline 
{\small{}{}Financial model }  & {\small{}{}Time-invariant}  & \multicolumn{6}{c}{{\small{}{}Time-varying}}\tabularnewline
 & $n^{\chi}$  & (i)  & $d$  & {\small{}{}$n^{\chi}$ }  & (ii)  & $d$  & {\small{}{}$n^{\chi}$}\tabularnewline
\hline 
{\small{}{}{}CAPM }  & {\small{}{}{}6,775 }  &  & {\small{}{}{}5 }  & {\small{}{}{}3,766 }  &  & $8$  & {\small{}{}{}3,004}\tabularnewline
{\small{}{}{}FF }  & {\small{}{}{}6,775 }  &  & {\small{}{}{}9 }  & {\small{}{}{}3,536 }  &  & $14$  & {\small{}{}{}2,780}\tabularnewline
{\small{}{}{}CAR }  & {\small{}{}{}6,775 }  &  & {\small{}{}{}11 }  & {\small{}{}{}3,468 }  &  & $17$  & {\small{}{}{}2,608}\tabularnewline
{\small{}{}{}5 FF}  & {\small{}{}{}6,775 }  &  & {\small{}{}{}13 }  & {\small{}{}{}2,957 }  &  & $20$  & {\small{}{}{}1,991}\tabularnewline
{\small{}{}{}HXZ}  & {\small{}{}{}6,775 }  &  & {\small{}{}{}11 }  & {\small{}{}{}3,344 }  &  & $17$  & {\small{}{}{}2,612}\tabularnewline
{\small{}{}{}FF and QMJ }  & {\small{}{}{}6,775 }  &  & {\small{}{}{}11}  & {\small{}{}{}3,365 }  &  & $17$  & {\small{}{}{}2,423}\tabularnewline
{\small{}{}{}FF and BAB }  & {\small{}{}{}6,775 }  &  & {\small{}{}{}11 }  & {\small{}{}{}3,224 }  &  & $17$  & {\small{}{}{}2,441}\tabularnewline
\hline 
\end{tabular}
\par\end{centering}
\smallskip{}

For each financial model of Table \ref{Tab_PureFinancial}, we report
the trimmed cross-sectional dimension $n^{\chi}$ for estimation from
monthly data. For the time-varying specifications, we give the dimension
$d$ of vector $x_{i,t}$ and $n^{\chi}$ for two sets of instruments:
(i) $Z_{t-1}=\left(1,divY_{t-1}\right)'$ and (ii) $Z_{t-1}=\left(1,divY_{t-1}\right)'$,
$Z_{i,t-1}=bm_{i,t-1}$. For the time-invariant specifications, we
have $d=K+1$ (see Table \ref{Tab_PureFinancial}). 
\end{table}

\begin{table}[H]
\textbf{\protect\caption{\textbf{\label{Tab_Results_timeInv_pureFin}Results for time-invariant
and time-varying financial models}}
}

\smallskip{}
\begin{centering}
{\small{}{}{}}%
\begin{tabular}{l|ccccc|cccccc}
\hline 
{{\small{}{}{}Financial }}  & \multicolumn{5}{c|}{{\small{}{}{}Panel A -}\emph{\small{}{}{} Time-invariant}} & \multicolumn{6}{c}{{\small{}{}{}Panel B - }\emph{\small{}{}{}Time-varying}}\tabularnewline
model  & {\small{}{}{}$\mu_{1}$}  & {\small{}{}{}$k$}  & {\small{}{}{}${\displaystyle \sum_{j=1}^{k}\mu_{j}}$}  & {\small{}{}{}$\mu_{k+1}$}  & {\small{}{}{}penalty} &  & {\small{}{}{}$\mu_{1}$}  & {\small{}{}{}$k$}  & {\small{}{}{}${\displaystyle \sum_{j=1}^{k}\mu_{j}}$}  & {\small{}{}{}$\mu_{k+1}$}  & {\small{}{}{}penalty}\tabularnewline
\hline 
{\small{}{}{}CAPM}  & {\small{}{}{}2.16\%}  & {\small{}{}{}2}  & {\small{}{}{}4.06\%}  & {\small{}{}{}1.47\%}  & {\small{}{}{}1.50\%}  & (i)  & {\small{}{}{}2.87\%}  & {\small{}{}{}1}  & {\small{}{}{}2.87\%}  & {\small{}{}{}1.79\%}  & {\small{}{}{}1.82\%}\tabularnewline
{\small{}{}{}}  & {\small{}{}{}}  & {\small{}{}{}}  & {\small{}{}{}}  & {\small{}{}{}}  & {\small{}{}{}}  & (ii)  & {\small{}{}{}3.00\%}  & {\small{}{}{}1}  & {\small{}{}{}3.00\%}  & {\small{}{}{}1.98\%}  & {\small{}{}{}2.00\%}\tabularnewline
{\small{}{}{}FF}  & {\small{}{}{}2.03\%}  & {\small{}{}{}1}  & {\small{}{}{}2.03\%}  & {\small{}{}{}1.16\%}  & {\small{}{}{}1.18\%}  & (i)  & {\small{}{}{}1.37\%}  & {\small{}{}{}0}  & {\small{}{}{}0.00\%}  & {\small{}{}{}1.37\%}  & {\small{}{}{}2.05\%}\tabularnewline
{\small{}{}{}}  & {\small{}{}{}}  & {\small{}{}{}}  & {\small{}{}{}}  & {\small{}{}{}}  & {\small{}{}{}}  & (ii)  & {\small{}{}{}1.53\%}  & {\small{}{}{}0}  & {\small{}{}{}0.00\%}  & {\small{}{}{}1.53\%}  & {\small{}{}{}2.17\%}\tabularnewline
{\small{}{}{}CAR}  & {\small{}{}{}2.03\%}  & {\small{}{}{}1}  & {\small{}{}{}2.03\%}  & {\small{}{}{}1.12\%}  & {\small{}{}{}1.15\%}  & (i)  & {\small{}{}{}1.34\%}  & {\small{}{}{}0}  & {\small{}{}{}0.00\%}  & {\small{}{}{}1.34\%}  & {\small{}{}{}2.05\%}\tabularnewline
{\small{}{}{}}  & {\small{}{}{}}  & {\small{}{}{}}  & {\small{}{}{}}  & {\small{}{}{}}  & {\small{}{}{}}  & (ii)  & {\small{}{}{}1.51\%}  & {\small{}{}{}0}  & {\small{}{}{}0.00\%}  & {\small{}{}{}1.51\%}  & {\small{}{}{}2.20\%}\tabularnewline
{\small{}{}{}5FF}  & {\small{}{}{}1.42\%}  & {\small{}{}{}0}  & {\small{}{}{}0.00\%}  & {\small{}{}{}1.42\%}  & {\small{}{}{}1.79\%}  & (i)  & {\small{}{}{}1.45\%}  & {\small{}{}{}0}  & {\small{}{}{}0.00\%}  & {\small{}{}{}1.45\%}  & {\small{}{}{}2.13\%}\tabularnewline
{\small{}{}{}}  & {\small{}{}{}}  & {\small{}{}{}}  & {\small{}{}{}}  & {\small{}{}{}}  & {\small{}{}{}}  & (ii)  & {\small{}{}{}1.81\%}  & {\small{}{}{}0}  & {\small{}{}{}0.00\%}  & {\small{}{}{}1.81\%}  & {\small{}{}{}2.37\%}\tabularnewline
{\small{}{}{}HXZ}  & {\small{}{}{}1.43\%}  & {\small{}{}{}0}  & {\small{}{}{}0.00\%}  & {\small{}{}{}1.43\%}  & {\small{}{}{}1.79\%}  & (i)  & {\small{}{}{}1.35\%}  & {\small{}{}{}0}  & {\small{}{}{}0.00\%}  & {\small{}{}{}1.35\%}  & {\small{}{}{}2.07\%}\tabularnewline
{\small{}{}{}}  & {\small{}{}{}}  & {\small{}{}{}}  & {\small{}{}{}}  & {\small{}{}{}}  & {\small{}{}{}}  & (ii)  & {\small{}{}{}1.54\%}  & {\small{}{}{}0}  & {\small{}{}{}0.00\%}  & {\small{}{}{}1.54\%}  & {\small{}{}{}2.20\%}\tabularnewline
{\small{}{}{}FF and QMJ}  & {\small{}{}{}1.39\%}  & {\small{}{}{}0}  & {\small{}{}{}0.00\%}  & {\small{}{}{}1.39\%}  & {\small{}{}{}1.79\%}  & (i)  & {\small{}{}{}1.33\%}  & {\small{}{}{}0}  & {\small{}{}{}0.00\%}  & {\small{}{}{}1.33\%}  & {\small{}{}{}2.07\%}\tabularnewline
{\small{}{}{}}  & {\small{}{}{}}  & {\small{}{}{}}  & {\small{}{}{}}  & {\small{}{}{}}  & {\small{}{}{}}  & (ii)  & {\small{}{}{}1.60\%}  & {\small{}{}{}0}  & {\small{}{}{}0.00\%}  & {\small{}{}{}1.60\%}  & {\small{}{}{}2.24\%}\tabularnewline
{\small{}{}{}FF and BAB}  & {\small{}{}{}1.64\%}  & {\small{}{}{}0}  & {\small{}{}{}0.00\%}  & {\small{}{}{}1.64\%}  & {\small{}{}{}1.79\%}  & (i)  & {\small{}{}{}1.40\%}  & {\small{}{}{}0}  & {\small{}{}{}0.00\%}  & {\small{}{}{}1.40\%}  & {\small{}{}{}2.09\%}\tabularnewline
{\small{}{}{}}  & {\small{}{}{}}  & {\small{}{}{}}  & {\small{}{}{}}  & {\small{}{}{}}  & {\small{}{}{}}  & (ii)  & {\small{}{}{}1.58\%}  & {\small{}{}{}0}  & {\small{}{}{}0.00\%}  & {\small{}{}{}1.58\%}  & {\small{}{}{}2.24\%}\tabularnewline
\hline 
\end{tabular}
\par\end{centering}
\medskip{}

The table shows the contribution of the first eigenvalue $\mu_{1}$
to the variance of normalised residuals, the number of omitted factors
$k$, the contributions of the first $k$, and of the $\left(k+1\right)$-th
eigenvalues, and the penalty term. Panels A and B report results for
time-invariant and time-varying financial models estimated from monthly
data, respectively. The time-varying specifications use two sets of
instruments: (i) $Z_{t-1}=\left(1,divY_{t-1}\right)'$ and (ii) $Z_{t-1}=\left(1,divY_{t-1}\right)'$,
$Z_{i,t-1}=bm_{i,t-1}$. 
\end{table}

\pagebreak{}

\begin{table}[H]
\textbf{\protect\caption{\textbf{\label{Tab_Results_pureMacro}Results for the macroeconomic
models}}
}

\smallskip{}
\begin{centering}
{\small{}{}{}}%
\begin{tabular}{l|cccccc}
\hline 
{{\small{}{}{}Macroeconomic model}}  & $n^{\chi}$  & {\small{}{}{}$\mu_{1}$}  & {\small{}{}{}$k$}  & {\small{}{}{}${\displaystyle \sum_{j}^{k}\mu_{j}}$}  & {\small{}{}{}$\mu_{k+1}$}  & {\small{}{}{}penalty}\tabularnewline
\hline 
{\small{}{}{}CCAPM}  & $6,707$  & {\small{}{}{}8.12\%}  & {\small{}{}{}1}  & {\small{}{}{}8.12\%}  & {\small{}{}{}6.24\%}  & {\small{}{}{}6.28\%} \tabularnewline
{\small{}{}{}EZ}  & $6,707$  & {\small{}{}{}3.07\%}  & {\small{}{}{}0}  & {\small{}{}{}0.00\%}  & {\small{}{}{}3.07\%}  & {\small{}{}{}3.74\%} \tabularnewline
{\small{}{}{}NDC and DC}  & $6,306$  & {\small{}{}{}8.07\%}  & {\small{}{}{}1}  & {\small{}{}{}8.06\%}  & {\small{}{}{}6.14\%}  & {\small{}{}{}6.17\%} \tabularnewline
{\small{}{}{}YO}  & $6,270$  & {\small{}{}{}3.38\%}  & {\small{}{}{}0}  & {\small{}{}{}0.00\%}  & {\small{}{}{}3.38\%}  & {\small{}{}{}3.76\%} \tabularnewline
{\small{}{}{}LVX}  & $6,707$  & {\small{}{}{}7.96\%}  & {\small{}{}{}1}  & {\small{}{}{}7.96\%}  & {\small{}{}{}6.09\%}  & {\small{}{}{}6.13\%} \tabularnewline
{\small{}{}{}CRR}  & $6,153$  & {\small{}{}{}6.42\%}  & {\small{}{}{}2}  & {\small{}{}{}11.30\%}  & {\small{}{}{}2.45\%}  & {\small{}{}{}2.48\%} \tabularnewline
\hline 
\end{tabular}
\par\end{centering}
\medskip{}

For each macroeconomic model of Table \ref{Tab_PureMacroeco}, we
report the trimmed cross-sectional dimension $n^{\chi}$ for time-invariant
specifications estimated from quarterly data. We further show the
contribution of the first eigenvalue $\mu_{1}$ to the variance of
normalised residuals, the number of omitted factors $k$, the contributions
of the first $k$, and of the $\left(k+1\right)$-th eigenvalues,
and the penalty term. 
\end{table}

\pagebreak{}

\newgeometry{top=0.5cm, bottom=0.5cm, left=2cm, right=2cm}\thispagestyle{empty}

\newgeometry{top=0.5cm, bottom=0.5cm, left=2cm, right=2cm}\thispagestyle{empty}

\newgeometry{top=0.5cm, bottom=0.5cm, left=2cm, right=2cm}\thispagestyle{empty}

\newgeometry{top=0.5cm, bottom=0.5cm, left=2cm, right=2cm}\thispagestyle{empty}

\newgeometry{top=0.5cm, bottom=0.5cm, left=2cm, right=2cm}\thispagestyle{empty}

\restoregeometry

\newpage{}

\appendix

\global\long\def\thesection{Appendix \arabic{section}}

\global\long\def\thesubsection{A.\arabic{section}.\arabic{subsection}}

\global\long\def\thesubsubsection{A.\arabic{section}.\arabic{subsection}.\arabic{subsubsection}}

\section{Regularity conditions \label{sec:Appendix_RegularityConditions}}

In this appendix, we list and comment additional assumptions used
in the proofs in \ref{sec:Proofs}. The error terms ${\displaystyle \left(\varepsilon_{i,t}\right)}$
are ${\displaystyle \varepsilon_{i,t}=u_{i,t}}$ under model ${\displaystyle {\cal M}_{1}}$,
and ${\displaystyle \varepsilon_{i,t}=\theta_{i}^{\prime}h_{t}+u_{i,t}}$
under model ${\displaystyle {\cal M}}_{2}$ (see Equation (\ref{FactirStructure_M2_MatrixNotat})).
Since models ${\cal M}_{1}\left(k\right)$ and ${\cal M}_{2}\left(k\right)$
are subsets of model ${\cal M}_{2}$, the assumptions stated for ${\cal M}_{2}$
also hold for ${\cal M}_{1}\left(k\right)$ and ${\cal M}_{2}\left(k\right)$,
for any $k\geq1$. We use $M$ as a generic constant in the assumptions.


\begin{Aassumption}\label{A_Ass_ErrTerm_u_M2} For a constant $M>0$
and for all $n,T\in\mathbb{N}$, we have: 
\[
\frac{1}{n^{2}T^{2}}\sum_{i,j}\sum_{t_{1},t_{2},t_{3},t_{4}}E\left[\left|E\left[u_{i,t_{1}}u_{i,t_{2}}u_{j,t_{3}}u_{j,t_{4}}\left|x_{i,\underline{T}},x_{j,\underline{T}},\gamma_{i},\gamma_{j}\right.\right]\right|\right]\leq M.
\]

\end{Aassumption}

\begin{Aassumption}\label{A_Ass_upperBound_Mom_uit} We have $E[\vert u_{i,t}\vert^{q}]\leq M$,
for all $i$, $t$, and some constants $q\geq8$ and $M>0$.\end{Aassumption}

\begin{Aassumption} \label{A_Ass_restricSerialCrossDep} Let $\delta=\delta_{n}\uparrow\infty$
be a diverging sequence such that $\sqrt{T}/\delta^{q-1}=o(1)$ and
$\delta\geq n^{\beta}$, for $\beta>2/q$. Let $e_{i,t}=u_{i,t}1\{\vert u_{i,t}\vert\leq\delta\}-E[u_{i,t}1\{\vert u_{i,t}\vert\leq\delta\}\vert\gamma_{i}]$.
Then: 
\[
\frac{1}{n^{k}}\sum_{i_{1},...,i_{k}}\sum_{t_{1},...,t_{k}}E\left[\vert E[e_{i_{1},t_{k}}e_{i_{1},t_{1}}e_{i_{2},t_{1}}e_{i_{2},t_{2}}e_{i_{3},t_{2}}\cdots e_{i_{k-1},t_{k-1}}e_{i_{k},t_{k-1}}e_{i_{k},t_{k}}\vert\gamma_{i_{1}},...,\gamma_{i_{k}}]\vert\right]\leq M^{k},
\]
for a sequence of integers $k=k_{n}\uparrow\infty$ and a constant
$M>0$, where indices $i_{1},...,i_{k}$ run from $1$ to $n$, and
indices $t_{1},...,t_{k}$ from $1$ to $T$. \end{Aassumption}

\begin{Aassumption}\label{A_Ass_regressor} There exists a constant
$M>0$ such that ${\displaystyle \left\Vert x_{i,t}\right\Vert \leq M}$,
$P$-a.s., for any $i$ and $t$.\end{Aassumption}

\begin{Aassumption}\label{A_Ass_H_theta} Under model $\mathcal{M}_{2}$,
a) there exists a constant $M>0$ such that ${\displaystyle \left\Vert h_{t}\right\Vert \leq M}$,
$P$-a.s., for all $t$. Moreover, b) ${\displaystyle \left\Vert \theta_{i}\right\Vert <M}$,
for all $i$.\end{Aassumption}

\begin{Aassumption} \label{A_Ass_RC1} Under model $\mathcal{M}_{2}$,
for a constant $M>0$ and for all $n,T\in\mathbb{N}$, we have: 
\[
\frac{1}{n^{2}T^{2}}\sum_{i,j}\sum_{t_{1},t_{2},t_{3},t_{4}}E\left[\Vert E[(x_{i,t_{1}}h_{t_{1}}')(x_{i,t_{2}}h_{t_{2}}')'(x_{j,t_{3}}h_{t_{3}}')(x_{j,t_{4}}h_{t_{4}}')'\vert\gamma_{i},\gamma_{j}]\Vert\right]\leq M.
\]
\end{Aassumption}

\begin{Aassumption} \label{A_Ass_missingAtRandom} The processes
$\left(I_{t}(\gamma)\right)$ and $(\varepsilon_{t}(\gamma))$ are
independent. \end{Aassumption}

\begin{Aassumption}\label{A_Ass_Dep_Obsevability} There exist constants
$\eta$, $\bar{\eta}\in(0,1]$ and $C_{1},C_{2},C_{3},C_{4}>0$ such
that, for all $\delta>0$ and $n,T\in\mathbb{N}$, we have: \\
 $a){\displaystyle \sup_{1\leq i\leq n}P\left[\Vert\frac{1}{T}\sum_{t}I_{i,t}\left(h_{t}h_{t}'-\Sigma_{h}\right)\Vert\geq\delta|\gamma_{i}\right]\leq C_{1}T\exp\{-C_{2}\delta^{2}T^{\eta}\}+C_{3}\delta^{-1}\exp\{-C_{4}T^{\bar{\eta}}\}.}$
\\
 Furthermore the same upper bound holds for \\
 $b){\displaystyle \sup_{1\leq i\leq n}P\left[\vert\frac{1}{T}\sum_{t}I_{i,t}-E[I_{i,t}|\gamma_{i}]\vert\geq\delta|\gamma_{i}\right]}$,\\
$c){\displaystyle \sup_{1\leq i\leq n}P\left[\vert\frac{1}{T}\sum_{t}\left(x_{i,t}x_{i,t}^{\prime}-E[x_{i,t}x_{i,t}^{\prime}|\gamma_{i}]\right)\vert\geq\delta|\gamma_{i}\right]}$.\end{Aassumption}

\begin{Aassumption}\label{A_Ass_Bounded_Obsevability} ${\displaystyle \inf_{1\leq i\leq n}E[I_{i,t}|\gamma_{i}]\geq M^{-1},}$
for all $n\in\mathbb{N}$ and a constant $M>0$.\end{Aassumption}

\begin{Aassumption}\label{A_Ass_CHI} The trimming constants $\chi_{1,T}$
and $\chi_{2,T}$ are such that ${\displaystyle \chi_{1,T}^{4}\chi_{2,\text{T}}^{2}=o\left(Tg\left(n,T\right)\right)}$.\end{Aassumption}

\begin{Aassumption} \label{A_Ass_boundII} We have ${\displaystyle \mu_{1}(W)=O_{p}(C_{n,T}^{-2})}$,
where $W=[w_{t,s}]$ is the $T\times T$ matrix with elements\linebreak{}
 ${\displaystyle w_{t,s}=\frac{1}{nT}\sum_{i}(I_{i,t}-\bar{I}_{t})(I_{i,s}-\bar{I}_{s})}$,
and ${\displaystyle \bar{I}_{t}=\frac{1}{n}\sum_{i}I_{i,t}}$. \end{Aassumption}

Assumption \ref{A_Ass_ErrTerm_u_M2} restricts serial dependence in
the bivariate process of error terms $\left(u_{i,t},u_{j,t}\right)$
of any two assets. It involves conditional expectations of products
of error terms $u_{i,t}$ for different dates and any pair of assets.
That assumption can be satisfied under weak serial dependence of the
errors $\left(u_{i,t},u_{j,t}\right)$, such as mixing, with mixing
size uniformly bounded across pairs $\left(i,j\right)$. Assumption
\ref{A_Ass_upperBound_Mom_uit} is an upper bound on higher-orders
moments of $u_{i,t}$, to control tail thickness. Assumption \ref{A_Ass_restricSerialCrossDep}
is a restriction on both serial and cross-sectional dependence of
the error terms and on the growth rates of $n$ and $T$. We use Assumptions
\ref{A_Ass_upperBound_Mom_uit} and \ref{A_Ass_restricSerialCrossDep}
to characterize the asymptotic behavior of the spectrum of the cross-sectional
variance-covariance matrix of errors under the rival models. Assumption
\ref{A_Ass_upperBound_Mom_uit} yields the so-called truncation and
centralization lemmas, which are used together with Assumption \ref{A_Ass_restricSerialCrossDep}
in the proof of \ref{Lemma_L1} building on \cite{Geman_1980}, \cite{Yin_Bai_Krishnaiah_1988}
and \cite{Bai_Yin_1993}. For those lemmas, we do not need a structure
on the error terms based on matrix transformations of i.i.d.\ random
variables as in \cite{Onatski_2010} and \cite{Ahn_Horenstein_2013}.
In \ref{sec:Appendix_CheckBlockDep}, we show that Assumptions \ref{A_Ass_ErrTerm_u_M2}
and \ref{A_Ass_restricSerialCrossDep} are satisfied under cross-sectional
block-dependence and time-series independence of the errors, provided
$n$ grows sufficiently faster than $T$. Under cross-sectional independence
of the errors, the condition $T/n=o\left(1\right)$ is enough as discussed
at the end of \ref{sec:Appendix_CheckBlockDep}. The arguments in
\cite{Yin_Bai_Krishnaiah_1988}, page 520, show that Assumption \ref{A_Ass_restricSerialCrossDep}
is also satisfied under i.i.d. $u_{i,t}$ and proportional asymptotics.
Assumptions \ref{A_Ass_regressor} and \ref{A_Ass_H_theta} require
upper bounds on regressor values, latent factors and factor loadings.
Assumption \ref{A_Ass_RC1} restricts serial dependence of the products
of latent factors and regressors. Recall that matrices $x_{i,t}h_{t}^{\prime}$
are zero-mean under Assumption \ref{A1-1}. In Assumption \ref{A_Ass_missingAtRandom},
we assume a missing-at-random design (\cite{Rubin_1976}), that is,
independence between unobservability and return generation. Another
design would require an explicit modeling of the link between the
unobservability mechanism and the return process of the continuum
of assets (\cite{Heckman_1979}); this would yield a nonlinear factor
structure. Assumption \ref{A_Ass_Dep_Obsevability} a) restricts the
serial dependence of the latent factors and the individual processes
of observability indicators. Specifically, Assumption \ref{A_Ass_Dep_Obsevability}
a) gives an upper bound for large deviation probabilities of the sample
average of zero-mean random matrices $h_{t}h_{t}'-\Sigma_{h}$, computed
over date with available observations for assets $i$, uniformly w.r.t.\ asset
$i$. It implies that the unbalanced sample moment of squared components
of the latent factor vector converges in probability to the corresponding
population moment at a rate $O_{p}(T^{-\eta/2}(\log T)^{c})$, for
some $c>0$. Assumptions \ref{A_Ass_Dep_Obsevability} b) and c) give
similar upper bounds for large-deviation probabilities of sample averages
of observability indicators and cross-moments of regressors uniformly
w.r.t. asset $i$. We use such assumptions to get the convergence
of time-series averages uniformly across assets as in GOS. Assumption
\ref{A_Ass_Bounded_Obsevability} implies that asymptotically the
fraction of the time period in which an asset return is observed is
bounded away from zero uniformly across assets, so that ${\displaystyle \tau_{i}=\plim_{T\rightarrow\infty}\tau_{i,T}=E[I_{i,t}|\gamma_{i}]^{-1}}$
is bounded uniformly across all assets as in GOS. Assumption \ref{A_Ass_CHI}
gives an upper bound on the divergence rate of the trimming constants.
Assumption \ref{A_Ass_boundII} controls the rate at which the largest
eigenvalue of the matrix with entries made of cross-sectional empirical
covariances of observability indicators vanishes to zero. The matrix
gathering those empirical covariances should not be associated to
an omitted factor structure.

\section{Proofs \label{sec:Proofs} \label{Appendix_Proof}}

We start by listing several results known from matrix theory. They
are used several times in the proofs.

(i) Weyl inequality: The singular-value version states that if $A$
and $B$ are $T\times n$ matrices, then \linebreak{}
 ${\displaystyle \mu_{i+j-1}[(A+B)(A+B)']^{1/2}\leq\mu_{i}(AA')^{1/2}+\mu_{j}(BB')^{1/2}}$,
for any $1\leq i,j\leq\min\{n,T\}$ such that $1\leq i+j\leq\min\{n,T\}+1$
(see Theorem 3.3.16 of \cite{Horn_Johnson_1985}). The Weyl inequality
for $i=k+1$ and $j=1$ yields: 
\begin{equation}
\mu_{k+1}[(A+B)(A+B)']^{1/2}\leq\mu_{k+1}(AA')^{1/2}+\mu_{1}(BB')^{1/2},\label{Weyl1}
\end{equation}
\begin{equation}
\mu_{k+1}[(A+B)(A+B)']^{1/2}\geq\mu_{k+1}(AA')^{1/2}-\mu_{1}(BB')^{1/2},\label{Weyl2}
\end{equation}
for any $T\times n$ matrices $A$ and $B$ and integer $k$ such
that $0\leq k\leq\min\{n,T\}-1$. We also use Weyl inequality for
eigenvalues: for any $T\times T$ symmetric matrices $A$ and $B$
we have $\mu_{i+j-1}(A+B)\leq\mu_{i}(A)+\mu_{j}(B)$, for any $1\leq i,j\leq T$
such that $i+j\leq T+1$ (see Theorem 8.4.11 in \cite{Bernstein_2009}).

(ii) Equality between largest eigenvalue and operator norm: The largest
eigenvalue $\mu_{1}(A)$ of a symmetric positive semi-definite matrix
$A$ is equal to its operator norm ${\displaystyle \Vert A\Vert_{op}=\underset{x:\Vert x\Vert=1}{\max}\Vert Ax\Vert}$.
Besides, ${\displaystyle \Vert A\Vert_{op}\leq\Vert A\Vert}$ for
any square matrix $A$, where $\Vert\cdot\Vert$ is the Frobenius
norm (see e.g. \cite{Meyer_2000}).

(iii) Inequalities for the eigenvalues of matrix products: If $A$
and $B$ are $m\times m$ positive semidefinite and positive definite
matrices, respectively, 
\begin{equation}
\mu_{k}\left(A\right)\mu_{m}\left(B\right)\leq\mu_{k}\left(AB\right)\leq\mu_{k}\left(A\right)\mu_{1}\left(B\right),\label{MO1}
\end{equation}
for $k=1,2,...,m$ (see Fact 8.19.17 in \cite{Bernstein_2009}).

(iv) Courant-Fischer min-max Theorem: If $A$ is a $T\times T$ symmetric
matrix, we have, for $k=1,...,T$, 
\begin{equation}
\mu_{k}(A)=\underset{\mathcal{G}:dim(\mathcal{G})=T-k+1}{\min}\ \underset{x\in\mathcal{G}:\Vert x\Vert=1}{\max}\ x'Ax,\label{CourantFischer_minMax}
\end{equation}
where the minimization is w.r.t. the $(T-k+1)$-dimensional linear
subspace $\mathcal{G}$ of $\mathbb{R}^{T}$ (see e.g.\ \cite{Bernstein_2009}).
The max-min formulation states: 
\begin{equation}
\mu_{k}(A)=\underset{\mathcal{G}:dim(\mathcal{G})=k}{\max}\ \underset{x\in\mathcal{G}:\Vert x\Vert=1}{\min}\ x'Ax,\label{CourantFisher_maxMin}
\end{equation}
where the maximization is w.r.t. the $k$-dimensional linear subspace
$\mathcal{G}$ of $\mathbb{R}^{T}$.

(v) Courant-Fischer formula: If $A$ is a $T\times T$ symmetric matrix,
we have, for $k=1,...,T$, 
\begin{equation}
\mu_{k}(A)=\underset{x\in\mathcal{F}_{k-1}^{\perp}:\Vert x\Vert=1}{\max}\ x'Ax,\label{R2-1}
\end{equation}
where $\mathcal{F}_{k}^{\perp}$ is the orthogonal complement of $\mathcal{F}_{k}$,
with $\mathcal{F}_{k}$ being the linear space spanned by the eigenvectors
associated to the $k$ largest eigenvalues of matrix $A$, and $\mathcal{F}_{0}\equiv\mathbb{R}^{T}$.

\subsection{Proof of \ref{Prop_ModelSelectionRule} \label{Appendix_ProofProp1}}

\noindent \textbf{a) }The OLS estimator of $\beta_{i}$ in matrix
notation is ${\displaystyle \hat{\beta}_{i}=\left(\tilde{X}_{i}^{\prime}\tilde{X}_{i}\right)^{-1}\tilde{X}_{i}^{\prime}\tilde{R}_{i}}$,
with $\tilde{X}_{i}=\boldsymbol{I}_{i}\odot X_{i}$ and $\tilde{R}_{i}=\boldsymbol{I}_{i}\odot R_{i}$,
where $\boldsymbol{I}_{i}$ is the $T\times1$ vector of indicators
$I_{i,t}$ for asset $i$, and $\odot$ is the Hadamard product. We
get the vector of residuals ${\displaystyle \hat{\varepsilon}_{i}=R_{i}-X_{i}\left(\tilde{X}_{i}^{\prime}\tilde{X}_{i}\right)^{-1}\tilde{X}_{i}^{\prime}\tilde{R}_{i}}$.
Then, we have ${\displaystyle \bar{\varepsilon}_{i}=\boldsymbol{I}_{i}\odot\hat{\varepsilon}_{i}}=M_{\tilde{X}_{i}}\tilde{R}_{i}=M_{\tilde{X}_{i}}\tilde{\varepsilon}_{i}$,
where $\tilde{\varepsilon}_{i}=\boldsymbol{I}_{i}\odot\varepsilon_{i}$
and $M_{\tilde{X}_{i}}=I_{T}-P_{\tilde{X}_{i}}$, with $P_{\tilde{X}_{i}}=\tilde{X}_{i}\left(\tilde{X}_{i}^{\prime}\tilde{X}_{i}\right)^{-1}\tilde{X}_{i}^{\prime}$.
Thus, under $\mathcal{M}_{1}$, we have the decomposition $\bold{1}_{i}^{\chi}\bar{\varepsilon}_{i}=\tilde{\varepsilon}_{i}-(1-\bold{1}_{i}^{\chi})\tilde{\varepsilon}_{i}-\bold{1}_{i}^{\chi}P_{\tilde{X}_{i}}\tilde{\varepsilon}_{i}$.
From Weyl inequality (\ref{Weyl1}) with $k=0$, and the inequality
between matrix norms, we get: 
\begin{equation}
\mu_{1}\left(\frac{1}{nT}\sum_{i}\bold{1}_{i}^{\chi}\bar{\varepsilon}_{i}\bar{\varepsilon}_{i}'\right)^{1/2}\leq\mu_{1}\left(\frac{1}{nT}\sum_{i}\tilde{\varepsilon}_{i}\tilde{\varepsilon}_{i}'\right)^{1/2}+I_{1}^{1/2}+I_{2}^{1/2},\label{up1}
\end{equation}
where: 
\begin{eqnarray}
I_{1}:=\Vert\frac{1}{nT}\sum_{i}(1-\bold{1}_{i}^{\chi})\tilde{\varepsilon}_{i}\tilde{\varepsilon}_{i}'\Vert,\qquad I_{2}:=\Vert\frac{1}{nT}\sum_{i}\bold{1}_{i}^{\chi}P_{\tilde{X}_{i}}\tilde{\varepsilon}_{i}\tilde{\varepsilon}_{i}'P_{\tilde{X}_{i}}\Vert.\label{defI}
\end{eqnarray}
We bound the largest eigenvalue of matrix ${\displaystyle \frac{1}{nT}\sum_{i}\tilde{\varepsilon}_{i}\tilde{\varepsilon}_{i}'}$
and the remainder terms $I_{1}$ and $I_{2}$ in the next two lemmas.

\begin{lemma} \label{Lemma_L1} Under model $\mathcal{M}_{1}$ and
Assumptions \ref{A1nT}, \ref{A_Ass_upperBound_Mom_uit}, \ref{A_Ass_restricSerialCrossDep},
\ref{A_Ass_missingAtRandom}, as $n,T\rightarrow\infty$, we have
${\displaystyle \mu_{1}\left(\frac{1}{nT}\sum_{i}\tilde{\varepsilon}_{i}\tilde{\varepsilon}_{i}'\right)=}$\linebreak{}
${\displaystyle O_{p}(C_{n,T}^{-2})}$. \end{lemma}

\begin{lemma} \label{Lemma_L2} Under model $\mathcal{M}_{1}$ and
Assumptions \ref{A1nT}, \ref{A_Ass_ErrTerm_u_M2}, \ref{A_Ass_upperBound_Mom_uit},
\ref{A_Ass_regressor}, \ref{A_Ass_Dep_Obsevability} b), c) and \ref{A_Ass_Bounded_Obsevability},
as $n,T\rightarrow\infty$, we have: (i) $I_{1}=O_{p}(T^{-\bar{b}})$,
for any $\bar{b}>0$; (ii) ${\displaystyle I_{2}=O_{p}(\chi_{1,T}^{4}\chi_{2,T}^{2}/T)}$.
\end{lemma}

From Inequality (\ref{up1}) and Lemmas 1 and 2, we get ${\displaystyle \xi=O_{p}(C_{n,T}^{-2})+O_{p}(\frac{\chi_{1,T}^{4}\chi_{2,T}^{2}}{T})-g(n,T).}$
Then, from Assumption \ref{A_Ass_CHI} on the trimming constants and
the properties of penalty function $g(n,T)$, \ref{Prop_ModelSelectionRule}(a)
follows.

\textbf{b)} Let us now consider the case $\mathcal{M}_{2}$. We have
$\bar{\varepsilon}_{i}=M_{\tilde{X}_{i}}\tilde{\varepsilon}_{i}$
and $\tilde{\varepsilon}_{i}=\tilde{H}_{i}\theta_{i}+\tilde{u}_{i}$,
where $\tilde{H}_{i}=\bold{I}_{i}\odot H$ and $H$ is the $T\times m$
matrix of latent factor values, with $m\geq1$. Hence, we have the
decomposition $\bold{1}_{i}^{\chi}\bar{\varepsilon}_{i}=\tilde{H}_{i}\theta_{i}+\tilde{u}_{i}-(1-\bold{1}_{i}^{\chi})\tilde{\varepsilon}_{i}-\bold{1}_{i}^{\chi}P_{\tilde{X}_{i}}\tilde{H}_{i}\theta_{i}-\bold{1}_{i}^{\chi}P_{\tilde{X}_{i}}\tilde{u}_{i}$.
By using Weyl inequality (\ref{Weyl2}) with $k=0$, and the inequality
between matrix norms, we get: 
\begin{equation}
\mu_{1}\left(\frac{1}{nT}\sum_{i}\bold{1}_{i}^{\chi}\bar{\varepsilon}_{i}\bar{\varepsilon}_{i}'\right)^{1/2}\geq\mu_{1}\left(\frac{1}{nT}\sum_{i}\tilde{H}_{i}\theta_{i}\theta_{i}'\tilde{H}_{i}'\right)^{1/2}-\mu_{1}\left(\frac{1}{nT}\sum_{i}\tilde{u}_{i}\tilde{u}_{i}'\right)^{1/2}-I^{1/2},\label{down1}
\end{equation}
where $I^{1/2}=I_{1}^{1/2}+I_{3}^{1/2}+I_{4}^{1/2}$, term $I_{1}$
is defined as in (\ref{defI}), and 
\[
I_{3}^{1/2}:=\Vert\frac{1}{nT}\sum_{i}\bold{1}_{i}^{\chi}P_{\tilde{X}_{i}}\tilde{H}_{i}\theta_{i}\theta_{i}'\tilde{H}_{i}'P_{\tilde{X}_{i}}\Vert^{1/2},\qquad I_{4}^{1/2}:=\Vert\frac{1}{nT}\sum_{i}\bold{1}_{i}^{\chi}P_{\tilde{X}_{i}}\tilde{u}_{i}\tilde{u}_{i}'P_{\tilde{X}_{i}}\Vert^{1/2}.
\]
By \ref{Lemma_L1} applied on $\tilde{u}_{i}$ instead of $\tilde{\varepsilon}_{i}$,
we have ${\displaystyle \mu_{1}\left(\frac{1}{nT}\sum_{i}\tilde{u}_{i}\tilde{u}_{i}'\right)=O_{p}(C_{n,T}^{-2})}$.
Moreover, from the next \ref{Lemma_L3} and Assumption \ref{A_Ass_CHI}
on the trimming constants, we get $I=o_{p}(g\left(n,T\right))$ under
$\mathcal{M}_{2}$.

\begin{lemma} \label{Lemma_L3} Under model $\mathcal{M}_{2}$ and
Assumptions \ref{A1nT}, \ref{A_Ass_upperBound_Mom_uit}, \ref{A_Ass_regressor},
\ref{A_Ass_H_theta} and \ref{A_Ass_RC1}, as $n,T\rightarrow\infty$,
we have: (i) $I_{1}=O_{p}(T^{-\bar{b}})$, for any $\bar{b}>0$; (ii)
${\displaystyle I_{3}=O_{p}(\chi_{1,T}^{4}\chi_{2,T}^{2}/T)}$; (iii)
${\displaystyle I_{4}=O_{p}(\chi_{1,T}^{4}\chi_{2,T}^{2}/T)}$. \end{lemma}

The next \ref{Lemma_L4} provides a lower bound for the first term
in the r.h.s. of Inequality (\ref{down1}).

\begin{lemma} \label{Lemma_L4} Under model $\mathcal{M}_{2}$ and
Assumptions \ref{ass_ftheta}, \ref{A_Ass_Dep_Obsevability} and \ref{A_Ass_Bounded_Obsevability},
we have ${\displaystyle \mu_{1}\left(\frac{1}{nT}\sum_{i}\tilde{H}_{i}\theta_{i}\theta_{i}'\tilde{H}_{i}'\right)\geq C}$,
w.p.a. $1$, for a constant $C>0$. \end{lemma}

Then, from Inequality (\ref{down1}) and \ref{Lemma_L4}, we get ${\displaystyle \xi\geq C/2}$,
w.p.a. $1$, and \ref{Prop_ModelSelectionRule}(b) follows.

\subsection{Proof of \ref{Prop_propgeneral} \label{Appendix_ProofProp3}}

We prove \ref{Prop_propgeneral} along similar lines as \ref{Prop_ModelSelectionRule}
by exploiting the Weyl inequalities (\ref{Weyl1}) and (\ref{Weyl2})
for a generic $k$.

\textbf{a)} Let us first consider the case $\mathcal{M}_{1}(k)$.
We have $\bar{\varepsilon}_{i}=M_{\tilde{X}_{i}}\tilde{\varepsilon}_{i}$
and $\tilde{\varepsilon}_{i}=\tilde{H}_{i}\theta_{i}+\tilde{u}_{i}$,
where $\tilde{H}_{i}=\bold{I}_{i}\odot H$ and $H$ is the $T\times k$
matrix of latent factor values. Then, $\bold{1}_{i}^{\chi}\bar{\varepsilon}_{i}=\tilde{H}_{i}\theta_{i}+\tilde{u}_{i}-(1-\bold{1}_{i}^{\chi})\tilde{\varepsilon}_{i}-\bold{1}_{i}^{\chi}P_{\tilde{X}_{i}}\tilde{H}_{i}\theta_{i}-\bold{1}_{i}^{\chi}P_{\tilde{X}_{i}}\tilde{u}_{i}$.
From Weyl inequalities (\ref{Weyl1}) and (\ref{Weyl2}), and the
inequality between matrix norms, we get: 
\begin{equation}
\mu_{k+1}\left(\frac{1}{nT}\sum_{i}\bold{1}_{i}^{\chi}\bar{\varepsilon}_{i}\bar{\varepsilon}_{i}'\right)^{1/2}\leq\mu_{k+1}\left(\frac{1}{nT}\sum_{i}\tilde{H}_{i}\theta_{i}\theta_{i}'\tilde{H}_{i}'\right)^{1/2}+\mu_{1}\left(\frac{1}{nT}\sum_{i}\tilde{u}_{i}\tilde{u}_{i}'\right)^{1/2}+I^{1/2},\label{up}
\end{equation}
where ${\displaystyle I^{1/2}=I_{1}^{1/2}+I_{3}^{1/2}+I_{4}^{1/2}}$
and terms $I_{1}$, $I_{3}$ and $I_{4}$ are defined as in the proof
of \ref{Prop_ModelSelectionRule}. Since model $\mathcal{M}_{1}(k)$
is included in model $\mathcal{M}_{2}$ for any $k\geq1$, we get
$I=o_{p}(g\left(n,T\right))$, from \ref{Lemma_L3} and Assumption
\ref{A_Ass_CHI} on the trimming constants. Moreover, ${\displaystyle \mu_{1}\left(\frac{1}{nT}\sum_{i}\tilde{u}_{i}\tilde{u}_{i}'\right)=O_{p}(C_{n,T}^{-2})}$
by \ref{Lemma_L1} with $\tilde{u}_{i}$ replacing $\tilde{\varepsilon}_{i}$.
The first term in the r.h.s. of (\ref{up}) is bounded by the next
lemma.

\begin{lemma} \label{lemma_L4} Under model $\mathcal{M}_{1}(k)$
and Assumptions \ref{A_Ass_H_theta} and \ref{A_Ass_boundII}, we
have ${\displaystyle \mu_{k+1}\left(\frac{1}{nT}\sum_{i}\tilde{H}_{i}\theta_{i}\theta_{i}'\tilde{H}_{i}'\right)=O_{p}(C_{n,T}^{-2})}$.
\end{lemma} The bound in \ref{lemma_L4} would be trivial in the
case $\tilde{H}_{i}=H$, i.e., with a balanced panel, because in that
case ${\displaystyle \mu_{k+1}\left(\frac{1}{nT}\sum_{i}H\theta_{i}\theta_{i}'H'\right)=0}$
under $\mathcal{M}_{1}(k)$.

From Inequality (\ref{up}) and \ref{lemma_L4}, we get $\xi=O_{p}(C_{n,T}^{-2})+o_{p}(g\left(n,T\right))-g(n,T)$.
Then, by the properties of $g(n,T)$, \ref{Prop_propgeneral}(a) follows.

\textbf{b)} Let us now consider the case $\mathcal{M}_{2}(k)$. We
have $\bar{\varepsilon}_{i}=M_{\tilde{X}_{i}}\tilde{\varepsilon}_{i}$
and $\tilde{\varepsilon}_{i}=\tilde{H}_{i}\theta_{i}+\tilde{u}_{i}$,
where $\tilde{H}_{i}=\bold{I}_{i}\odot H$ and $H$ is the $T\times m$
matrix of latent factor values, with $m\geq k+1$. By similar arguments
as in part \textbf{a)}, using Weyl inequalities (\ref{Weyl1}) and
(\ref{Weyl2}), and the inequality between matrix norms, we get: 
\begin{equation}
\mu_{k+1}\left(\frac{1}{nT}\sum_{i}\bold{1}_{i}^{\chi}\bar{\varepsilon}_{i}\bar{\varepsilon}_{i}'\right)^{1/2}\geq\mu_{k+1}\left(\frac{1}{nT}\sum_{i}\tilde{H}_{i}\theta_{i}\theta_{i}'\tilde{H}_{i}'\right)^{1/2}-\mu_{1}\left(\frac{1}{nT}\sum_{i}\tilde{u}_{i}\tilde{u}_{i}'\right)^{1/2}-I^{1/2}.\label{down}
\end{equation}
As in part \textbf{a)} we have ${\displaystyle \mu_{1}\left(\frac{1}{nT}\sum_{i}\tilde{u}_{i}\tilde{u}_{i}'\right)=O_{p}(C_{n,T}^{-2})}$
and $I=o_{p}(g(n,T))$.

\begin{lemma} \label{lemma_L5} Under model $\mathcal{M}_{2}(k)$
and Assumptions \ref{ass_ftheta}(i), \ref{ass3}, \ref{A_Ass_Dep_Obsevability}
and \ref{A_Ass_Bounded_Obsevability}, we have \linebreak{}
${\displaystyle \mu_{k+1}\left(\frac{1}{nT}\sum_{i}\tilde{H}_{i}\theta_{i}\theta_{i}'\tilde{H}_{i}'\right)\geq C}$,
w.p.a. $1$, for a constant $C>0$. \end{lemma}

Then, from Inequality (\ref{down}) and \ref{lemma_L5}, we get ${\displaystyle \xi\geq C/2}$,
w.p.a. $1$, and \ref{Prop_propgeneral}(b) follows.

\subsection{Proof of \ref{Prop_choice} \label{Appendix_ProofProp4}}

Let us define the events $A_{k}=\{\xi(k)\geq0\}$, $k=0,...,k_{0}-1$,
and $A_{k_{0}}=\{\xi(k_{0})<0\}$. We have ${\displaystyle P[\hat{k}=k_{0}]=P[\{A_{0}\cap A_{1}\cap...\cap A_{k_{0}-1}\}\cap A_{k_{0}}]}$.
For generic events $B$ and $C$, we have $P[B\cap C]=P[B]+P[C]-P[B\cup C]$,
and we conclude that $P[B\cap C]\rightarrow1$ if both $P[B]$ and
$P[C]$ converge to 1 since $P[B\cup C]\geq P[B]$ and $P[B\cup C]\geq P[C]$.
Applying repeatedly this argument to the probability $P[\{A_{0}\cap A_{1}\cap...\cap A_{k_{0}-1}\}\cap A_{k_{0}}]$
yields $P[\hat{k}=k_{0}]\rightarrow1$ since $P[A_{k}]\rightarrow1$,
$k=0,...,k_{0}-1$, and $P[A_{k_{0}}]\rightarrow1$, under $\mathcal{M}_{1}(k_{0})$
from Proposition 2.

\subsection{Proof of \ref{Lemma_L1}}

We prove: 
\begin{equation}
\underset{n,T\rightarrow\infty}{\lim\sup}\ \mu_{1}\left(\frac{1}{n}\tilde{\mathcal{E}}\tilde{\mathcal{E}}'\right)\leq C,\ a.s.,\label{TP}
\end{equation}
for a constant $C<\infty$, where $\tilde{\mathcal{E}}$ is the $T\times n$
matrix with elements $\tilde{\varepsilon}_{i,t}=I_{i,t}\varepsilon_{i,t}$.
Then, the statement of \ref{Lemma_L1} follows. To show (\ref{TP}),
we follow similar arguments as in \cite{Geman_1980}, \cite{Yin_Bai_Krishnaiah_1988},
and \cite{Bai_Yin_1993}.

We first establish suitable versions of the so-called truncation and
centralization lemmas. We denote by $\Xi$ and $E$ the $T\times n$
matrices with elements $(\xi_{i,t})$ and $(e_{i,t})$, where $\xi_{i,t}=\varepsilon_{i,t}1\{\vert\varepsilon_{i,t}\vert\leq\delta\}$
and $e_{i,t}=\xi_{i,t}-E[\xi_{i,t}\vert\gamma_{i}]$, and $\delta=\delta_{n}\uparrow\infty$
is a diverging sequence as in Assumption \ref{A_Ass_restricSerialCrossDep}.
Let us define matrices $\tilde{E}$ and $\tilde{\Xi}$ with elements
$\left(I_{i,t}e_{i,t}\right)$ and $\left(I_{i,t}\xi_{i,t}\right)$
by analogy to $\tilde{\mathcal{E}}$. \ref{Ltruncation} shows that
we can substitute the truncated $\xi_{i,t}$ and $I_{i,t}\xi_{i,t}$
for $\varepsilon_{i,t}$ and $I_{i,t}\varepsilon_{i,t}$, and \ref{Lcentralization}
shows that we can substitute the centered $I_{i,t}e_{i,t}$ for the
$I_{i,t}\xi_{i,t}$ to show boundedness of the largest eigenvalue
in (\ref{TP}). We prove Lemmas 7 and 8 in the supplementary material.

\begin{lemma} \label{Ltruncation} Under Assumption \ref{A_Ass_upperBound_Mom_uit},
if $\delta=\delta_{n}$ is such that $\delta\geq n^{\beta}$ for $\beta>2/q$,
then: (i) $P\left(\mathcal{E}\neq\Xi\ i.o.\right)=0$, and (ii) $P\left(\tilde{\mathcal{E}}\neq\tilde{\Xi}\ i.o.\right)=0$,
where $i.o.$ means infinitely often for $n=1,2,...$. \end{lemma}

\begin{lemma} \label{Lcentralization} Under Assumption \ref{A_Ass_upperBound_Mom_uit},
if $\delta=\delta_{n}\uparrow\infty$ such that $\sqrt{T}/\delta^{q-1}=o(1)$,
then: 
\[
\mu_{1}\left(\frac{1}{n}\tilde{{\Xi}}\tilde{{\Xi}}'\right)=\mu_{1}\left(\frac{1}{n}\tilde{E}\tilde{E}'\right)+o(1),\quad a.s.
\]
\end{lemma}

\noindent From \ref{Ltruncation}(ii) and \ref{Lcentralization},
condition (\ref{TP}) is implied by: 
\begin{equation}
\underset{n,T\rightarrow\infty}{\lim\sup}\ \mu_{1}\left(\frac{1}{n}\tilde{E}\tilde{E}'\right)\leq C,\ a.s.,\label{TP1}
\end{equation}
for a constant $C<\infty$. Now, we use that the upper bound (\ref{TP1})
is implied by the condition: 
\begin{equation}
\sum_{n=1}^{\infty}E\left[\left(\mu_{1}\left(\frac{1}{n}\tilde{E}\tilde{E}'\right)/C\right)^{k}\right]<\infty,\label{cond}
\end{equation}
for an increasing sequence of integers $k=k_{n}\uparrow\infty$. To
prove the validity of condition (\ref{cond}), we use that: 
\begin{eqnarray*}
\mu_{1}\left(\frac{1}{n}\tilde{E}\tilde{E}'\right)^{k} & \leq Tr\left[\left(\frac{1}{n}\tilde{E}\tilde{E}'\right)^{k}\right] & =\frac{1}{n^{k}}\sum_{i_{1},...,i_{k}}\sum_{t_{1},...,t_{k}}\tilde{e}_{i_{1},t_{k}}\tilde{e}_{i_{1},t_{1}}\tilde{e}_{i_{2},t_{1}}\tilde{e}_{i_{2},t_{2}}\tilde{e}_{i_{3},t_{2}}\cdots\tilde{e}_{i_{k-1},t_{k-1}}\tilde{e}_{i_{k},t_{k-1}}\tilde{e}_{i_{k},t_{k}},
\end{eqnarray*}
for any integer $k$, where in the summation the indices $i_{1}$,
..., $i_{k}$ run from $1$ to $n$, and indices $t_{1}$, ..., $t_{k}$
run from $1$ to $T$. Therefore, from Assumption \ref{A_Ass_missingAtRandom}:
\begin{eqnarray*}
E\left[\mu_{1}\left(\frac{1}{n}\tilde{{E}}\tilde{{E}}'\right)^{k}\right]\leq\frac{1}{n^{k}}\sum_{i_{1},...,i_{k}}\sum_{t_{1},...,t_{k}}E\left[\vert E[e_{i_{1},t_{k}}e_{i_{1},t_{1}}e_{i_{2},t_{1}}e_{i_{2},t_{2}}e_{i_{3},t_{2}}\cdots e_{i_{k-1},t_{k-1}}e_{i_{k},t_{k-1}}e_{i_{k},t_{k}}\vert\gamma_{i_{1}},...,\gamma_{i_{k}}]\vert\right].
\end{eqnarray*}
Then, we get ${\displaystyle E\left[\mu_{1}\left(\frac{1}{n}\tilde{{E}}\tilde{{E}}'\right)^{k}\right]\leq M}{}^{k}$,
for the sequence $k=k_{n}$ defined in Assumption \ref{A_Ass_restricSerialCrossDep}.
Condition (\ref{cond}) holds for any $C>M$, and the conclusion follows.

\subsection{Proof of \ref{Lemma_L2}}

i) We have ${\displaystyle I_{1}^{2}=}\Vert\frac{1}{nT}\sum_{i}(1-\bold{1}_{i}^{\chi})\tilde{\varepsilon}_{i}\tilde{\varepsilon}_{i}'\Vert^{2}=\frac{1}{n^{2}T^{2}}\sum_{i,j}(1-\bold{1}_{i}^{\chi})(1-\bold{1}_{j}^{\chi})(\tilde{\varepsilon}_{i}'\tilde{\varepsilon}_{j})^{2}=\frac{1}{n^{2}T^{2}}\sum_{i,j}\sum_{t_{1},t_{2}}(1-\bold{1}_{i}^{\chi})(1-\bold{1}_{j}^{\chi})I_{i,t_{1}}I_{j,t_{1}}I_{i,t_{2}}I_{j,t_{2}}\varepsilon_{i,t_{1}}\varepsilon_{j,t_{1}}\varepsilon_{i,t_{2}}\varepsilon_{j,t_{2}}.$
By the Cauchy-Schwarz inequality: 
\begin{eqnarray*}
E[I_{1}^{2}]\leq\frac{1}{n^{2}T^{2}}\sum_{i,j}\sum_{t_{1},t_{2}}E[1-\bold{1}_{i}^{\chi}]^{1/4}E[1-\bold{1}_{j}^{\chi}]^{1/4}E[\varepsilon_{i,t_{1}}^{8}]^{1/8}E[\varepsilon_{j,t_{1}}^{8}]^{1/8}E[\varepsilon_{i,t_{2}}^{8}]^{1/8}E[\varepsilon_{j,t_{2}}^{8}]^{1/8}.
\end{eqnarray*}
Now, we have $E[\varepsilon_{i,t}^{8}]\leq M$ from Assumption \ref{A_Ass_upperBound_Mom_uit}
and ${\displaystyle E[1-\bold{1}_{i}^{\chi}]=P[\bold{1}_{i}^{\chi}=0]=O(T^{-\bar{b}})}$
for any $\bar{b}>0$, uniformly in $i$ and $t$ from Assumptions
\ref{A_Ass_regressor}, \ref{A_Ass_Dep_Obsevability}c) and \ref{A_Ass_Bounded_Obsevability}
(see Lemma 7 in GOS). Then, ${\displaystyle I_{1}=O_{p}(T^{-\bar{b}})}$
for any $\bar{b}>0$.

ii) We have: 
\begin{eqnarray*}
I_{2}^{2} & = & \Vert\frac{1}{nT}\sum_{i}\bold{1}_{i}^{\chi}P_{\tilde{X}_{i}}\tilde{\varepsilon}_{i}\tilde{\varepsilon}_{i}'P_{\tilde{X}_{i}}\Vert^{2}=\frac{1}{n^{2}T^{2}}\sum_{i,j}\bold{1}_{i}^{\chi}\bold{1}_{j}^{\chi}Tr\left[P_{\tilde{X}_{i}}\tilde{\varepsilon}_{i}\tilde{\varepsilon}_{i}'P_{\tilde{X}_{i}}P_{\tilde{X}_{j}}\tilde{\varepsilon}_{j}\tilde{\varepsilon}_{j}'P_{\tilde{X}_{j}}\right]\\
 & = & \frac{1}{n^{2}T^{2}}\sum_{i,j}\bold{1}_{i}^{\chi}\bold{1}_{j}^{\chi}\frac{\tau_{i,T}^{2}\tau_{j,T}^{2}}{\tau_{T,ij}^{2}}Tr\left[\hat{Q}_{x,i}^{-1}\left(\frac{{\tilde{X}_{i}}'\tilde{\varepsilon}_{i}}{\sqrt{T}}\right)\left(\frac{\tilde{\varepsilon}_{i}'{\tilde{X}_{i}}}{\sqrt{T}}\right)\hat{Q}_{x,i}^{-1}\hat{Q}_{x,ij}\hat{Q}_{x,j}^{-1}\left(\frac{{\tilde{X}_{j}}'\tilde{\varepsilon}_{j}}{\sqrt{T}}\right)\right.\left.\left(\frac{\tilde{\varepsilon}_{j}'{\tilde{X}_{j}}}{\sqrt{T}}\right)\hat{Q}_{x,j}^{-1}\hat{Q}_{x,ji}\right],
\end{eqnarray*}
where ${\displaystyle \hat{Q}_{x,ij}=\frac{1}{T_{i,j}}\sum_{t}I_{i,t}I_{j,t}x_{i,t}x_{j,t}'}$
and $\tau_{ij,T}=T/T_{ij}$. By using $Tr(AB')\leq\Vert A\Vert\Vert B\Vert$,
$\bold{1}_{i}^{\chi}\Vert\hat{Q}_{x,i}^{-1}\Vert\leq C\chi_{1,T}^{2}$,
$\bold{1}_{i}^{\chi}\tau_{i,T}\leq\chi_{2,T}$, $\Vert x_{i,t}\Vert\leq M$
(Assumption \ref{A_Ass_regressor}), $\tau_{ij,T}\geq1$, for all
$i$ and $t$, we get: 
\begin{eqnarray*}
I_{2}^{2} & \leq & \frac{C\chi_{1,T}^{8}\chi_{2,T}^{4}}{n^{2}T^{2}}\sum_{i,j}\Vert\frac{\tilde{\varepsilon}_{i}'{\tilde{X}_{i}}}{\sqrt{T}}\Vert^{2}\Vert\frac{\tilde{\varepsilon}_{j}'{\tilde{X}_{j}}}{\sqrt{T}}\Vert^{2}\\
 & = & \frac{C\chi_{1,T}^{8}\chi_{2,T}^{4}}{n^{2}T^{4}}\sum_{i,j}\sum_{t_{1},t_{2},t_{3},t_{4}}I_{i,t_{1}}I_{i,t_{2}}I_{j,t_{3}}I_{j,t_{4}}\varepsilon_{i,t_{1}}\varepsilon_{i,t_{2}}\varepsilon_{j,t_{3}}\varepsilon_{j,t_{4}}x_{i,t_{1}}'x_{i,t_{2}}x_{j,t_{3}}'x_{j,t_{4}}.
\end{eqnarray*}
Thus: 
\begin{eqnarray*}
 &  & E[I_{2}^{2}\vert I_{i,\underline{T}},I_{j,\underline{T}},x_{i,\underline{T}},x_{j,\underline{T}},\gamma_{i},\gamma_{j}]\\
 & \leq & \frac{C\chi_{1,T}^{8}\chi_{2,T}^{4}}{n^{2}T^{4}}\sum_{i,j}\sum_{t_{1},t_{2},t_{3},t_{4}}\Vert x_{i,t_{1}}\Vert\Vert x_{i,t_{2}}\Vert\Vert x_{j,t_{3}}\Vert\Vert x_{j,t_{4}}\Vert\vert E[\varepsilon_{i,t_{1}}\varepsilon_{i,t_{2}}\varepsilon_{j,t_{3}}\varepsilon_{j,t_{4}}\vert x_{i,\underline{T}},x_{j,\underline{T}},\gamma_{i},\gamma_{j}]\vert.
\end{eqnarray*}
Hence ${\displaystyle E[I_{2}^{2}]\leq\frac{CM^{5}\chi_{1,T}^{8}\chi_{2,T}^{4}}{T^{2}},}$
from Assumptions \ref{A_Ass_ErrTerm_u_M2} and \ref{A_Ass_regressor}.
It follows ${\displaystyle E[I_{2}^{2}]=O(\frac{\chi_{1,T}^{8}\chi_{2,T}^{4}}{T^{2}})}$,
which implies ${\displaystyle I_{2}=O_{p}(\frac{\chi_{1,T}^{4}\chi_{2,T}^{2}}{T})}$.

\subsection{Proof of \ref{Lemma_L3}}

i) The proof of \ref{Lemma_L3}(i) is the same as that of \ref{Lemma_L2}(i),
since the bound $E[\vert\varepsilon_{i,t}\vert^{8}]\leq M$ applies
under $\mathcal{M}_{2}$ as well (Assumptions \ref{A_Ass_upperBound_Mom_uit}
and \ref{A_Ass_H_theta}).

ii) The proof of \ref{Lemma_L3}(ii) is similar to that of \ref{Lemma_L2}(ii),
by replacing $\tilde{\varepsilon}_{i}$ with $\tilde{H}_{i}\theta_{i}$
and using Assumption \ref{A_Ass_RC1}. We have: 
\begin{eqnarray*}
I_{2}^{2} & = & \Vert\frac{1}{nT}\sum_{i}\bold{1}_{i}^{\chi}P_{\tilde{X}_{i}}\tilde{H}_{i}\theta_{i}\theta_{i}'\tilde{H}_{i}'P_{\tilde{X}_{i}}\Vert^{2}=\frac{1}{n^{2}T^{2}}\sum_{i,j}\bold{1}_{i}^{\chi}\bold{1}_{j}^{\chi}Tr\left[P_{\tilde{X}_{i}}\tilde{H}_{i}\theta_{i}\theta_{i}'\tilde{H}_{i}'P_{\tilde{X}_{i}}P_{\tilde{X}_{j}}\tilde{H}_{j}\theta_{j}\theta_{j}'\tilde{H}_{j}'P_{\tilde{X}_{j}}\right]\\
 & = & \frac{1}{n^{2}T^{2}}\sum_{i,j}\bold{1}_{i}^{\chi}\bold{1}_{j}^{\chi}\frac{\tau_{i,T}^{2}\tau_{j,T}^{2}}{\tau_{T,ij}^{2}}Tr\left[\hat{Q}_{x,i}^{-1}\left(\frac{{\tilde{X}_{i}}'\tilde{H}_{i}}{\sqrt{T}}\right)\theta_{i}\theta_{i}'\left(\frac{\tilde{H}_{i}'{\tilde{X}_{i}}}{\sqrt{T}}\right)\hat{Q}_{x,i}^{-1}\hat{Q}_{x,ij}\hat{Q}_{x,j}^{-1}\left(\frac{{\tilde{X}_{j}}'\tilde{H}_{j}}{\sqrt{T}}\right)\right.\\
 &  & \hspace{7cm}\left.\theta_{j}\theta_{j}'\left(\frac{\tilde{H}_{j}'{\tilde{X}_{j}}}{\sqrt{T}}\right)\hat{Q}_{x,j}^{-1}\hat{Q}_{x,ji}\right].
\end{eqnarray*}
By using $Tr(AB')\leq\Vert A\Vert\Vert B\Vert$, $\bold{1}_{i}^{\chi}\Vert\hat{Q}_{x,i}^{-1}\Vert\leq C\chi_{1,T}^{2}$,
$\bold{1}_{i}^{\chi}\tau_{i,T}\leq\chi_{2,T}$, $\Vert\theta_{i}\Vert\leq M$,
$\Vert x_{i,t}\Vert\leq M$, $\tau_{ij,T}\geq1$, for all $i$ and
$t$, we get: 
\begin{eqnarray*}
I_{2}^{2} & \leq & \frac{C\chi_{1,T}^{8}\chi_{2,T}^{4}}{n^{2}T^{2}}\sum_{i,j}\Vert\frac{\tilde{H}_{i}'{\tilde{X}_{i}}}{\sqrt{T}}\Vert^{2}\Vert\frac{\tilde{H}_{j}'{\tilde{X}_{j}}}{\sqrt{T}}\Vert^{2}\\
 & = & \frac{C\chi_{1,T}^{8}\chi_{2,T}^{4}}{n^{2}T^{4}}\sum_{i,j}\sum_{t_{1},t_{2},t_{3},t_{4}}I_{i,t_{1}}I_{i,t_{2}}I_{j,t_{3}}I_{j,t_{4}}h_{t_{1}}'h_{t_{2}}x_{i,t_{1}}'x_{i,t_{2}}h_{t_{3}}'h_{t_{4}}x_{j,t_{3}}'x_{j,t_{4}}.
\end{eqnarray*}
Thus: 
\begin{eqnarray*}
E[I_{2}^{2}\vert I_{\underline{T},i},I_{\underline{T},j},\gamma_{i},\gamma_{j}] & \leq & \frac{C\chi_{1,T}^{8}\chi_{2,T}^{4}}{n^{2}T^{4}}\sum_{i,j}\sum_{t_{1},t_{2},t_{3},t_{4}}\vert E[h_{t_{1}}'h_{t_{2}}x_{i,t_{1}}'x_{i,t_{2}}h_{t_{3}}'h_{t_{4}}x_{j,t_{3}}'x_{j,t_{4}}\vert\gamma_{i},\gamma_{j}]\vert.
\end{eqnarray*}
Hence ${\displaystyle E[I_{2}^{2}]\leq\frac{CM\chi_{1,T}^{8}\chi_{2,T}^{4}}{T^{2}},}$
from Assumption \ref{A_Ass_RC1}. It follows ${\displaystyle E[I_{2}^{2}]=O(\frac{\chi_{1,T}^{8}\chi_{2,T}^{4}}{T^{2}})}$,
which implies ${\displaystyle I_{2}=O_{p}(\frac{\chi_{1,T}^{4}\chi_{2,T}^{2}}{T})}$.

iii) The proof of \ref{Lemma_L3}(iii) is the same as that of \ref{Lemma_L2}(ii),
by replacing $\tilde{\varepsilon}_{i}$ with $\tilde{u}_{i}$.

\subsection{Proof of \ref{Lemma_L4}}

We have ${\displaystyle \mu_{1}\left(\frac{1}{nT}\sum_{i}\tilde{H}_{i}\theta_{i}\theta_{i}'\tilde{H}_{i}'\right)=\underset{x\in\mathbb{R}^{T}:\Vert x\Vert=1}{\max}\ x'\left(\frac{1}{nT}\sum_{i}\tilde{H}_{i}\theta_{i}\theta_{i}'\tilde{H}_{i}'\right)x.}$
From Assumption \ref{ass_ftheta} (i), matrix ${\displaystyle \frac{1}{T}{H'H}=\frac{1}{T}\sum_{t}h_{t}h_{t}'}$
is positive definite w.p.a. $1$. Thus, for any $a\in\mathbb{R}^{m}$
with $\Vert a\Vert=1$, the vector $x(a)\in\mathbb{R}^{T}$ defined
by ${\displaystyle x(a)=\frac{1}{\sqrt{T}}Ha[a'(H'H/T)a]^{-1/2}}$
is such that $\Vert x(a)\Vert=1$, w.p.a. $1$. Therefore: 
\begin{eqnarray*}
\mu_{1}\left(\frac{1}{nT}\sum_{i}\tilde{H}_{i}\theta_{i}\theta_{i}'\tilde{H}_{i}'\right) & \geq & \underset{a\in\mathbb{R}^{m}:\Vert a\Vert=1}{\max}\ x(a)'\left(\frac{1}{nT}\sum_{i}\tilde{H}_{i}\theta_{i}\theta_{i}'\tilde{H}_{i}'\right)x(a)\\
 & = & \underset{a\in\mathbb{R}^{m}:\Vert a\Vert=1}{\max}\ \frac{{\displaystyle a'\left[\frac{1}{n}\sum_{i}(H'\tilde{H}_{i}/T)\theta_{i}\theta_{i}'(\tilde{H}_{i}'H/T)\right]a}}{{\displaystyle a'(H'H/T)a}}\\
 & = & \underset{a\in\mathbb{R}^{m}:\Vert a\Vert=1}{\max}\ \frac{{\displaystyle a'\left[\frac{1}{n}\sum_{i}\tau_{i,T}^{-2}\left(\frac{1}{T_{i}}\sum_{t}I_{i,t}h_{t}h_{t}'\right)\theta_{i}\theta_{i}'\left(\frac{1}{T_{i}}\sum_{t}I_{i,t}h_{t}h_{t}'\right)\right]a}}{{\displaystyle a'\left(\frac{1}{T}\sum_{t}h_{t}h_{t}'\right)a}}.
\end{eqnarray*}
We have ${\displaystyle a'\left(\frac{1}{T}\sum_{t}h_{t}h_{t}'\right)a\leq\mu_{1}\left(\frac{1}{T}\sum_{t}h_{t}h_{t}'\right)}$,
for any $a\in\mathbb{R}^{m}$ such that $\Vert a\Vert=1$, and from
Assumption \ref{ass_ftheta} (i), we have ${\displaystyle \mu_{1}\left(\frac{1}{T}\sum_{t}h_{t}h_{t}'\right)\leq2\mu_{1}(\Sigma_{h})}$
w.p.a. $1$. Moreover, from the proof of Lemma 3 in GOS, under Assumptions
\ref{A_Ass_Dep_Obsevability} and \ref{A_Ass_Bounded_Obsevability},
and $n=O\left(T^{\bar{\gamma}}\right),\ \bar{\gamma}>0,$ we have
${\displaystyle \underset{1\leq i\leq n}{\sup}\ \Vert\frac{1}{T_{i}}\sum_{t}I_{i,t}h_{t}h_{t}'-\Sigma_{h}\Vert=o_{p}(1)}$,
${\displaystyle \underset{1\leq i\leq n}{\sup}\ \vert\tau_{i,T}-\tau_{i}\vert=o_{p}(1)}$,
and $1\leq\tau_{i}\leq M$, for all $i$. It follows: 
\begin{eqnarray*}
\mu_{1}\left(\frac{1}{nT}\sum_{i}\tilde{H}_{i}\theta_{i}\theta_{i}'\tilde{H}_{i}'\right) & \geq & C\underset{a\in\mathbb{R}^{m}:\Vert a\Vert=1}{\max}\ a'\Sigma_{h}\left(\frac{1}{n}\sum_{i}\theta_{i}\theta_{i}'\right)\Sigma_{h}a=C\mu_{1}\left(\Sigma_{h}\left(\frac{1}{n}\sum_{i}\theta_{i}\theta_{i}'\right)\Sigma_{h}\right),
\end{eqnarray*}
w.p.a. $1$, for a constant $C>0$. From inequality (\ref{MO1}) for
the eigenvalues of a matrix product applied twice, we have ${\displaystyle \mu_{1}\left(\Sigma_{h}\left(\frac{1}{n}\sum_{i}\theta_{i}\theta_{i}'\right)\Sigma_{h}\right)\geq\mu_{1}\left(\frac{1}{n}\sum_{i}\theta_{i}\theta_{i}'\right)\mu_{m}(\Sigma_{h})^{2}}$.
From Assumption \ref{ass_ftheta}, the conclusion follows.

\subsection{Proof of \ref{lemma_L4}}

We start with the case $k=1$, and then extend the arguments to the
case $k\geq2$.

\textbf{a)} When $k=1$, let us consider matrix ${\displaystyle \tilde{A}=\frac{1}{nT}\sum_{i}\theta_{i}^{2}\tilde{H}_{i}\tilde{H}_{i}'=(\tilde{a}_{t,s})}$
with elements \linebreak{}
${\displaystyle \tilde{a}_{t,s}=\frac{1}{nT}\sum_{i}I_{i,t}I_{i,s}\theta_{i}^{2}h_{t}h_{s}}{\displaystyle =:a_{t,s}h_{t}h_{s}.}$
Further, define matrices $A=(a_{t,s})$ and $D=diag(h_{t}:t=1,...,T)$.
Then $\tilde{A}=DAD$, and both $\tilde{A}$ and $A$ are positive
semidefinite matrices. In the first step of the proof, we show that:
\begin{equation}
\mu_{2}(\tilde{A})\leq M^{2}\mu_{2}(A),\label{ineqM2}
\end{equation}
where $M$ is the constant in Assumption \ref{A_Ass_H_theta} a).

Let $\mathcal{G}$ be a linear subspace of $\mathbb{R}^{T}$ and consider
the maximization problem $\underset{x\in\mathcal{G}:\Vert x\Vert=1}{\max}\ x'\tilde{A}x=$\linebreak{}
${\displaystyle \underset{x\in\mathcal{G}:\Vert x\Vert=1}{\max}\ x'DADx.}$
For $x\in\mathcal{G}$ such that $\Vert x\Vert=1$, define $y=Dx$.
Then, $y\in D(\mathcal{G})$ (the image of space $\mathcal{G}$ under
the linear mapping defined by matrix $D$) and $\Vert y\Vert^{2}\leq\Vert h\Vert_{\infty,T}^{2}\Vert x\Vert^{2}=\Vert h\Vert_{\infty,T}^{2}\leq M^{2},$
where ${\displaystyle \Vert h\Vert_{\infty,T}=\underset{t=1,...,T}{\max}\ \vert h_{t}\vert\leq M}$
under Assumption \ref{A_Ass_H_theta} a). Then: 
\begin{eqnarray}
\underset{x\in\mathcal{G}:\Vert x\Vert=1}{\max}\ x'\tilde{A}x\leq\underset{y\in D(\mathcal{G}):\Vert y\Vert\leq M}{\max}\ y'Ay=M^{2}\ \underset{y\in D(\mathcal{G}):\Vert y\Vert=1}{\max}\ y'Ay.\label{ineq1}
\end{eqnarray}
Suppose that $h_{t}\neq0$ for all $t=1,...,T$ (an event of probability
$1$). Then $D$ corresponds to a one-to-one linear mapping. Let $\mathcal{F}_{1}$
be the eigenspace associated to the largest eigenvalue of matrix $A$,
and define $\mathcal{G}=D^{-1}(\mathcal{F}_{1}^{\perp})$, which is
a linear subspace of $\mathbb{R}^{T}$ with dimension $T-1$. Then,
from Inequality (\ref{ineq1}) we get: 
\begin{eqnarray}
\underset{x\in D^{-1}(\mathcal{F}_{1}^{\perp}):\Vert x\Vert=1}{\max}\ x'\tilde{A}x\leq M^{2}\ \underset{y\in\mathcal{F}_{1}^{\perp}:\Vert y\Vert=1}{\max}\ y'Ay.\label{ineq22}
\end{eqnarray}
From the Courant-Fisher min-max theorem (\ref{CourantFischer_minMax}),
we have: $\mu_{2}(\tilde{A})\leq\underset{x\in D^{-1}(\mathcal{F}_{1}^{\perp}):\Vert x\Vert=1}{\max}\ x'\tilde{A}x,$
and, from the Courant-Fisher formula (\ref{R2-1}), we have: $\mu_{2}(A)=\underset{y\in\mathcal{F}_{1}^{\perp}:\Vert y\Vert=1}{\max}\ y'Ay.$
Then, Inequality (\ref{ineq22}) implies bound (\ref{ineqM2}).

Finally, let us bound $\mu_{2}(A)$. By writing ${\displaystyle A=\frac{1}{nT}(B+C)(B+C)'}$,
where $B=\left(b_{t,i}\right)$ and $C=\left(c_{t,i}\right)$ are
$T\times n$ matrices with elements $b_{t,i}=\theta_{i}\bar{I}_{t}$
and $c_{t,i}=\theta_{i}(I_{i,t}-\bar{I}_{t})$, the Weyl inequality
(\ref{Weyl2}) implies ${\displaystyle \mu_{2}(A)^{1/2}\leq\mu_{2}\left(\frac{1}{nT}BB'\right)^{1/2}+\mu_{1}\left(\frac{1}{nT}CC'\right)^{1/2}=\mu_{1}\left(\frac{1}{nT}CC'\right)^{1/2},}$
since matrix $BB'$ has rank $1$. Now ${\displaystyle \frac{1}{nT}CC'=\frac{1}{nT}\tilde{C}D\tilde{C}^{\prime}}$,
where the elements of the $T\times n$ matrix $\tilde{C}$ are ${\displaystyle \tilde{c}_{t,i}=I_{i,t}-\bar{I}_{t}}$
and $D$ is a $n\times n$ diagonal matrix with elements $\theta_{i}^{2}$.
From Assumption \ref{A_Ass_H_theta}b), we have ${\displaystyle \mu_{1}\left(\frac{1}{nT}CC^{\prime}\right)\leq M^{2}\mu_{1}\left(W\right),}$
where the elements of matrix ${\displaystyle W=\frac{1}{nT}\tilde{C}\tilde{C}^{\prime}}$
are ${\displaystyle {\displaystyle w_{t,s}=\frac{1}{nT}}\sum_{i}\left(I_{i,t}-\bar{I}_{t}\right)\left(I_{i,s}-\bar{I}_{s}\right)}$.
Thus, from Assumption \ref{A_Ass_boundII}, we get ${\displaystyle \mu_{2}(A)=O_{p}(C_{n,T}^{-2})}$.
From bound (\ref{ineqM2}), the conclusion follows.

\textbf{b)} Let us now consider the case $k\geq1$. Consider the matrix
${\displaystyle \tilde{A}=\frac{1}{nT}\sum_{i}\tilde{H}_{i}\theta_{i}\theta_{i}'\tilde{H}_{i}'=(\tilde{a}_{t,s})}$
with elements ${\displaystyle \tilde{a}_{t,s}=\frac{1}{nT}\sum_{i}I_{i,t}I_{i,s}\theta_{i}'h_{t}\theta_{i}'h_{s}=\sum_{m,l}\left(\frac{1}{nT}\sum_{i}I_{i,t}I_{i,s}\theta_{i,m}\theta_{i,l}\right)h_{t,m}h_{s,l}=:\sum_{m,l}a_{t,s}^{(m,l)}h_{t,m}h_{s,l},}$
where summation w.r.t. $m,l$ is from $1$ to $k$. Then, we have
$\tilde{A}=\sum_{m,l}D^{(m)}A^{(m,l)}D^{(l)}=DAD^{\prime},$ where
$A^{(m,l)}=[a_{t,s}^{(m,l)}]$, $D^{(m)}=diag(h_{t,m}:t=1,...,T)$,
the $T\times(Tk)$ matrix $D$ is defined by $D=[D^{(1)}:...:D^{(k)}]$
and $A$ is the $(Tk)\times(Tk)$ block matrix with blocks $A^{(m,l)}$.

\begin{lemma} \label{lemma_block} Let ${\displaystyle \left(\begin{array}{cc}
A & B\\
B' & D
\end{array}\right)}$ be a positive definite (or semi-definite) block matrix. Then, $\left(\begin{array}{cc}
A & B\\
B' & D
\end{array}\right)\leq2\left(\begin{array}{cc}
A & 0\\
0 & D
\end{array}\right),$ where the inequality is w.r.t. the ranking of symmetric matrices.
\end{lemma}

By repeated application of \ref{lemma_block}, we get: $A\leq2^{k-1}\left(\begin{array}{ccc}
A^{(1,1)}\\
 & \ddots\\
 &  & A^{(k,k)}
\end{array}\right).$ This implies $\tilde{A}\leq2^{k-1}\sum_{m}D^{(m)}A^{(m,m)}D^{(m)}.$
Since two symmetric matrices are ranked if, and only if, their corresponding
eigenvalues are ranked, we get: 
\begin{equation}
\mu_{k+1}(\tilde{A})\leq2^{k-1}\mu_{k+1}\left(\sum_{m}D^{(m)}A^{(m,m)}D^{(m)}\right).\label{ineq2}
\end{equation}
Moreover, we use the next lemma.

\begin{lemma} \label{lemma_sum} For $k$ symmetric matrices $A_{1}$,
$A_{2}$, ... $A_{k}$, $\mu_{k+1}(A_{1}+...+A_{k})\leq\mu_{2}(A_{1})+...+\mu_{2}(A_{k}).$\end{lemma}

From Inequality (\ref{ineq2}) and \ref{lemma_sum}, we get: $\mu_{k+1}(\tilde{A})\leq2^{k-1}\sum_{m}\mu_{2}\left(D^{(m)}A^{(m,m)}D^{(m)}\right).$
By using the arguments deployed for the case $k=1$ in part \textbf{a}),
we have $\mu_{2}(D^{(m)}A^{(m,m)}D^{(m)})\leq M^{2}\mu_{2}(A^{(m,m)}).$
Therefore, we get $\mu_{k+1}(\tilde{A})\leq2^{k-1}M^{2}\sum_{m}\mu_{2}(A^{(m,m)}).$
As in part \textbf{a}), the Weyl inequality and Assumptions \ref{A_Ass_H_theta}b)
and \ref{A_Ass_boundII} imply ${\displaystyle \mu_{2}(A^{(m,m})\leq M^{2}\mu_{1}(W^ {})=O_{p}(C_{n,T}^{-2})}$.
Thus $\mu_{k+1}(\tilde{A})=O_{p}(C_{n,T}^{-2})$.

\subsection{Proof of \ref{lemma_L5}}

From the Courant-Fisher max-min Theorem (\ref{CourantFisher_maxMin}),
we have: 
\begin{equation}
\mu_{k+1}\left(\frac{1}{nT}\sum_{i}\tilde{H}_{i}\theta_{i}\theta_{i}'\tilde{H}_{i}'\right)=\underset{\mathcal{G}:\dim(\mathcal{G})=k+1}{\max}\ \underset{x\in\mathcal{G}:\Vert x\Vert=1}{\min}\ x'\left(\frac{1}{nT}\sum_{i}\tilde{H}_{i}\theta_{i}\theta_{i}'\tilde{H}_{i}'\right)x,\label{CF1}
\end{equation}
where the maximization is w.r.t. the linear $(k+1)$-dimensional subspace
$\mathcal{G}$ of $\mathbb{R}^{T}$. From Assumption \ref{ass_ftheta}
(i), under model $\mathcal{M}_{2}(k)$ matrix $H/\sqrt{T}$ has full
column-rank equal to $m$, w.p.a. $1$, with $m\geq k+1$. Thus, for
any linear subspace $\mathbb{A}$ of $\mathbb{R}^{m}$ with dimension
$k+1$, the set ${\displaystyle \mathcal{G}_{\mathbb{A}}:=\left\{ x\in\mathbb{R}^{T}:\ x=\frac{1}{\sqrt{T}}Ha,\ a\in\mathbb{A}\right\} }$
is a linear subspace of $\mathbb{R}^{T}$ of dimension $k+1$. We
deduce from (\ref{CF1}): 
\begin{eqnarray*}
\mu_{k+1}\left(\frac{1}{nT}\sum_{i}\tilde{H}_{i}\theta_{i}\theta_{i}'\tilde{H}_{i}'\right) & \geq & \underset{\mathbb{A}:\dim(\mathbb{A})=k+1}{\max}\ \underset{x\in\mathcal{G}_{\mathbb{A}}:\Vert x\Vert=1}{\min}\ x'\left(\frac{1}{nT}\sum_{i}\tilde{H}_{i}\theta_{i}\theta_{i}'\tilde{H}_{i}'\right)x\\
 & = & \underset{\mathbb{A}:\dim(\mathbb{A})=k+1}{\max}\ \underset{a\in\mathbb{A}:\Vert a\Vert=1}{\min}\ \frac{{\displaystyle a'\left(\frac{1}{n}\sum_{i}\frac{H'\tilde{H}_{i}}{T}\theta_{i}\theta_{i}'\frac{\tilde{H}_{i}'H}{T}\right)a}}{{\displaystyle a'\left(\frac{1}{T}H'H\right)a}}.
\end{eqnarray*}
By similar arguments as in the proof of \ref{Lemma_L4}, under Assumptions
\ref{A_Ass_Dep_Obsevability} and \ref{A_Ass_Bounded_Obsevability},
we get the inequality: 
\begin{eqnarray*}
\mu_{k+1}\left(\frac{1}{nT}\sum_{i}\tilde{H}_{i}\theta_{i}\theta_{i}'\tilde{H}_{i}'\right) & \geq & C\underset{\mathbb{A}:\dim(\mathbb{A})=k+1}{\max}\ \underset{a\in\mathbb{A}:\Vert a\Vert=1}{\min}\ a'\Sigma_{h}\left(\frac{1}{n}\sum_{i}\theta_{i}\theta_{i}'\right)\Sigma_{h}a,
\end{eqnarray*}
w.p.a. $1$. By the max-min Theorem, the r.h.s. is such that: 
\[
\underset{\mathbb{A}:\dim(\mathbb{A})=k+1}{\max}\ \underset{a\in\mathbb{A}:\Vert a\Vert=1}{\min}\ a'\Sigma_{h}\left(\frac{1}{n}\sum_{i}\theta_{i}\theta_{i}'\right)\Sigma_{h}a=\mu_{k+1}\left(\Sigma_{h}\left(\frac{1}{n}\sum_{i}\theta_{i}\theta_{i}'\right)\Sigma_{h}\right).
\]
Moreover, from inequality (\ref{MO1}) for the eigenvalues of product
matrices applied twice, we have \break ${\displaystyle \mu_{k+1}\left(\Sigma_{h}\left(\frac{1}{n}\sum_{i}\theta_{i}\theta_{i}'\right)\Sigma_{h}\right)\geq\mu_{k+1}\left(\left(\frac{1}{n}\sum_{i}\theta_{i}\theta_{i}'\right)\right)\mu_{m}(\Sigma_{h})^{2}}$.
Then, from Assumptions \ref{ass_ftheta} (i) and \ref{ass3}, the
conclusion follows.

\section{Check of Assumptions \ref{A_Ass_ErrTerm_u_M2} and \ref{A_Ass_restricSerialCrossDep}
under block dependence\label{sec:Appendix_CheckBlockDep}}

In this appendix, we verify that the high-level Assumptions \ref{A_Ass_ErrTerm_u_M2}
and \ref{A_Ass_restricSerialCrossDep} on serial and cross-sectional
dependences of error terms are satisfied under a block-dependence
structure in a serially i.i.d.\ framework.

\begin{BDassumption} \label{Ass_BD} The error terms $u_{t}(\gamma)$
are i.i.d.\ over time with $E[u_{t}(\gamma)]=0$, for all $\gamma\in[0,1]$.
For any $n$, there exists a partition of the interval $[0,1]$ into
$b_{n}\leq n$ subintervals of approximate length $B_{n}=O(1/b_{n})$,
such that $u_{t}(\gamma)$ and $u_{t}(\gamma')$ are independent if
$\gamma$ and $\gamma'$ belong to different subintervals, and $b_{n}^{-1}=O(n^{-\alpha})$
as $n\rightarrow\infty$, where $\alpha\in(0,1]$. \end{BDassumption}

\begin{BDassumption} \label{Ass_BD2} The error terms $(u_{t}(\gamma))$,
the factors $(f_{t})$, and the instruments $(Z_{t})$, $(Z_{t}(\gamma))$,
$\gamma\in[0,1]$, are mutually independent. \end{BDassumption}

The block-dependence structure as in Assumption \ref{Ass_BD} is satisfied
for instance when there are unobserved industry-specific factors independent
among industries and over time, as in \cite{Ang_Liu_Schwarz_2010}.
In empirical applications, blocks can match industrial sectors. Then,
the number $b_{n}$ of blocks amounts to a couple of dozens, and the
number of assets $n$ amounts to a couple of thousands. There are
approximately $nB_{n}$ assets in each block, when $n$ is large.
In the asymptotic analysis, Assumption \ref{Ass_BD} requires that
the number of independent blocks grows with $n$ fast enough. Within
blocks, covariances do not need to vanish asymptotically.

\begin{lemma} \label{Lemma_BD1} Under Assumptions \ref{A_Ass_upperBound_Mom_uit}
and \ref{Ass_BD}: (i) Assumption \ref{A_Ass_ErrTerm_u_M2} holds.
(ii) Assumption \ref{A_Ass_restricSerialCrossDep} holds if $n\geq T^{\bar{\gamma}}$
and: 
\begin{equation}
\alpha>4/q,\quad\bar{\gamma}>\frac{1}{\alpha-4/q}.\label{condition_BD}
\end{equation}
\end{lemma}

The conditions in (\ref{condition_BD}) provide a restriction on the
relative growth rate of the cross-sectional and time-series dimensions
in terms of: (i) the strength of cross-sectional dependence (via $\alpha$),
and (ii) the existence of higher-order moments of the error terms
(via $q$). We can have $\bar{\gamma}$ (arbitrarily) close to $1$,
if cross-sectional dependence is sufficiently weak and the tails of
the errors are sufficiently thin. These conditions are compatible
with $T/n=o(1)$.

\newpage{}

\setcounter{page}{1}

\begin{center}
\textbf{\Large{}{}{}{}{}SUPPLEMENTARY MATERIALS}{\Large{}{}{}{}{}
} 
\par\end{center}

\begin{center}
\textbf{A diagnostic criterion for approximate factor structure} 
\par\end{center}

\begin{center}
Patrick Gagliardini, Elisa Ossola and Olivier Scaillet 
\par\end{center}

These supplementary materials provide the proofs of technical Lemmas
7-11 (Appendix 4), the verification that conditional independence
implies Assumption \ref{A1-1} (Appendix 5), the link with \cite{Stock_Watson_2002b}
(Appendix 6), and the results of Monte-Carlo experiments (Appendix
7).

\section{Proofs of technical Lemmas}

\subsection{Proof of \ref{Ltruncation}}

We follow the arguments in the proof of Lemma 2.2 in \cite{Yin_Bai_Krishnaiah_1988}.
From the conditions $\delta\geq n^{\beta}$ and $T\leq C_{2}n$, we
have: 
\begin{eqnarray*}
P(\mathcal{E}\neq\Xi\ i.o.) & \leq & \underset{k\rightarrow\infty}{\lim}\ \sum_{m=k}^{\infty}P\left(\underset{2^{m-1}\leq n<2^{m}}{\bigcup}\ \bigcup_{i=1}^{n}\bigcup_{t=1}^{T}\{\vert\varepsilon_{i,t}\vert>\delta\}\right)\\
 & \leq & \underset{k\rightarrow\infty}{\lim}\ \sum_{m=k}^{\infty}P\left(\bigcup_{i=1}^{2^{m}}\bigcup_{t=1}^{C_{2}2^{m}}\{\vert\varepsilon_{i,t}\vert>2^{(m-1)\beta}\}\right)\\
 & \leq & \underset{k\rightarrow\infty}{\lim}\ \sum_{m=k}^{\infty}C_{2}2^{2m}P\left(\vert\varepsilon_{i,t}\vert>2^{(m-1)\beta}\right).
\end{eqnarray*}
Thus, part (i) follows from the summability condition: 
\begin{equation}
\sum_{m=1}^{\infty}2^{2m}P\left(\vert\varepsilon_{i,t}\vert>2^{(m-1)\beta}\right)<\infty.\label{sum-1}
\end{equation}
To prove the summability condition (\ref{sum-1}), we use the Chebyshev
inequality and Assumption \ref{A_Ass_upperBound_Mom_uit}. We have
${\displaystyle P\left(\vert\varepsilon_{i,t}\vert>2^{(m-1)\beta}\right)\leq E[\vert\varepsilon_{i,t}\vert^{q}]/2^{(m-1)\beta q}\leq M/2^{(m-1)\beta q}}$.
Therefore, we get: 
\[
\sum_{m=1}^{\infty}2^{2m}P\left(\vert\varepsilon_{i,t}\vert>2^{(m-1)\beta}\right)\leq M\sum_{m=1}^{\infty}\frac{2^{2m}}{2^{(m-1)\beta q}}=M2^{\beta q}\sum_{m=1}^{\infty}\frac{1}{2^{(\beta q-2)m}}<\infty,
\]
since $q\beta>2$.

Part (ii) is a straightforward consequence of part (i), since $P(\tilde{\mathcal{E}}\neq\tilde{\Xi}\ i.o.)\leq P(\mathcal{E}\neq\Xi\ i.o.)$.

\subsection{Proof of \ref{Lcentralization}}

We follow the arguments in \cite{Bai_Yin_1993}, p. 1278. We use the
von Neumann inequality (\cite{Neumann_1937}): for any $n\times T$
matrices $A$ and $B$, 
\begin{equation}
tr(A'B)\leq\sum_{k=1}^{T}\mu_{k}(A'A)^{1/2}\mu_{k}(B'B)^{1/2}.\label{vN}
\end{equation}
We have: 
\begin{eqnarray*}
\left[\mu_{1}^{1/2}\left(\frac{1}{n}\tilde{\Xi}\tilde{\Xi}'\right)-\mu_{1}^{1/2}\left(\frac{1}{n}\tilde{E}\tilde{E}'\right)\right]^{2} & \leq & \sum_{k=1}^{T}\left[\mu_{k}^{1/2}\left(\frac{1}{n}\tilde{\Xi}\tilde{\Xi}'\right)-\mu_{k}^{1/2}\left(\frac{1}{n}\tilde{E}\tilde{E}'\right)\right]^{2}\\
 & = & tr\left(\frac{1}{n}\tilde{\Xi}\tilde{\Xi}'\right)+tr\left(\frac{1}{n}\tilde{E}\tilde{E}'\right)-2\sum_{k=1}^{T}\mu_{k}^{1/2}\left(\frac{1}{n}\tilde{\Xi}\tilde{\Xi}'\right)\mu_{k}^{1/2}\left(\frac{1}{n}\tilde{E}\tilde{E}'\right).
\end{eqnarray*}
The last term in the r.h.s. is bounded by the von Neumann inequality
(\ref{vN}): 
\begin{eqnarray}
\left[\mu_{1}^{1/2}\left(\frac{1}{n}\tilde{\Xi}\tilde{\Xi}'\right)-\mu_{1}^{1/2}\left(\frac{1}{n}\tilde{E}\tilde{E}'\right)\right]^{2} & \leq & tr\left(\frac{1}{n}\tilde{\Xi}\tilde{\Xi}'\right)+tr\left(\frac{1}{n}\tilde{E}\tilde{E}'\right)-2\frac{1}{n}tr\left(\tilde{\Xi}\tilde{E}'\right)\nonumber \\
 & = & \frac{1}{n}tr\left[(\tilde{\Xi}-\tilde{E})(\tilde{\Xi}-\tilde{E})'\right].\label{vN1}
\end{eqnarray}
The elements of matrix $\tilde{\Xi}-\tilde{E}$ are $I_{i,t}E[\varepsilon_{i,t}1\{\vert\varepsilon_{i,t}\vert\leq\delta\}\vert\gamma_{i}]$.
By the zero-mean property of the errors $\varepsilon_{i,t}$, the
Minkowski inequality and Assumption \ref{A_Ass_upperBound_Mom_uit},
we have: 
\begin{eqnarray*}
\vert E[\varepsilon_{i,t}1\{\vert\varepsilon_{i,t}\vert\leq\delta\}]\vert & = & \vert E[\varepsilon_{i,t}1\{\vert\varepsilon_{i,t}\vert>\delta\}]\vert\leq E[\vert\varepsilon_{i,t}\vert^{q}]^{1/q}P[\vert\varepsilon_{i,t}\vert>\delta]^{1/\bar{q}},
\end{eqnarray*}
where $1/q+1/\bar{q}=1$, with $q$ defined in Assumption \ref{A_Ass_upperBound_Mom_uit}.
By the Chebyshev inequality and Assumption \ref{A_Ass_upperBound_Mom_uit},
we get: 
\begin{eqnarray*}
E[\vert\varepsilon_{i,t}\vert^{q}]^{1/q}P[\vert\varepsilon_{i,t}\vert>\delta]^{1/\bar{q}} & \leq & E[\vert\varepsilon_{i,t}\vert^{q}]^{1/q}\left(\frac{E[\vert\varepsilon_{i,t}\vert^{q}]}{\delta^{q}}\right)^{1/\bar{q}}=\frac{E[\vert\varepsilon_{i,t}\vert^{q}]}{\delta^{q-1}}\leq\frac{M}{\delta^{q-1}}.
\end{eqnarray*}
Thus, we get: 
\begin{equation}
\frac{1}{n}tr\left[(\tilde{\Xi}-\tilde{E})(\tilde{\Xi}-\tilde{E})'\right]=\frac{1}{n}\sum_{i}\sum_{t}I_{i,t}E[\varepsilon_{i,t}1\{\vert\varepsilon_{i,t}\vert\leq\delta\}]^{2}\leq T\frac{M^{2}}{\delta^{2(q-1)}}.\label{vN2}
\end{equation}
From inequalities (\ref{vN1}) and (\ref{vN2}), we get ${\displaystyle \vert\mu_{1}^{1/2}\left(\frac{1}{n}\tilde{\Xi}\tilde{\Xi}'\right)-\mu_{1}^{1/2}\left(\frac{1}{n}\tilde{E}\tilde{E}'\right)\vert\leq\sqrt{T}\frac{M}{\delta^{q-1}}}$.
Since the sequence $\delta=\delta_{n}$ is such that $\sqrt{T}/\delta^{q-1}=o(1)$,
the conclusion follows.

\subsection{Proof of \ref{lemma_block}}

We have: 
\[
2\left(\begin{array}{cc}
A & 0\\
0 & D
\end{array}\right)-\left(\begin{array}{cc}
A & B\\
B' & D
\end{array}\right)=\left(\begin{array}{cc}
A & -B\\
-B' & D
\end{array}\right),
\]
and: 
\[
\left(\begin{array}{cc}
x_{1}' & x_{2}'\end{array}\right)\left(\begin{array}{cc}
A & -B\\
-B' & D
\end{array}\right)\left(\begin{array}{c}
x_{1}\\
x_{2}
\end{array}\right)=\left(\begin{array}{cc}
x_{1}' & -x_{2}'\end{array}\right)\left(\begin{array}{cc}
A & B\\
B' & D
\end{array}\right)\left(\begin{array}{c}
x_{1}\\
-x_{2}
\end{array}\right)\geq0,
\]
for all $x=(x_{1}',x_{2}')'$.

\subsection{Proof of \ref{lemma_sum}}

By repeated application of the Weyl inequality for eigenvalues (see
\ref{sec:Proofs} (i) ) we have: 
\begin{eqnarray*}
\mu_{k+1}(A_{1}+...+A_{k}) & \leq & \mu_{k}(A_{1}+...+A_{k-1})+\mu_{2}(A_{k})\\
 & \leq & \mu_{k-1}(A_{1}+...+A_{k-2})+\mu_{2}(A_{k-1})+\mu_{2}(A_{k})\\
 &  & \cdots\\
 & \leq & \mu_{2}(A_{1})+...+\mu_{2}(A_{k}).
\end{eqnarray*}

\subsection{Proof of \ref{Lemma_BD1}}

\subsubsection{Proof of Part (i)}

By the serial independence of the error terms, we have: 
\begin{eqnarray*}
 &  & \frac{1}{n^{2}T^{2}}\sum_{i,j}\sum_{t_{1},t_{2},t_{3},t_{4}}E\left[u_{i,t_{1}}u_{i,t_{2}}u_{j,t_{3}}u_{j,t_{4}}\left|x_{i,\underline{T}},x_{j,\underline{T}},\gamma_{i},\gamma_{j}\right.\right]\\
 & = & \frac{1}{n^{2}T^{2}}\sum_{i,j}\sum_{t_{1},t_{2},t_{3},t_{4}}E\left[u_{i,t_{1}}u_{i,t_{2}}u_{j,t_{3}}u_{j,t_{4}}\left|\gamma_{i},\gamma_{j}\right.\right]\\
 & = & \frac{1}{n^{2}T^{2}}\sum_{i,j}\sum_{t_{1}}E\left[u_{i,t_{1}}u_{i,t_{1}}u_{j,t_{1}}u_{j,t_{1}}\left|\gamma_{i},\gamma_{j}\right.\right]\\
 &  & +\frac{1}{n^{2}T^{2}}\sum_{i,j}\sum_{t_{1}\neq t_{2}}E\left[u_{i,t_{1}}u_{i,t_{1}}\left|\gamma_{i}\right.\right]E\left[u_{j,t_{2}}u_{j,t_{2}}\left|\gamma_{j}\right.\right]\\
 &  & +\frac{1}{n^{2}T^{2}}\sum_{i,j}\sum_{t_{1}\neq t_{2}}E\left[u_{i,t_{1}}u_{j,t_{1}}\left|\gamma_{i},\gamma_{j}\right.\right]E\left[u_{i,t_{2}}u_{j,t_{2}}\left|\gamma_{i},\gamma_{j}\right.\right].
\end{eqnarray*}
The conclusion follows by taking absolute values and expectation,
and using the triangular inequality, the Cauchy-Schwarz inequality
and Assumption \ref{A_Ass_upperBound_Mom_uit}.

\subsubsection{Proof of Part (ii)}

Here, we treat $\varepsilon_{i,t}$ as a random variable but not through
the random draw $\gamma_{i}$. This avoids the notational burden coming
from conditional expectations. We show directly the inequality $\break{\displaystyle \frac{1}{n^{k}}\sum_{i_{1},...,i_{k}}\sum_{t_{1},...,t_{k}}\vert E[e_{i_{1},t_{k}}e_{i_{1},t_{1}}e_{i_{2},t_{1}}e_{i_{2},t_{2}}e_{i_{3},t_{2}}\cdots e_{i_{k-1},t_{k-1}}e_{i_{k},t_{k-1}}e_{i_{k},t_{k}}]\vert\leq M^{k}}$,
which implies Assumption \ref{A_Ass_restricSerialCrossDep}. Under
Assumption \ref{Ass_BD}, there are $b=b_{n}$ blocks of approximate
size $d=d_{n}$, where $bd=O(n)$.

1) Let $\omega>0$ be such that $E[\varepsilon_{i,t}^{2}]\leq\omega^{2}$,
for all $i$ and $t$, and define $\phi_{i,t}=e_{i,t}/\omega$. The
scaled $\phi_{i,t}$ are such that $E[\phi_{i,t}]=0$, $E[\phi_{i,t}^{2}]\leq1$,
and $E[\vert\phi_{i,t}\vert^{r-2}]=O(\delta^{r-2})$, for all $r\geq3$,
uniformly in $i$ and $t$. Note that $\phi_{i,t}$ is a (nonlinear)
transformation of $\varepsilon_{i,t}$. Hence, the variables $\phi_{i,t}$
have the same block dependence structure as the variables $\varepsilon_{i,t}$.
Moreover: 
\begin{eqnarray}
 &  & \frac{1}{n^{k}}\sum_{i_{1},...,i_{k}}\sum_{t_{1},...,t_{k}}\vert E[e_{i_{1},t_{k}}e_{i_{1},t_{1}}e_{i_{2},t_{1}}\varepsilon_{i_{2},t_{2}}e_{i_{3},t_{2}}\cdots e_{i_{k-1},t_{k-1}}e_{i_{k},t_{k-1}}e_{i_{k},t_{k}}]\vert\nonumber \\
 & \leq & \omega^{2k}\frac{1}{n^{k}}\sum_{i_{1},...,i_{k}}\sum_{t_{1},...,t_{k}}\vert E[\phi_{i_{1},t_{k}}\phi_{i_{1},t_{1}}\phi_{i_{2},t_{1}}\phi_{i_{2},t_{2}}\phi_{i_{3},t_{2}}\cdots\phi_{i_{k-1},t_{k-1}}\phi_{i_{k},t_{k-1}}\phi_{i_{k},t_{k}}]\nonumber \\
 & =: & \omega^{2k}I_{k}.\label{wIk}
\end{eqnarray}
Let us now bound $I_{k}$.

2) For $m=1,...,k$, let $\mathcal{C}_{m}$ denote the set of $k$-tuples
$(i_{1},...,i_{k})$ such that indices $i_{1}$, ..., $i_{k}$ belong
to $m$ different blocks. Let $N_{m}$ denote the number of different
$2k$-tuples $(i_{1},...,i_{k})$, $(t_{1},...,t_{k})$ such that
$(i_{1},...,i_{k})\in\mathcal{C}_{m}$ and the expectation $E[\phi_{i_{1},t_{k}}\phi_{i_{1},t_{1}}\phi_{i_{2},t_{1}}\phi_{i_{2},t_{2}}\phi_{i_{3},t_{2}}\cdots\phi_{i_{k-1},t_{k-1}}\phi_{i_{k},t_{k-1}}\phi_{i_{k},t_{k}}]$
does not vanish. Moreover, let $Q_{m}$ be an upper bound for such
a non vanishing expectation. Then: 
\begin{eqnarray}
I_{k} & \leq & \frac{1}{n^{k}}\sum_{m=1}^{k}N_{m}Q_{m}.\label{sum}
\end{eqnarray}

3) We need upper bounds for $N_{m}$ and $Q_{m}$, for $m=1,2,...,k$,
and any integer $k$. 
\begin{itemize}
\item $m=1$: The number of $k$-tuples $(i_{1},...,i_{k})$ with all indices
in the same block is $O(bd^{k})$. Indeed, we can select the block
among $b$ alternatives, and we have $O(d^{k})$ possibilities to
select the indices within the block. Then, $N_{1}=O(bd^{k}T^{k})$.
Moreover, by the Cauchy-Schwarz inequality, 
\[
E\left[\phi_{i_{1},t_{k}}\phi_{i_{1},t_{1}}\phi_{i_{2},t_{1}}\phi_{i_{2},t_{2}}\phi_{i_{3},t_{2}}\cdots\phi_{i_{k-1},t_{k-1}}\phi_{i_{k},t_{k-1}}\phi_{i_{k},t_{k}}\right]\leq\sup_{i,t}E[\vert\phi_{i,t}\vert^{2k}]=O(\delta^{2k-2}).
\]
Thus, $Q_{1}=O(\delta^{2(k-1)})$. 
\item $m=k$: The number of $k$-tuples $(i_{1},...,i_{k})$ with indices
in $k$ different blocks is $O(b^{k}d^{k})$. For such a $k$-tuple:
\[
E\left[\phi_{i_{1},t_{k}}\phi_{i_{1},t_{1}}\phi_{i_{2},t_{1}}\phi_{i_{2},t_{2}}\cdots\phi_{i_{k},t_{k-1}}\phi_{i_{k},t_{k}}\right]=E\left[\phi_{i_{1},t_{k}}\phi_{i_{1},t_{1}}\right]E\left[\phi_{i_{2},t_{1}}\phi_{i_{2},t_{2}}\right]\cdots E\left[\phi_{i_{k},t_{k-1}}\phi_{i_{k},t_{k}}\right].
\]
Hence, the indices $t_{1}$, ... $t_{k}$ must be all equal for this
expectation not to vanish. Then, $N_{k}=O(b^{k}d^{k}T)$ and $Q_{k}\leq1$.
\footnote{For $k>b$, there are no $k$-tuples $(i_{1},...,i_{k})$ with indices
in $k$ different blocks, and $N_{k}=0$. The upper bound $N_{k}=O(b^{k}d^{k}T)$
trivially holds also in this case. However, this case will not occur
with our choice of sequence $k$, since (\ref{rates}) implies $k=o(b)$,
see below.} 
\item $m=2$: The number $N_{2}$ is ${\displaystyle O(b^{2})\times\stirling{k}{2}\times O(d^{k})\times O(T^{k-1})}$,
where ${\displaystyle \stirling{k}{2}=2^{k-1}-1}$ is the number of
different ways in which we can divide $k$ objects into two (non-empty)
groups (a Stirling number of the second kind). Indeed, $O(b^{2})$
is a bound for the number of different ways to select the two distinct
blocks. Then, for each $j=1,...,k$ we select whether index $i_{j}$
is in the first or the second block; we have ${\displaystyle \stirling{k}{2}}$
different possibilities. Once we have fixed the blocks, we have $O(d^{k})$
alternatives to select the indices. By block dependence, the expectation
$E[\phi_{i_{1},t_{k}}\phi_{i_{1},t_{1}}\phi_{i_{2},t_{1}}\phi_{i_{2},t_{2}}\cdots\phi_{i_{k},t_{k-1}}\phi_{i_{k},t_{k}}]$
can be splitted into two expectations, and at least a pair of indices
in the $k$-tuple $(t_{1},...,t_{k})$ must be equal for the expectation
not to vanish. Hence the term $O(T^{k-1})$.

Suppose the expectation $E[\phi_{i_{1},t_{k}}\phi_{i_{1},t_{1}}\phi_{i_{2},t_{1}}\phi_{i_{2},t_{2}}\cdots\phi_{i_{k},t_{k-1}}\phi_{i_{k},t_{k}}]$
is splitted into two expectations, with $r_{1}$ indices $i_{j}$
in the first block, and $r_{2}$ indices in the second block, $r_{1}+r_{2}=k$.
Then, $E[\phi_{i_{1},t_{k}}\phi_{i_{1},t_{1}}\phi_{i_{2},t_{1}}\phi_{i_{2},t_{2}}\cdots\phi_{i_{k},t_{k-1}}\phi_{i_{k},t_{k}}]=O(\delta^{2(r_{1}-1)})\times O(\delta^{2(r_{1}-1)})=O(\delta^{2(k-2)})$.
Hence, $Q_{2}=O(\delta^{2(k-2)})$.
\item Generic $m$: We have 
\begin{eqnarray}
N_{m} & = & O(b^{m})\times\stirling{k}{m}\times O(d^{k})\times O(T^{k-m+1}),\label{Nm}\\
Q_{m} & = & O(\delta^{2(k-m)}),\label{Qm}
\end{eqnarray}
where the Stirling number of the second kind ${\displaystyle \stirling{k}{m}=\frac{1}{m!}\sum_{j=0}^{m}(-1)^{m-j}\binom{m}{j}j^{k}}$
gives the number of different ways in which we can divide $k$ objects
into $m$ (non-empty) groups (see e.g. \cite{Rennie_Dobson_1969})
and ${\displaystyle \binom{k}{m}}$ is a binomial coefficient. 
\end{itemize}
From bounds (\ref{sum}), (\ref{Nm}) and (\ref{Qm}), and using $d=O(n/b)$,
we get: 
\begin{eqnarray}
I_{k} & \leq & \frac{const}{n^{k}}\sum_{m=1}^{k}b^{m}d^{k}\stirling{k}{m}T^{k-m+1}\delta^{2(k-m)}\nonumber \\
 & = & const\times T\sum_{m=1}^{k}\stirling{k}{m}(\delta^{2}T/b)^{k-m}.\label{Ik}
\end{eqnarray}

4) We exploit the following upper bound for the Stirling numbers of
the second kind (see \cite{Rennie_Dobson_1969}, Theorem 3) ${\displaystyle \stirling{k}{m}\leq\frac{1}{2}\binom{k}{m}m^{k-m}.}$
Then, we get: ${\displaystyle \sum_{m=1}^{k}\stirling{k}{m}(\delta^{2}T/b)^{k-m}\leq}$\\
 ${\displaystyle \frac{1}{2}\sum_{m=1}^{k}m^{k-m}\binom{k}{m}(\delta^{2}T/b)^{k-m}\leq\frac{1}{2}\sum_{m=0}^{k}\binom{k}{m}(k\delta^{2}T/b)^{k-m}{\displaystyle =\frac{1}{2}(1+k\delta^{2}T/b)^{k},}}$
from the binomial theorem. Thus, from (\ref{Ik}), we get: 
\begin{equation}
I_{k}\leq constT(1+k\delta^{2}T/b)^{k}.\label{B2}
\end{equation}

5) Assume that the sequence $k=k_{n}\uparrow\infty$ is such that:
\begin{equation}
k\delta^{2}T/b=o(1),\quad T=O(e^{k}).\label{rates}
\end{equation}
From (\ref{B2}) and (\ref{rates}), we get $I_{k}\leq(2e)^{k}$.
Then, from (\ref{wIk}): 
\begin{eqnarray*}
\frac{1}{n^{k}}\sum_{i_{1},...,i_{k}}\sum_{t_{1},...,t_{k}}\vert E[e_{i_{1},t_{k}}e_{i_{1},t_{1}}e_{i_{2},t_{1}}\varepsilon_{i_{2},t_{2}}e_{i_{3},t_{2}}\cdots e_{i_{k-1},t_{k-1}}e_{i_{k},t_{k-1}}e_{i_{k},t_{k}}]\vert & \leq & (2e\omega)^{k},
\end{eqnarray*}
i.e., the bound in Assumption \ref{A_Ass_restricSerialCrossDep} holds
with $C=2e\omega$.

6) Let us now verify the compatibility of the different rates, i.e.,
that we can choose sequences $\delta=n^{\beta}$ and $k=c\log(n)$,
$\beta,c>0$, such that $\sqrt{T}/\delta^{q-1}=o(1)$, $\beta>2/q,$
and they match conditions (\ref{rates}). Let $n\geq T^{\bar{\gamma}}$
and $b\geq n^{\alpha}$, with $\bar{\gamma}>1$ and $\alpha\in(0,1]$.
Condition $T=O(e^{k})$ is satisfied if $c\geq1/\bar{\gamma}$. Condition
$k\delta^{2}T/b=o(1)$ implies: 
\begin{equation}
\beta<\frac{1}{2}(\alpha-1/\bar{\gamma}).\label{condrates1}
\end{equation}
Condition $\sqrt{T}/\delta^{q-1}=o(1)$ implies ${\displaystyle \beta>\frac{1}{2\bar{\gamma}(q-1)}.}$
The latter inequality is implied by 
\begin{equation}
\beta>\frac{2}{q},\label{condrates2}
\end{equation}
since $\bar{\gamma}>1$ and $q\geq8$ in Assumption \ref{A_Ass_upperBound_Mom_uit}.
Then, there exists a power $\beta>0$ satisfying conditions (\ref{condrates1})
and (\ref{condrates2}) if, and only if, ${\displaystyle \frac{1}{2}(\alpha-1/\bar{\gamma})>\frac{2}{q}}$,
which corresponds to Condition (\ref{condition_BD}). This condition
clarifies the link between the behaviour of expectations of products
of error terms and the assumption of a bounded largest eigenvalue
used for example in \cite{Chamberlain_Rothschild_1983} p.\ 1294
for arbitrage pricing theory.

\section{Verification that conditional independence implies \protect \protect
\protect \protect \protect \\
 Assumption \ref{A1-1}}

Let us verify that Assumption \ref{A1-1} is true if the latent factors
are independent of the lagged stock-specific instruments, conditional
on the observable factors and the lagged common instruments.

We have: 
\begin{eqnarray*}
h_{t}\perp\{Z_{i,t-1},i=1,...\}\ \vert\ f_{t},Z_{t-1} & \Rightarrow & h_{t}\perp\{\tilde{x}_{i,t},i=1,...\}\ \vert\ f_{t},Z_{t-1}\\
 & \Rightarrow & h_{t}\perp\{\tilde{x}_{i,t},i=1,...\}\ \vert\ x_{t}\\
 & \Rightarrow & EL[h_{t}\vert{x}_{i,t},i=1,...]=EL[h_{t}\vert x_{t}],
\end{eqnarray*}
where $A\perp B\vert C$ denotes independence of $A$ and $B$ conditional
on $C$.

\section{Link with \cite{Stock_Watson_2002b} \label{sec:LinkStockWatson}}

We consider the EM algorithm proposed by \cite{Stock_Watson_2002b}
applied to residuals $\hat{\varepsilon}_{i,t}$: 
\[
\tilde{\tilde{\varepsilon}}_{i,t}=\begin{cases}
\hat{\varepsilon}_{i,t}, & \text{if }I_{i,t}=1,\\
\hat{\theta}_{i}\hat{h}_{t}, & \text{if }I_{i,t}=0.
\end{cases}
\]
Let us define the criterion ${\displaystyle \xi^{SW}=\mu_{1}\left(\frac{\tilde{\tilde{\varepsilon}}\tilde{\tilde{\varepsilon}}^{\prime}}{nT}\right)-\frac{1}{nT}\sum_{i}\sum_{t}\left(1-I_{i,t}\right)\left(\hat{\theta}_{i}\hat{h}_{t}\right)^{2}-g\left(n,T\right).}$
Below we show that $\xi^{SW}$ is the penalized difference of the
EM criteria under the two rival models. Comparing the criteria $\xi$
and $\xi^{SW}$ gives the following link: ${\displaystyle \frac{1}{nT}\sum_{i}\sum_{t}\left(1-I_{i,t}\right)\left(\hat{\theta}_{i}\hat{h}_{t}\right)^{2}=\frac{1}{nT}\left\Vert \tilde{\varepsilon}-\tilde{\tilde{\varepsilon}}\right\Vert ^{2}.}$

To study the EM algorithm, we work as if the true error terms $\varepsilon_{i,t}$
are observed when $I_{i,t}=1$. This error is replaced by the residual
$\hat{\varepsilon}_{i,t}$. We consider the $j$th iteration of the
algorithm. Let $\tilde{\zeta}=\left(\tilde{\Theta},\tilde{H}\right)$
denotes the estimates of $\Theta$ and $H$ obtained from the $\left(j-1\right)$th
iteration, and let ${\displaystyle Q\left(\zeta,\tilde{\zeta}\right)=E_{\tilde{\zeta}}\left[{\cal L}\left(\zeta\right)|\varepsilon\right],}$
where ${\displaystyle {\cal L}\left(\zeta\right)=\frac{1}{nT}\sum_{i}\sum_{t}\left(\varepsilon_{i,t}^{*}-\theta_{i}h_{t}\right)^{2},}$
and $E_{\tilde{\zeta}}\left[\cdot|\varepsilon\right]$ denotes conditional
expectation given the panel of observations under parameter $\tilde{\zeta}$.
We study $Q\left(\zeta,\tilde{\zeta}\right)$ under the two models.
Under both ${\cal M}_{1}$ and ${\cal M}_{2}$, we consider a pseudo
model for the innovations such that $u_{i,t}\sim i.i.d.\left(0,\sigma_{i,t}^{2}\right).$ 
\begin{itemize}
\item Under ${\cal M}_{1}$: we get 
\[
{\displaystyle Q_{0}\left(\zeta,\tilde{\zeta}\right)=E\left[\frac{1}{nT}\sum_{i}\sum_{t}\left(\varepsilon_{i,t}^{*}\right)^{2}|\varepsilon\right]=\frac{1}{nT}\sum_{i}\sum_{t}E\left[\left(\varepsilon_{i,t}^{*}\right)^{2}|\varepsilon\right].}
\]
We have 
\[
E\left[\varepsilon_{i,t}^{*}|\varepsilon\right]=\begin{cases}
\varepsilon_{i,t}, & \text{if }I_{i,t}=1,\\
0, & \text{if }I_{i,t}=0,
\end{cases}\text{ and }V\left[\varepsilon_{i,t}^{*}|\varepsilon\right]=\begin{cases}
0, & \text{if }I_{i,t}=1,\\
\sigma_{i,t}^{2}, & \text{if }I_{i,t}=0.
\end{cases}
\]
and ${\displaystyle E\left[\left(\text{\ensuremath{\varepsilon}}_{i,t}^{*}\right)^{2}|\varepsilon\right]=I_{i,t}\text{\ensuremath{\varepsilon}}_{i,t}^{2}+\left(1-I_{i,t}\right)\sigma_{i,t}^{2}.}$
Thus, 
\begin{eqnarray*}
Q_{0}=Q_{0}\left(\text{\ensuremath{\zeta}},\tilde{\zeta}\right)=\frac{1}{nT}\sum_{i}\sum_{t}I_{i,t}\varepsilon_{i,t}^{2}+\frac{1}{nT}\sum_{i}\sum_{t}\left(1-I_{i,t}\right)\sigma_{i,t}^{2}.
\end{eqnarray*}
\item Under ${\cal M}_{2}$: we get 
\begin{eqnarray*}
Q_{1}\left(\zeta,\tilde{\zeta}\right) & = & E_{\text{\ensuremath{\tilde{\zeta}}}}\left[\frac{1}{nT}\sum_{i}\sum_{t}\left(\varepsilon_{i,t}^{*}-\theta_{i}h_{t}\right)^{2}|\varepsilon\right]\\
 & = & \frac{1}{nT}\sum_{i}\sum_{t}E_{\tilde{\zeta}}\left[\left(\varepsilon_{i,t}^{*}-\theta_{i}h_{t}\right)^{2}|\varepsilon\right]\\
 & = & \frac{1}{nT}\sum_{i}\sum_{t}V_{\tilde{\zeta}}\left[\varepsilon_{i,t}^{*}|\varepsilon\right]+\frac{1}{nT}\sum_{i}\sum_{t}\left(E_{\tilde{\zeta}}\left[\varepsilon_{i,t}^{*}|\varepsilon\right]-\theta_{i}h_{t}\right)^{2}.
\end{eqnarray*}
We have 
\[
\tilde{\varepsilon}_{i,t}:=E_{\tilde{\zeta}}\left[\varepsilon_{i,t}^{*}|\varepsilon\right]=\begin{cases}
\varepsilon_{i,t}, & \text{if }I_{i,t}=1,\\
\tilde{\theta}_{i}\tilde{h}_{t}, & \text{if }I_{i,t}=0,
\end{cases}\text{ and }V\left[\varepsilon_{i,t}^{*}|\varepsilon\right]=\begin{cases}
0, & \text{if }I_{i,t}=1,\\
\sigma_{i,t}^{2}, & \text{if }I_{i,t}=0.
\end{cases}
\]
Thus, ${\displaystyle Q_{1}\left(\text{\ensuremath{\zeta}},\tilde{\zeta}\right)=\frac{1}{nT}\sum_{i}\sum_{t}\left(\tilde{\tilde{\varepsilon}}_{i,t}-\theta_{i}h_{t}\right)^{2}+\frac{1}{nT}\sum_{i}\sum_{t}\left(1-I_{i,t}\right)\sigma_{i,t}^{2},}$
and the values of $\zeta$ that minimize $Q_{1}\left(\zeta,\tilde{\zeta}\right)$
can be calculated by ${\displaystyle \min_{\zeta}\frac{1}{nT}\sum_{i}\sum_{t}\left(\tilde{\tilde{\varepsilon}}_{i,t}-\theta_{i}h_{t}\right)^{2}}$.
This minimization problem reduces to the usual PCA on data $\tilde{\tilde{\varepsilon}}$:
${\displaystyle {\displaystyle \min_{\zeta}\frac{1}{nT}\sum_{i}\sum_{t}\left(\tilde{\tilde{\varepsilon}}_{i,t}-\theta_{i}h_{t}\right)^{2}}=\frac{1}{nT}\sum_{i}\sum_{t}\tilde{\tilde{\varepsilon}}_{i,t}^{2}-\mu_{1}\left(\frac{\tilde{\tilde{\varepsilon}}\tilde{\tilde{\varepsilon}}^{\prime}}{nT}\right).}$
Therefore, at convergence with $\hat{\zeta}=\tilde{\zeta}$, we have
\begin{eqnarray*}
Q_{1}\left(\hat{\zeta},\tilde{\zeta}\right) & = & \frac{1}{nT}\sum_{i}\sum_{t}\tilde{\tilde{\varepsilon}}_{i,t}^{2}-\mu_{1}\left(\frac{\tilde{\tilde{\varepsilon}}\tilde{\tilde{\varepsilon}}^{\prime}}{nT}\right)+\frac{1}{nT}\sum_{i}\sum_{t}\left(1-I_{i,t}\right)\sigma_{i,t}^{2}\\
 & = & \frac{1}{nT}\sum_{i}\sum_{t}I_{i,t}\varepsilon_{i,t}^{2}+\frac{1}{nT}\sum_{i}\sum_{t}\left(1-I_{i,t}\right)\left(\hat{\theta}_{i}\hat{h}_{t}\right)^{2}\\
 &  & \qquad-\mu_{1}\left(\frac{\tilde{\tilde{\varepsilon}}\tilde{\tilde{\varepsilon}}^{\prime}}{nT}\right)+\frac{1}{nT}\sum_{i}\sum_{t}\left(1-I_{i,t}\right)\sigma_{i,t}^{2}.
\end{eqnarray*}
\end{itemize}
Finally, the difference of the two EM criteria is 
\begin{eqnarray*}
Q_{0}-Q_{1}\left(\hat{\zeta},\hat{\zeta}\right) & = & \mu_{1}\left(\frac{\tilde{\tilde{\varepsilon}}\tilde{\tilde{\varepsilon}}^{\prime}}{nT}\right)-\frac{1}{nT}\sum_{i}\sum_{t}\left(1-I_{i,t}\right)\left(\hat{\theta}_{i}\hat{h}_{t}\right)^{2},
\end{eqnarray*}
which gives the criterion $\xi^{SW}$ after penalization.


\section{Monte-Carlo experiments \label{sec:App_MC_experiments}}

From the core text, we know that our selection procedure is equivalent
to the penalized least-squares strategy of \cite{Bai_Ng_2002} when
we use the same penalty term. The first part of our Monte-Carlo experiments
in Section \ref{subsec:SimLATENTFactorsOnly} aims to show, as expected,
that the penalisation introduced in Section \ref{sec:Implementation}
delivers a performance similar to the one observed in the literature
with other penalisations in the presence of latent factors only. We
investigate settings with $n$ and $T$ of comparable sizes as well
as $n$ much larger than $T$, as covered by our theory, starting
with balanced panels. Then we investigate how our selection procedure
performs with unbalanced panels. The main result is that the performance
is similar when the operative sizes of an unbalanced panel, i.e.,
the cross-sectional dimension $n^{\chi}$ and the average $\bar{T}_{i}$
of the time-series dimensions $T_{i}$, $i=1,...,n^{\chi}$, obtained
after trimming, are close to the sizes $n$ and $T$ of a balanced
panel. The second part of our Monte-Carlo experiments in Section \ref{subsec:SimObservable=000026LATENTFactors}
aims to extend the performance study to settings where we face observable
factors, and apply the selection procedure to residuals. We show that
the performance is close to the one obtained without observable factors.

\subsection{Simulations with $r$ latent factors\label{subsec:SimLATENTFactorsOnly}}

In this section, we perform simulation exercises in the presence of
$r$ latent factors only. We consider both balanced and unbalanced
panels in the study of the properties of our diagnostic criterion.
In the balanced case, we consider the simulation design in \cite{Ahn_Horenstein_2013}.
We generate $S=1,000$ datasets of dimension $n\times T$ so that,
at each simulation $s=1,...,S$, 
\begin{equation}
R_{i,t}^{s}=\sum_{j=1}^{r}b_{i,j}^{s}f_{j,t}^{s}+\varepsilon_{i,t}^{s},\ \text{with }\varepsilon_{i,t}^{s}=\sqrt{\frac{1-\rho^{2}}{1+2J\beta^{2}}}e_{i,t}^{s},\ \text{for }i=1,..,n,\ \text{and }t=1,...,T,\label{DGP_AH}
\end{equation}
where ${\displaystyle e_{i,t}^{s}=\rho e_{i,t-1}^{s}+v_{i,t}^{s}+\sum_{h=\max\left(i-J,1\right)}^{i-1}\beta v_{h,t}^{s}+\sum_{h=i+1}^{\min\left(i+J,n\right)}\beta v_{h,t}^{s}}$,
and the error term $v_{i,t}^{s}$, the factor loading $b_{i,j}^{s}$,
and the factor $f_{j,t}^{s}$ are drawn from standardized normal distributions.
\cite{Bai_Ng_2002} and \cite{Onatski_2010} use a similar data generating
process (DGP) in their simulation exercises. The DGP in Equation \eqref{DGP_AH}
depends on three parameters: (i) $\rho$ measures the magnitude of
the time-series correlation in the idiosyncratic errors $e_{i,t}^{s}$,
(ii) $\beta$ measures the magnitude of the cross-sectional correlation
between the errors $e_{i,t}^{s}$, (iii) $J$ defines the number of
units $i$ that are cross-correlated. At each simulation, we compute
our diagnostic criterion on the standardized $R_{i,t}^{s}$, when
$r=0$ and $r=3$. The trimming levels do not affect the number of
assets $n$ in the simulations since the panel is balanced and $T$
is sufficiently large. In order to understand how the criterion works
in practice, we consider several covariances structures, i.e., several
combinations of parameters $\left(\rho,\beta,J\right)$, and several
combinations of the cross-sectional dimension $n$ and the time-series
dimension $T$. Table \ref{MCtab_BalCase_LatentFactorONLY} reports
the selection probability of the correct model estimated from the
simulated datasets when $r=0$, i.e., $Pr(\xi<0\vert{\cal M}_{1})$.
The selection probabilities are close to $100\%$ for most combinations
of the cross-sectional sample size $n$ and the time-series dimension
$T$. Table \ref{MCtab_BalCase_LatentFactorONLY}, Panels A-D, compares
the selection probabilities when the magnitude of the time-series
correlation changes in the error structure. Table \ref{MCtab_BalCase_LatentFactorONLY},
Panels E-G contains the results when the magnitude of the cross-sectional
correlation increases throught parameters $\beta$ and $J$. The increase
of the correlation in the cross-section affects the selection probabilities
when the ratio $T/n$ is too far from zero. In Panels F and G, the
selection probability falls to zero when $n$ is much smaller than
$T$. In Panel G, we can explain $Pr(\xi<0\vert{\cal M}_{1})=85.30$,
when $n=T=150$ by the cross-sectional correlation being confused
with a common latent factor when $n$ is too small. The magnitude
of the cross-sectional correlation in the errors has a larger effect
on the selection probabilities than the presence of the time-series
correlation in the errors. Indeed, we always select correctly the
model in Panels B-D. Table \ref{MCtab_BalCase_LatentFactorONLY} also
reports the $x\%$ of the replications that result in overestimation
w.r.t. the $y\%$ of the replications that result in underestimation
of the number of factors for $r=3$. The criterion $\xi\left(k\right)$
introduced in Equation \eqref{xik} estimates the correct number of
unobservable factors for any combinations of $n$ and $T$ in the
different designs for the error structure. The penalty function is
based on a Gaussian reference model, namely ${\displaystyle g(n_{\chi},T)=c\frac{\left(\sqrt{n_{\chi}}+\sqrt{T}\right)^{2}}{n_{\chi}T}\ln\left(\frac{n_{\chi}T}{\left(\sqrt{n_{\chi}}+\sqrt{T}\right)^{2}}\right)}$,
where $c$ is a data-driven constant. The constant is selected as
in \citet{Alessi_Barigozzi_Capasso_2010} (see also \cite{Hallin_Liska_2007}
in the general dynamic factor model). The procedure for selecting
the constant $c$ relies on the behavior of the variance of the selected
number of factors $\hat{k}_{j}\left(c\right)$ computed across cross-sectional
subsamples $j=1,...,33$ of increasing size $\left(n_{j},T\right)$,
for a whole interval of values of the constant $c$. If the panel
is balanced, we choose the smallest value of $c$ in the second stability
interval of the variance (i.e., the second interval of values $c$
for which the variance of $\hat{k}_{j}\left(c\right)$ is insensitive
to $c$ as advocated by \citet{Alessi_Barigozzi_Capasso_2010}). The
criterion $\xi\left(k\right)$ thus compares well with the other selection
methods proposed by the literature, as expected. The results in Panels
A and B are comparable with Tables 1 and 2 in \cite{Onatski_2010}
and Figure 1 in \cite{Ahn_Horenstein_2013}. The criterion $\xi\left(k\right)$
performs equally well as the maximization criteria proposed by \cite{Bai_Ng_2002},
\cite{Ahn_Horenstein_2013}, and \cite{Onatski_2010} when the error
terms $e_{i,t}^{s}$ are i.i.d. and $r=3$. However, the maximization
criteria in Tables 2 and 3 in \cite{Onatski_2010} performs worse
than the criterion $\xi\left(k\right)$ in the estimation of the number
of factors when the idiosyncratic terms are time-serially correlated
(see also Figure 1 Panel B in \cite{Ahn_Horenstein_2013}). When the
error terms are cross-sectionally correlated, we get similar results
as in Figure 1 Panel C in \cite{Ahn_Horenstein_2013}, namely that
large dimensions of $n$ and $T$ allow the criterion to work better.
A similar result is also achieved when the error terms are serially
and cross-sectionally correlated (see Figure 1 Panel D in \cite{Ahn_Horenstein_2013}). 

For unbalanced panels, we explore the properties of the diagnostic
criterion using a simulation design that mimics the empirical features
of our data. We simulate a matrix of observability indicators $I^{s}\in\mathbb{R}^{n\times T}$
as follows. We fix an integer $\min\left(T_{i}\right)\leq T$. For
each $i$, we draw $T_{i}$ from a uniform distribution on an interval
of integers between $\min\left(T_{i}\right)$ and $T$. The $T_{i}$
ones for asset $i$ ($I_{i,t}^{s}=1$) are set for consecutive dates
starting from a random date $t_{0,i}$. The draws across individuals
$i$ are independent. We generate $R_{i,t}^{s}$ in Equation \eqref{DGP_AH}
if $I_{i,t}^{s}=1,\,\text{for }i=1,...,n,\,\text{and }\,t=1,...,T$.
In this framework, the trimming approach is not needed. We keep the
cross-sectional dimensions equal to $n=150,500,1500,3000,6000.$ In
Table \ref{MCTab_TSizeUNBcase}, we report the mean of the averages
$\bar{T}_{i}$ of the time-series sizes $T_{i}$, $i=1,...,n$, across
simulations, as well as the min and the max of the $\bar{T}_{i}$.
The choice of the $\min\left(T_{i}\right)$ in the simulation approach
defines the amount of missing values in the simulated sample. Since
the effective time-series sizes are smaller than $T$, we need to
compensate for this in the penalty function of the diagnostic criterion.
Otherwise, we have a tendency to underpenalize and to diagnose a too
large number of omitted factors. Thus, we select a constant $c$ larger
than the constant selected for balanced panels, namely, we choose
the smallest value of $c$ of the third (instead of the second) stability
interval of the variance of the selected number of factors.   Tables
\ref{MCtab_UnBalCase_LatentFactorONLY60}-\ref{MCtab_UnBalCase_LatentFactorONLY240}
report the results for several levels of $\min\left(T_{i}\right)$.
In general, the unbalanced property of the dataset does not seem to
deteriorate the selection probabilities of the correct model when
$r=0$ or $r=3$. We observe a deterioration of the probability when
the time-series size is too short w.r.t. the cross-sectional dimension
and the amount of missing values is high (see Panel D in Table \ref{MCtab_UnBalCase_LatentFactorONLY60}).

\subsection{Simulations with one observable factor and $r$ latent factors \label{subsec:SimObservable=000026LATENTFactors}}

In this section, we gauge the impact of estimation error coming from
using residuals instead of the true errors in the diagnostic criterion.
We repeat the experiments described in the previous section, but with
one observable factor and $r$ latent factors in the DGP, so that,
at each simulation $s=1,...,S$, 
\begin{equation}
R_{i,t}^{s}=B_{i}F_{t}+\sum_{j=1}^{r}b_{i,j}^{s}f_{j,t}^{s}+\varepsilon_{i,t}^{s},\ \text{with }\varepsilon_{i,t}^{s}=\sqrt{\frac{1-\rho^{2}}{1+2J\beta^{2}}}e_{i,t}^{s},\ \text{for }i=1,..,n,\ \text{and }t=1,...,T,\label{DGP_AH-1}
\end{equation}
where $B_{i}$ is the factor loading of the single observable factor
$F_{t}$, both initially drawn from standardized normal distributions.
At each simulation, we estimate a one-factor model on $R_{i,t}^{s}$,
and we compute the diagnostic criterion on the standardized residuals,
when $r=0$ and $r=3$. In Table \ref{MCtab_BalCase_OneOBSLatentFactor},
we get the results for the balanced case. The performance of the criterion
$\xi\left(k\right)$ is similar to what we get in Table \ref{MCtab_BalCase_LatentFactorONLY}.
Thus, the presence of observable factors does not corrode the performance
of $\xi\left(k\right)$. We conclude similarly for the unbalanced
case through comparing the simulation results reported in Tables \ref{MCtab_UnBalCase_OneObsLatentFactor60}-\ref{MCtab_UnBalCase_OneObsLatentFactor240}
with the ones of Tables \ref{MCtab_UnBalCase_LatentFactorONLY60}-\ref{MCtab_UnBalCase_LatentFactorONLY240}.
We only observe a moderate tendency to underestimate the number of
latent factors when $n$ is too small. 



\newgeometry{top=1cm, bottom=1cm, left=2cm, right=2cm}\thispagestyle{empty}
\begin{sidewaystable}
\textbf{\protect\caption{\textbf{\label{MCtab_BalCase_LatentFactorONLY}Monte Carlo replications
of DGP \eqref{DGP_AH} on balanced panels with $r=0$ or $3$ latent
factors only}}
}
\begin{centering}
{\scriptsize{}}%
\begin{tabular}{c|ccccc|ccccc|ccccc|ccccc}
\hline 
{\scriptsize{}{}$T$ } & \multicolumn{5}{c|}{{\scriptsize{}{}150}} & \multicolumn{5}{c|}{{\scriptsize{}{}500}} & \multicolumn{5}{c|}{{\scriptsize{}{}150}} & \multicolumn{5}{c}{{\scriptsize{}{}500}}\tabularnewline
\hline 
{\scriptsize{}{}$n$ } & {\scriptsize{}{}150 } & {\scriptsize{}{}500 } & {\scriptsize{}{}1,500 } & {\scriptsize{}{}3,000 } & {\scriptsize{}{}6,000 } & {\scriptsize{}{}150 } & {\scriptsize{}{}500 } & {\scriptsize{}{}1,500 } & {\scriptsize{}{}3,000 } & {\scriptsize{}{}6,000 } & {\scriptsize{}{}150 } & {\scriptsize{}{}500 } & {\scriptsize{}{}1,500 } & {\scriptsize{}{}3,000 } & {\scriptsize{}{}6,000 } & {\scriptsize{}{}150 } & {\scriptsize{}{}500 } & {\scriptsize{}{}1,500 } & {\scriptsize{}{}3,000 } & {\scriptsize{}{}6,000 }\tabularnewline
\hline 
 & \multicolumn{10}{c|}{{\scriptsize{}{}Panel A: i.i.d. errors}\textbf{\scriptsize{}{} $\left(\rho=0,\beta=0,J=0\right)$}} & \multicolumn{10}{c}{{\scriptsize{}{}Panel B: serially correlated errors}\textbf{\scriptsize{}{}
$\left(\rho=0.3,\beta=0,J=0\right)$}}\tabularnewline
\hline 
{\scriptsize{}{}$Pr(\xi<0\vert{\cal M}_{1})$ } & {\scriptsize{}100} & {\scriptsize{}100} & {\scriptsize{}100} & {\scriptsize{}100} & {\scriptsize{}100} & {\scriptsize{}100} & {\scriptsize{}100} & {\scriptsize{}100} & {\scriptsize{}100} & {\scriptsize{}100} & {\scriptsize{}100} & {\scriptsize{}100} & {\scriptsize{}100} & {\scriptsize{}100} & {\scriptsize{}100} & {\scriptsize{}100} & {\scriptsize{}100} & {\scriptsize{}100} & {\scriptsize{}100} & {\scriptsize{}100}\tabularnewline
{\scriptsize{}{}$Pr(\hat{k}>3)/\ Pr(\hat{k}<3)$} & {\scriptsize{}0/0} & {\scriptsize{}0/0} & {\scriptsize{}0/0} & {\scriptsize{}0/0} & {\scriptsize{}0/0} & {\scriptsize{}0/0} & {\scriptsize{}0/0} & {\scriptsize{}0/0} & {\scriptsize{}0/0} & {\scriptsize{}0/0} & {\scriptsize{}0/0} & {\scriptsize{}0/0} & {\scriptsize{}0/0} & {\scriptsize{}0/0} & {\scriptsize{}0/0} & {\scriptsize{}0/0} & {\scriptsize{}0/0} & {\scriptsize{}0/0} & {\scriptsize{}0/0} & {\scriptsize{}0/0}\tabularnewline
\hline 
 & \multicolumn{10}{c|}{{\scriptsize{}{}Panel C: serially correlated errors}\textbf{\scriptsize{}{}
$\left(\rho=0.5,\beta=0,J=0\right)$}} & \multicolumn{10}{c}{{\scriptsize{}{}Panel D: serially correlated errors}\textbf{\scriptsize{}{}
$\left(\rho=0.7,\beta=0,J=0\right)$}}\tabularnewline
\hline 
{\scriptsize{}{}$Pr(\xi<0\vert{\cal M}_{1})$ } & {\scriptsize{}100} & {\scriptsize{}100} & {\scriptsize{}100} & {\scriptsize{}100} & {\scriptsize{}100} & {\scriptsize{}100} & {\scriptsize{}100} & {\scriptsize{}100} & {\scriptsize{}100} & {\scriptsize{}100} & {\scriptsize{}100} & {\scriptsize{}100} & {\scriptsize{}100} & {\scriptsize{}100} & {\scriptsize{}100} & {\scriptsize{}100} & {\scriptsize{}100} & {\scriptsize{}100} & {\scriptsize{}100} & {\scriptsize{}100}\tabularnewline
{\scriptsize{}{}$Pr(\hat{k}>3)/\ Pr(\hat{k}<3)$} & {\scriptsize{}0/0} & {\scriptsize{}0/0} & {\scriptsize{}0/0} & {\scriptsize{}0/0} & {\scriptsize{}0/0} & {\scriptsize{}0/0} & {\scriptsize{}0/0} & {\scriptsize{}0/0} & {\scriptsize{}0/0} & {\scriptsize{}0/0} & {\scriptsize{}0/0} & {\scriptsize{}0/0} & {\scriptsize{}0/0} & {\scriptsize{}0/0} & {\scriptsize{}0/0} & {\scriptsize{}0/0} & {\scriptsize{}0/0} & {\scriptsize{}0/0} & {\scriptsize{}0/0} & {\scriptsize{}0/0}\tabularnewline
\hline 
 & \multicolumn{10}{c|}{{\scriptsize{}{}Panel E: cross-sectionally correlated errors}\textbf{\scriptsize{}{}
$\left(\rho=0,\beta=0.2,J=5\right)$}} & \multicolumn{10}{c}{{\scriptsize{}{}Panel F: cross-sectionally correlated errors}\textbf{\scriptsize{}{}
$\left(\rho=0,\beta=0.2,J=10\right)$}}\tabularnewline
\hline 
{\scriptsize{}{}$Pr(\xi<0\vert{\cal M}_{1})$ } & {\scriptsize{}100} & {\scriptsize{}100} & {\scriptsize{}100} & {\scriptsize{}100} & {\scriptsize{}100} & {\scriptsize{}100} & {\scriptsize{}100} & {\scriptsize{}100} & {\scriptsize{}100} & {\scriptsize{}100} & {\scriptsize{}1.3} & {\scriptsize{}100} & {\scriptsize{}100} & {\scriptsize{}100} & {\scriptsize{}100} & {\scriptsize{}0} & {\scriptsize{}99.23} & {\scriptsize{}100} & {\scriptsize{}100} & {\scriptsize{}100}\tabularnewline
{\scriptsize{}{}$Pr(\hat{k}>3)/\ Pr(\hat{k}<3)$} & {\scriptsize{}0/0} & {\scriptsize{}0/0} & {\scriptsize{}0/0} & {\scriptsize{}0/0} & {\scriptsize{}0/0} & {\scriptsize{}0/0} & {\scriptsize{}0/0} & {\scriptsize{}0/0} & {\scriptsize{}0/0} & {\scriptsize{}0/0} & {\scriptsize{}0/0} & {\scriptsize{}0/0} & {\scriptsize{}0/0} & {\scriptsize{}0/0} & {\scriptsize{}0/0} & {\scriptsize{}0/0} & {\scriptsize{}0/0} & {\scriptsize{}0/0} & {\scriptsize{}0/0} & {\scriptsize{}0/0}\tabularnewline
\hline 
 & \multicolumn{10}{c|}{{\scriptsize{}{}Panel G: cross-sectionally correlated errors}\textbf{\scriptsize{}{}
$\left(\rho=0,\beta=0.5,J=5\right)$}} & \multicolumn{10}{c}{{\scriptsize{}{}Panel H: serially and cross-sectionally correlated
errors}\textbf{\scriptsize{}{} $\left(\rho=0.2,\beta=0.2,J=5\right)$}}\tabularnewline
\hline 
{\scriptsize{}{}$Pr(\xi<0\vert{\cal M}_{1})$ } & {\scriptsize{}85.30} & {\scriptsize{}100} & {\scriptsize{}100} & {\scriptsize{}100} & {\scriptsize{}100} & {\scriptsize{}0} & {\scriptsize{}100} & {\scriptsize{}100} & {\scriptsize{}100} & {\scriptsize{}100} & {\scriptsize{}100} & {\scriptsize{}100} & {\scriptsize{}100} & {\scriptsize{}100} & {\scriptsize{}100} & {\scriptsize{}100} & {\scriptsize{}100} & {\scriptsize{}100} & {\scriptsize{}100} & {\scriptsize{}100}\tabularnewline
{\scriptsize{}{}$Pr(\hat{k}>3)/\ Pr(\hat{k}<3)$} & {\scriptsize{}0/0} & {\scriptsize{}0/0} & {\scriptsize{}0/0} & {\scriptsize{}0/0} & {\scriptsize{}0/0} & {\scriptsize{}0/0} & {\scriptsize{}0/0} & {\scriptsize{}0/0} & {\scriptsize{}0/0} & {\scriptsize{}0/0} & {\scriptsize{}0/0} & {\scriptsize{}0/0} & {\scriptsize{}0/0} & {\scriptsize{}0/0} & {\scriptsize{}0/0} & {\scriptsize{}0/0} & {\scriptsize{}0/0} & {\scriptsize{}0/0} & {\scriptsize{}0/0} & {\scriptsize{}0/0}\tabularnewline
\hline 
 & \multicolumn{10}{c|}{{\scriptsize{}{}Panel I: serially and cross-sectionally correlated
errors}\textbf{\scriptsize{}{} $\left(\rho=0.5,\beta=0.2,J=5\right)$}} & \multicolumn{10}{c}{{\scriptsize{}{}Panel L: serially and cross-sectionally correlated
errors}\textbf{\scriptsize{}{} $\left(\rho=0.2,\beta=0.5,J=5\right)$}}\tabularnewline
\hline 
{\scriptsize{}{}$Pr(\xi<0\vert{\cal M}_{1})$ } & {\scriptsize{}100} & {\scriptsize{}100} & {\scriptsize{}100} & {\scriptsize{}100} & {\scriptsize{}100} & {\scriptsize{}99.99} & {\scriptsize{}100} & {\scriptsize{}100} & {\scriptsize{}100} & {\scriptsize{}100} & {\scriptsize{}75.9} & {\scriptsize{}100} & {\scriptsize{}100} & {\scriptsize{}100} & {\scriptsize{}100} & {\scriptsize{}0} & {\scriptsize{}100} & {\scriptsize{}100} & {\scriptsize{}100} & {\scriptsize{}100}\tabularnewline
{\scriptsize{}{}$Pr(\hat{k}>3)/\ Pr(\hat{k}<3)$} & {\scriptsize{}0/0} & {\scriptsize{}0/0} & {\scriptsize{}0/0} & {\scriptsize{}0/0} & {\scriptsize{}0/0} & {\scriptsize{}0/0} & {\scriptsize{}0/0} & {\scriptsize{}0/0} & {\scriptsize{}0/0} & {\scriptsize{}0/0} & {\scriptsize{}0/0} & {\scriptsize{}0/0} & {\scriptsize{}0/0} & {\scriptsize{}0/0} & {\scriptsize{}0/0} & {\scriptsize{}0/0} & {\scriptsize{}0/0} & {\scriptsize{}0/0} & {\scriptsize{}0/0} & {\scriptsize{}0/0}\tabularnewline
\hline 
\end{tabular}\medskip{}
\par\end{centering}
The table contains the probabilities of correct model selection estimated
from the simulated balanced dataset when $r=0$ (first row in each
panel). For $r=3,$ the table reports the percentages $x\%/y\%$ of
the replications that result in overestimation / underestimation of
the number of factors (second row in each panel). The $\left(100-x-y\right)\%$
of the replications is the probability of correct estimation of the
number of unobservable factors. The simulations are based on the DGP
in Equation (\ref{DGP_AH}). Panels A-L contain the results assuming
different combinations of parameters $\rho,\beta$ and $J$. 
\end{sidewaystable}

\restoregeometry

\begin{table}[H]
\textbf{\footnotesize{}{}{}{}{}{}\protect\protect\caption{\textbf{\footnotesize{}\label{MCTab_TSizeUNBcase}Operative time-series
sample size}}
}{\footnotesize \par}

\medskip{}

\centering{}{\scriptsize{}}%
\begin{tabular}{c|ccc}
\hline 
{\scriptsize{}trimming level , $T=150$} & {\scriptsize{}$\min\left(T_{i}\right)=60$ } & {\scriptsize{}$\min\left(T_{i}\right)=120$ } & {\scriptsize{}$\min\left(T_{i}\right)=240$ }\tabularnewline
\hline 
{\scriptsize{}mean$\left(\bar{T}_{i}\right)$ } & {\scriptsize{}105} & {\scriptsize{}135} & {\scriptsize{}-}\tabularnewline
{\scriptsize{}$\min\left(\bar{T}_{i}\right)$ } & {\scriptsize{}98} & {\scriptsize{}133} & {\scriptsize{}-}\tabularnewline
{\scriptsize{}$\max\left(\bar{T}_{i}\right)$ } & {\scriptsize{}113} & {\scriptsize{}138} & {\scriptsize{}-}\tabularnewline
\hline 
\end{tabular}{\scriptsize \par}

{\scriptsize{}\centering{}}%
\begin{tabular}{c|ccc}
\hline 
{\scriptsize{}trimming level , $T=500$} & {\scriptsize{}$\min\left(T_{i}\right)=60$ } & {\scriptsize{}$\min\left(T_{i}\right)=120$ } & {\scriptsize{}$\min\left(T_{i}\right)=240$ }\tabularnewline
\hline 
{\scriptsize{}mean$\left(\bar{T}_{i}\right)$ } & {\scriptsize{}279} & {\scriptsize{}310} & {\scriptsize{}370}\tabularnewline
{\scriptsize{}$\min\left(\bar{T}_{i}\right)$ } & {\scriptsize{}237} & {\scriptsize{}277} & {\scriptsize{}348}\tabularnewline
{\scriptsize{}$\max\left(\bar{T}_{i}\right)$ } & {\scriptsize{}300} & {\scriptsize{}346} & {\scriptsize{}395}\tabularnewline
\hline 
\end{tabular}{\scriptsize \par}
\end{table}

\newgeometry{top=1cm, bottom=1cm, left=2cm, right=2cm}\thispagestyle{empty}
\begin{sidewaystable}
\textbf{\protect\caption{\textbf{\label{MCtab_UnBalCase_LatentFactorONLY60}Monte Carlo replications
of DGP \eqref{DGP_AH} on unbalanced panels with $\min\left(T_{i}\right)=60$
and $r=0$ or $3$ latent factors only}}
}
\begin{centering}
{\scriptsize{}}%
\begin{tabular}{c|ccccc|ccccc|ccccc|ccccc}
\hline 
{\scriptsize{}{}$T$ } & \multicolumn{5}{c|}{{\scriptsize{}{}150}} & \multicolumn{5}{c|}{{\scriptsize{}{}500}} & \multicolumn{5}{c|}{{\scriptsize{}{}150}} & \multicolumn{5}{c}{{\scriptsize{}{}500}}\tabularnewline
\hline 
{\scriptsize{}{}$n$ } & {\scriptsize{}{}150 } & {\scriptsize{}{}500 } & {\scriptsize{}{}1,500 } & {\scriptsize{}{}3,000 } & {\scriptsize{}{}6,000 } & {\scriptsize{}{}150 } & {\scriptsize{}{}500 } & {\scriptsize{}{}1,500 } & {\scriptsize{}{}3,000 } & {\scriptsize{}{}6,000 } & {\scriptsize{}{}150 } & {\scriptsize{}{}500 } & {\scriptsize{}{}1,500 } & {\scriptsize{}{}3,000 } & {\scriptsize{}{}6,000 } & {\scriptsize{}{}150 } & {\scriptsize{}{}500 } & {\scriptsize{}{}1,500 } & {\scriptsize{}{}3,000 } & {\scriptsize{}{}6,000 }\tabularnewline
\hline 
 & \multicolumn{10}{c|}{{\scriptsize{}{}Panel A: i.i.d. errors}\textbf{\scriptsize{}{} $\left(\rho=0,\beta=0,J=0\right)$}} & \multicolumn{10}{c}{{\scriptsize{}{}Panel B: serially correlated errors}\textbf{\scriptsize{}{}
$\left(\rho=0.3,\beta=0,J=0\right)$}}\tabularnewline
\hline 
{\scriptsize{}{}$Pr(\xi<0\vert{\cal M}_{1})$ } & {\scriptsize{}100} & {\scriptsize{}100} & {\scriptsize{}100} & {\scriptsize{}100} & {\scriptsize{}100} & {\scriptsize{}100} & {\scriptsize{}100} & {\scriptsize{}100} & {\scriptsize{}100} & {\scriptsize{}100} & {\scriptsize{}100} & {\scriptsize{}100} & {\scriptsize{}100} & {\scriptsize{}100} & {\scriptsize{}100} & {\scriptsize{}100} & {\scriptsize{}100} & {\scriptsize{}100} & {\scriptsize{}100} & {\scriptsize{}100}\tabularnewline
{\scriptsize{}{}$Pr(\hat{k}>3)/\ Pr(\hat{k}<3)$} & {\scriptsize{}0/68.8} & {\scriptsize{}0/5.2} & {\scriptsize{}0/0} & {\scriptsize{}0/0} & {\scriptsize{}0/0} & {\scriptsize{}0/0} & {\scriptsize{}0/0} & {\scriptsize{}0/0} & {\scriptsize{}0/0} & {\scriptsize{}0/0} & {\scriptsize{}0/67.4} & {\scriptsize{}0/2.4} & {\scriptsize{}0/0} & {\scriptsize{}0/0} & {\scriptsize{}0/0} & {\scriptsize{}0.5/0} & {\scriptsize{}0/0} & {\scriptsize{}0/0} & {\scriptsize{}0/0} & {\scriptsize{}0.2/0}\tabularnewline
\hline 
 & \multicolumn{10}{c|}{{\scriptsize{}{}Panel C: serially correlated errors}\textbf{\scriptsize{}{}
$\left(\rho=0.5,\beta=0,J=0\right)$}} & \multicolumn{10}{c}{{\scriptsize{}{}Panel D: serially correlated errors}\textbf{\scriptsize{}{}
$\left(\rho=0.7,\beta=0,J=0\right)$}}\tabularnewline
\hline 
{\scriptsize{}{}$Pr(\xi<0\vert{\cal M}_{1})$ } & {\scriptsize{}100} & {\scriptsize{}100} & {\scriptsize{}100} & {\scriptsize{}100} & {\scriptsize{}100} & {\scriptsize{}100} & {\scriptsize{}100} & {\scriptsize{}100} & {\scriptsize{}100} & {\scriptsize{}100} & {\scriptsize{}100} & {\scriptsize{}100} & {\scriptsize{}83.40} & {\scriptsize{}0} & {\scriptsize{}0} & {\scriptsize{}100} & {\scriptsize{}100} & {\scriptsize{}100} & {\scriptsize{}100} & {\scriptsize{}100}\tabularnewline
{\scriptsize{}{}$Pr(\hat{k}>3)/\ Pr(\hat{k}<3)$} & {\scriptsize{}0/65.6} & {\scriptsize{}0/5.2} & {\scriptsize{}0/0} & {\scriptsize{}0/0} & {\scriptsize{}0/0} & {\scriptsize{}0/0} & {\scriptsize{}0/0} & {\scriptsize{}0/0} & {\scriptsize{}0/0} & {\scriptsize{}0.1/0} & {\scriptsize{}0/69.8} & {\scriptsize{}0/3.6} & {\scriptsize{}0.2/0} & {\scriptsize{}0/0} & {\scriptsize{}0/0} & {\scriptsize{}0/0} & {\scriptsize{}0/0} & {\scriptsize{}1/0} & {\scriptsize{}0/0} & {\scriptsize{}0.3/0}\tabularnewline
\hline 
 & \multicolumn{10}{c|}{{\scriptsize{}{}Panel E: cross-sectionally correlated errors}\textbf{\scriptsize{}{}
$\left(\rho=0,\beta=0.2,J=5\right)$}} & \multicolumn{10}{c}{{\scriptsize{}{}Panel F: cross-sectionally correlated errors}\textbf{\scriptsize{}{}
$\left(\rho=0,\beta=0.2,J=10\right)$}}\tabularnewline
\hline 
{\scriptsize{}{}$Pr(\xi<0\vert{\cal M}_{1})$ } & {\scriptsize{}100} & {\scriptsize{}100} & {\scriptsize{}100} & {\scriptsize{}100} & {\scriptsize{}100} & {\scriptsize{}100} & {\scriptsize{}100} & {\scriptsize{}100} & {\scriptsize{}100} & {\scriptsize{}100} & {\scriptsize{}70.4} & {\scriptsize{}100} & {\scriptsize{}100} & {\scriptsize{}100} & {\scriptsize{}100} & {\scriptsize{}100} & {\scriptsize{}100} & {\scriptsize{}100} & {\scriptsize{}100} & {\scriptsize{}100}\tabularnewline
{\scriptsize{}{}$Pr(\hat{k}>3)/\ Pr(\hat{k}<3)$} & {\scriptsize{}0/68.6} & {\scriptsize{}0/4.4} & {\scriptsize{}0/0} & {\scriptsize{}0/0} & {\scriptsize{}0/0} & {\scriptsize{}0/0} & {\scriptsize{}1/0} & {\scriptsize{}0/0} & {\scriptsize{}0/0} & {\scriptsize{}0.1/0} & {\scriptsize{}0/66.2} & {\scriptsize{}0.2/5.2} & {\scriptsize{}0/0} & {\scriptsize{}0/0} & {\scriptsize{}0/0} & {\scriptsize{}0/0} & {\scriptsize{}0/0} & {\scriptsize{}0/0} & {\scriptsize{}0/0} & {\scriptsize{}0.4/0}\tabularnewline
\hline 
 & \multicolumn{10}{c|}{{\scriptsize{}{}Panel G: cross-sectionally correlated errors}\textbf{\scriptsize{}{}
$\left(\rho=0,\beta=0.5,J=5\right)$}} & \multicolumn{10}{c}{{\scriptsize{}{}Panel H: serially and cross-sectionally correlated
errors}\textbf{\scriptsize{}{} $\left(\rho=0.2,\beta=0.2,J=5\right)$}}\tabularnewline
\hline 
{\scriptsize{}{}$Pr(\xi<0\vert{\cal M}_{1})$ } & {\scriptsize{}99.8} & {\scriptsize{}100} & {\scriptsize{}100} & {\scriptsize{}100} & {\scriptsize{}100} & {\scriptsize{}100} & {\scriptsize{}100} & {\scriptsize{}100} & {\scriptsize{}100} & {\scriptsize{}100} & {\scriptsize{}100} & {\scriptsize{}100} & {\scriptsize{}100} & {\scriptsize{}100} & {\scriptsize{}100} & {\scriptsize{}100} & {\scriptsize{}100} & {\scriptsize{}100} & {\scriptsize{}100} & {\scriptsize{}100}\tabularnewline
{\scriptsize{}{}$Pr(\hat{k}>3)/\ Pr(\hat{k}<3)$} & {\scriptsize{}0/67.6} & {\scriptsize{}0/6} & {\scriptsize{}0/0} & {\scriptsize{}0/0} & {\scriptsize{}0/0} & {\scriptsize{}0/0} & {\scriptsize{}0.5/0} & {\scriptsize{}0/0} & {\scriptsize{}0/0} & {\scriptsize{}0.2/0} & {\scriptsize{}0/67.6} & {\scriptsize{}0/3.8} & {\scriptsize{}0/0} & {\scriptsize{}0/0} & {\scriptsize{}0/0} & {\scriptsize{}0/0} & {\scriptsize{}0/0} & {\scriptsize{}0/0} & {\scriptsize{}0/0} & {\scriptsize{}0.4/0}\tabularnewline
\hline 
 & \multicolumn{10}{c|}{{\scriptsize{}{}Panel I: serially and cross-sectionally correlated
errors}\textbf{\scriptsize{}{} $\left(\rho=0.5,\beta=0.2,J=5\right)$}} & \multicolumn{10}{c}{{\scriptsize{}{}Panel L: serially and cross-sectionally correlated
errors}\textbf{\scriptsize{}{} $\left(\rho=0.2,\beta=0.5,J=5\right)$}}\tabularnewline
\hline 
{\scriptsize{}{}$Pr(\xi<0\vert{\cal M}_{1})$ } & {\scriptsize{}100} & {\scriptsize{}100} & {\scriptsize{}100} & {\scriptsize{}100} & {\scriptsize{}100} & {\scriptsize{}100} & {\scriptsize{}100} & {\scriptsize{}100} & {\scriptsize{}100} & {\scriptsize{}100} & {\scriptsize{}99.2} & {\scriptsize{}100} & {\scriptsize{}100} & {\scriptsize{}100} & {\scriptsize{}100} & {\scriptsize{}100} & {\scriptsize{}100} & {\scriptsize{}100} & {\scriptsize{}100} & {\scriptsize{}100}\tabularnewline
{\scriptsize{}{}$Pr(\hat{k}>3)/\ Pr(\hat{k}<3)$} & {\scriptsize{}0/68.6} & {\scriptsize{}0/4.8} & {\scriptsize{}0/0} & {\scriptsize{}0/0} & {\scriptsize{}0/0} & {\scriptsize{}0/0} & {\scriptsize{}0.5/0} & {\scriptsize{}0/0} & {\scriptsize{}0/0} & {\scriptsize{}0.3/0} & {\scriptsize{}0/64} & {\scriptsize{}0/4.4} & {\scriptsize{}0/0} & {\scriptsize{}0/0} & {\scriptsize{}0/0} & {\scriptsize{}0/0} & {\scriptsize{}0/0} & {\scriptsize{}0/0} & {\scriptsize{}0/0} & {\scriptsize{}0.3/0}\tabularnewline
\hline 
\end{tabular}\medskip{}
\par\end{centering}
The table contains the probabilities of correct model selection estimated
from the simulated unbalanced dataset when $r=0$ (first row in each
panel). For $r=3,$ the table reports the percentages $x\%/y\%$ of
the replications that result in overestimation / underestimation of
the number of factors (second row in each panel). The $\left(100-x-y\right)\%$
of the replications is the probability of correct estimation of the
number of unobservable factors. The simulations are based on the DGP
in Equation (\ref{DGP_AH}) for unbalanced panels with\textbf{ $\min\left(T_{i}\right)=60$
}. Panels A-L contain the results assuming different combinations
of parameters $\rho,\beta$ and $J$. 
\end{sidewaystable}

\newgeometry{top=1cm, bottom=1cm, left=2cm, right=2cm}\thispagestyle{empty}
\begin{sidewaystable}
\textbf{\protect\caption{\textbf{\label{MCtab_UnBalCase_LatentFactorONLY120}Monte Carlo replications
of DGP \eqref{DGP_AH} on unbalanced panels with $\min\left(T_{i}\right)=120$
and $r=0$ or $3$ latent factors only}}
}
\begin{centering}
{\scriptsize{}}%
\begin{tabular}{c|ccccc|ccccc|ccccc|ccccc}
\hline 
{\scriptsize{}{}$T$ } & \multicolumn{5}{c|}{{\scriptsize{}{}150}} & \multicolumn{5}{c|}{{\scriptsize{}{}500}} & \multicolumn{5}{c|}{{\scriptsize{}{}150}} & \multicolumn{5}{c}{{\scriptsize{}{}500}}\tabularnewline
\hline 
{\scriptsize{}{}$n$ } & {\scriptsize{}{}150 } & {\scriptsize{}{}500 } & {\scriptsize{}{}1,500 } & {\scriptsize{}{}3,000 } & {\scriptsize{}{}6,000 } & {\scriptsize{}{}150 } & {\scriptsize{}{}500 } & {\scriptsize{}{}1,500 } & {\scriptsize{}{}3,000 } & {\scriptsize{}{}6,000 } & {\scriptsize{}{}150 } & {\scriptsize{}{}500 } & {\scriptsize{}{}1,500 } & {\scriptsize{}{}3,000 } & {\scriptsize{}{}6,000 } & {\scriptsize{}{}150 } & {\scriptsize{}{}500 } & {\scriptsize{}{}1,500 } & {\scriptsize{}{}3,000 } & {\scriptsize{}{}6,000 }\tabularnewline
\hline 
 & \multicolumn{10}{c|}{{\scriptsize{}{}Panel A: i.i.d. errors}\textbf{\scriptsize{}{} $\left(\rho=0,\beta=0,J=0\right)$}} & \multicolumn{10}{c}{{\scriptsize{}{}Panel B: serially correlated errors}\textbf{\scriptsize{}{}
$\left(\rho=0.3,\beta=0,J=0\right)$}}\tabularnewline
\hline 
{\scriptsize{}{}$Pr(\xi<0\vert{\cal M}_{1})$ } & {\scriptsize{}100} & {\scriptsize{}100} & {\scriptsize{}100} & {\scriptsize{}100} & {\scriptsize{}100} & {\scriptsize{}100} & {\scriptsize{}100} & {\scriptsize{}100} & {\scriptsize{}100} & {\scriptsize{}100} & {\scriptsize{}100} & {\scriptsize{}100} & {\scriptsize{}100} & {\scriptsize{}100} & {\scriptsize{}100} & {\scriptsize{}100} & {\scriptsize{}100} & {\scriptsize{}100} & {\scriptsize{}100} & {\scriptsize{}100}\tabularnewline
{\scriptsize{}{}$Pr(\hat{k}>3)/\ Pr(\hat{k}<3)$} & {\scriptsize{}0/42} & {\scriptsize{}0/0} & {\scriptsize{}0/0} & {\scriptsize{}0/0} & {\scriptsize{}0/0} & {\scriptsize{}0/12.4} & {\scriptsize{}0.1/0} & {\scriptsize{}0.8/0} & {\scriptsize{}0/0} & {\scriptsize{}0/0} & {\scriptsize{}0/42.3} & {\scriptsize{}0/0} & {\scriptsize{}0/0} & {\scriptsize{}0/0} & {\scriptsize{}0/0} & {\scriptsize{}0/13.8} & {\scriptsize{}0.1/0} & {\scriptsize{}1.0/0} & {\scriptsize{}0/0} & {\scriptsize{}0.2/0}\tabularnewline
\hline 
 & \multicolumn{10}{c|}{{\scriptsize{}{}Panel C: serially correlated errors}\textbf{\scriptsize{}{}
$\left(\rho=0.5,\beta=0,J=0\right)$}} & \multicolumn{10}{c}{{\scriptsize{}{}Panel D: serially correlated errors}\textbf{\scriptsize{}{}
$\left(\rho=0.7,\beta=0,J=0\right)$}}\tabularnewline
\hline 
{\scriptsize{}{}$Pr(\xi<0\vert{\cal M}_{1})$ } & {\scriptsize{}100} & {\scriptsize{}100} & {\scriptsize{}100} & {\scriptsize{}100} & {\scriptsize{}100} & {\scriptsize{}100} & {\scriptsize{}100} & {\scriptsize{}100} & {\scriptsize{}100} & {\scriptsize{}100} & {\scriptsize{}100} & {\scriptsize{}100} & {\scriptsize{}100} & {\scriptsize{}100} & {\scriptsize{}79} & {\scriptsize{}100} & {\scriptsize{}100} & {\scriptsize{}100} & {\scriptsize{}100} & {\scriptsize{}100}\tabularnewline
{\scriptsize{}{}$Pr(\hat{k}>3)/\ Pr(\hat{k}<3)$} & {\scriptsize{}0/45.4} & {\scriptsize{}0/0} & {\scriptsize{}0/0} & {\scriptsize{}0/0} & {\scriptsize{}0/0} & {\scriptsize{}0/12.2} & {\scriptsize{}0.1/0} & {\scriptsize{}1.3/0} & {\scriptsize{}0/0} & {\scriptsize{}0.1/0} & {\scriptsize{}0/43.6} & {\scriptsize{}0/0} & {\scriptsize{}0/0} & {\scriptsize{}0/0} & {\scriptsize{}0/0} & {\scriptsize{}0/13.2} & {\scriptsize{}0.3/0} & {\scriptsize{}1.5/0} & {\scriptsize{}0/0} & {\scriptsize{}0.3/0}\tabularnewline
\hline 
 & \multicolumn{10}{c|}{{\scriptsize{}{}Panel E: cross-sectionally correlated errors}\textbf{\scriptsize{}{}
$\left(\rho=0,\beta=0.2,J=5\right)$}} & \multicolumn{10}{c}{{\scriptsize{}{}Panel F: cross-sectionally correlated errors}\textbf{\scriptsize{}{}
$\left(\rho=0,\beta=0.2,J=10\right)$}}\tabularnewline
\hline 
{\scriptsize{}{}$Pr(\xi<0\vert{\cal M}_{1})$ } & {\scriptsize{}100} & {\scriptsize{}100} & {\scriptsize{}100} & {\scriptsize{}100} & {\scriptsize{}100} & {\scriptsize{}100} & {\scriptsize{}100} & {\scriptsize{}100} & {\scriptsize{}100} & {\scriptsize{}100} & {\scriptsize{}7.10} & {\scriptsize{}100} & {\scriptsize{}100} & {\scriptsize{}100} & {\scriptsize{}100} & {\scriptsize{}20} & {\scriptsize{}100} & {\scriptsize{}100} & {\scriptsize{}100} & {\scriptsize{}100}\tabularnewline
{\scriptsize{}{}$Pr(\hat{k}>3)/\ Pr(\hat{k}<3)$} & {\scriptsize{}0/43.1} & {\scriptsize{}0/0} & {\scriptsize{}0/0} & {\scriptsize{}0/0} & {\scriptsize{}0/0} & {\scriptsize{}0/12.8} & {\scriptsize{}0.3/0} & {\scriptsize{}1.6/0} & {\scriptsize{}0/0} & {\scriptsize{}0.1/0} & {\scriptsize{}0/42.9} & {\scriptsize{}0/0} & {\scriptsize{}0.1/0} & {\scriptsize{}0/0} & {\scriptsize{}0/0} & {\scriptsize{}0/11.9} & {\scriptsize{}0.3/0} & {\scriptsize{}1.2/0} & {\scriptsize{}0/0} & {\scriptsize{}0.4/0}\tabularnewline
\hline 
 & \multicolumn{10}{c|}{{\scriptsize{}{}Panel G: cross-sectionally correlated errors}\textbf{\scriptsize{}{}
$\left(\rho=0,\beta=0.5,J=5\right)$}} & \multicolumn{10}{c}{{\scriptsize{}{}Panel H: serially and cross-sectionally correlated
errors}\textbf{\scriptsize{}{} $\left(\rho=0.2,\beta=0.2,J=5\right)$}}\tabularnewline
\hline 
{\scriptsize{}{}$Pr(\xi<0\vert{\cal M}_{1})$ } & {\scriptsize{}93.50} & {\scriptsize{}100} & {\scriptsize{}100} & {\scriptsize{}100} & {\scriptsize{}100} & {\scriptsize{}95} & {\scriptsize{}100} & {\scriptsize{}100} & {\scriptsize{}100} & {\scriptsize{}100} & {\scriptsize{}100} & {\scriptsize{}100} & {\scriptsize{}100} & {\scriptsize{}100} & {\scriptsize{}100} & {\scriptsize{}100} & {\scriptsize{}100} & {\scriptsize{}100} & {\scriptsize{}100} & {\scriptsize{}100}\tabularnewline
{\scriptsize{}{}$Pr(\hat{k}>3)/\ Pr(\hat{k}<3)$} & {\scriptsize{}0/42} & {\scriptsize{}0/0} & {\scriptsize{}0/0} & {\scriptsize{}0/0} & {\scriptsize{}0/0} & {\scriptsize{}0/15.8} & {\scriptsize{}0.1/0} & {\scriptsize{}0.7/0} & {\scriptsize{}0/0} & {\scriptsize{}0.2/0} & {\scriptsize{}0/42.1} & {\scriptsize{}0/0} & {\scriptsize{}0/0} & {\scriptsize{}0/0} & {\scriptsize{}0/0} & {\scriptsize{}0/13} & {\scriptsize{}0/0} & {\scriptsize{}2/0} & {\scriptsize{}0/0} & {\scriptsize{}0.4/0}\tabularnewline
\hline 
 & \multicolumn{10}{c|}{{\scriptsize{}{}Panel I: serially and cross-sectionally correlated
errors}\textbf{\scriptsize{}{} $\left(\rho=0.5,\beta=0.2,J=5\right)$}} & \multicolumn{10}{c}{{\scriptsize{}{}Panel L: serially and cross-sectionally correlated
errors}\textbf{\scriptsize{}{} $\left(\rho=0.2,\beta=0.5,J=5\right)$}}\tabularnewline
\hline 
{\scriptsize{}{}$Pr(\xi<0\vert{\cal M}_{1})$ } & {\scriptsize{}100} & {\scriptsize{}100} & {\scriptsize{}100} & {\scriptsize{}100} & {\scriptsize{}100} & {\scriptsize{}100} & {\scriptsize{}100} & {\scriptsize{}100} & {\scriptsize{}100} & {\scriptsize{}100} & {\scriptsize{}88.10} & {\scriptsize{}100} & {\scriptsize{}100} & {\scriptsize{}100} & {\scriptsize{}100} & {\scriptsize{}90.2} & {\scriptsize{}100} & {\scriptsize{}100} & {\scriptsize{}100} & {\scriptsize{}100}\tabularnewline
{\scriptsize{}{}$Pr(\hat{k}>3)/\ Pr(\hat{k}<3)$} & {\scriptsize{}0/42.4} & {\scriptsize{}0/0} & {\scriptsize{}0/0} & {\scriptsize{}0/0} & {\scriptsize{}0/0} & {\scriptsize{}0/10.9} & {\scriptsize{}0.1/0} & {\scriptsize{}1.3/0} & {\scriptsize{}0/0} & {\scriptsize{}0.3/0} & {\scriptsize{}0/44.7} & {\scriptsize{}0/0} & {\scriptsize{}0/0} & {\scriptsize{}0/0} & {\scriptsize{}0/0} & {\scriptsize{}0/13.2} & {\scriptsize{}0.2/0} & {\scriptsize{}0.8/0} & {\scriptsize{}0/0} & {\scriptsize{}0.3/0}\tabularnewline
\hline 
\end{tabular}\medskip{}
\par\end{centering}
The table contains the probabilities of correct model selection estimated
from the simulated unbalanced dataset when $r=0$ (first row in each
panel). For $r=3,$ the table reports the percentages $x\%/y\%$ of
the replications that result in overestimation / underestimation of
the number of factors (second row in each panel). The $\left(100-x-y\right)\%$
of the replications is the probability of correct estimation of the
number of unobservable factors. The simulations are based on the DGP
in Equation (\ref{DGP_AH}) for unbalanced panels with\textbf{ $\min\left(T_{i}\right)=120$}.
Panels A-L contain the results assuming different combinations of
parameters $\rho,\beta$ and $J$. 
\end{sidewaystable}

\newgeometry{top=1cm, bottom=1cm, left=2cm, right=2cm}\thispagestyle{empty}
\begin{sidewaystable}
\textbf{\protect\caption{\textbf{\label{MCtab_UnBalCase_LatentFactorONLY240}Monte Carlo replications
of DGP \eqref{DGP_AH} on unbalanced panels with $\min\left(T_{i}\right)=240$
and $r=0$ or $3$ latent factors only}}
}
\begin{centering}
{\scriptsize{}}%
\begin{tabular}{c|ccccc|ccccc}
\hline 
{\scriptsize{}{}$T$ } & \multicolumn{5}{c|}{{\scriptsize{}{}500}} & \multicolumn{5}{c}{{\scriptsize{}{}500}}\tabularnewline
\hline 
{\scriptsize{}{}$n$ } & {\scriptsize{}{}150 } & {\scriptsize{}{}500 } & {\scriptsize{}{}1,500 } & {\scriptsize{}{}3,000 } & {\scriptsize{}{}6,000 } & {\scriptsize{}{}150 } & {\scriptsize{}{}500 } & {\scriptsize{}{}1,500 } & {\scriptsize{}{}3,000 } & {\scriptsize{}{}6,000 }\tabularnewline
\hline 
 & \multicolumn{5}{c|}{{\scriptsize{}{}Panel A: i.i.d. errors}\textbf{\scriptsize{}{} $\left(\rho=0,\beta=0,J=0\right)$}} & \multicolumn{5}{c}{{\scriptsize{}{}Panel B: serially correlated errors}\textbf{\scriptsize{}{}
$\left(\rho=0.3,\beta=0,J=0\right)$}}\tabularnewline
\hline 
{\scriptsize{}{}$Pr(\xi<0\vert{\cal M}_{1})$ } & {\scriptsize{}100} & {\scriptsize{}100} & {\scriptsize{}100} & {\scriptsize{}100} & {\scriptsize{}100} & {\scriptsize{}100} & {\scriptsize{}100} & {\scriptsize{}100} & {\scriptsize{}100} & {\scriptsize{}100}\tabularnewline
{\scriptsize{}{}$Pr(\hat{k}>3)/\ Pr(\hat{k}<3)$} & {\scriptsize{}0/1.2} & {\scriptsize{}0/0} & {\scriptsize{}0.8/0} & {\scriptsize{}0/0} & {\scriptsize{}0/0} & {\scriptsize{}0/1.8} & {\scriptsize{}0.2/0} & {\scriptsize{}0.4/0} & {\scriptsize{}0/0} & {\scriptsize{}0.2/0}\tabularnewline
\hline 
 & \multicolumn{5}{c|}{{\scriptsize{}{}Panel C: serially correlated errors}\textbf{\scriptsize{}{}
$\left(\rho=0.5,\beta=0,J=0\right)$}} & \multicolumn{5}{c}{{\scriptsize{}{}Panel D: serially correlated errors}\textbf{\scriptsize{}{}
$\left(\rho=0.7,\beta=0,J=0\right)$}}\tabularnewline
\hline 
{\scriptsize{}{}$Pr(\xi<0\vert{\cal M}_{1})$ } & {\scriptsize{}100} & {\scriptsize{}100} & {\scriptsize{}100} & {\scriptsize{}100} & {\scriptsize{}100} & {\scriptsize{}100} & {\scriptsize{}100} & {\scriptsize{}100} & {\scriptsize{}100} & {\scriptsize{}100}\tabularnewline
{\scriptsize{}{}$Pr(\hat{k}>3)/\ Pr(\hat{k}<3)$} & {\scriptsize{}0/1.8} & {\scriptsize{}0/0} & {\scriptsize{}0.6/0} & {\scriptsize{}0/0} & {\scriptsize{}0.1/0} & {\scriptsize{}0/2} & {\scriptsize{}0.2/0} & {\scriptsize{}0.4/0} & {\scriptsize{}0/0} & {\scriptsize{}0.3/0}\tabularnewline
\hline 
 & \multicolumn{5}{c|}{{\scriptsize{}{}Panel E: cross-sectionally correlated errors}\textbf{\scriptsize{}{}
$\left(\rho=0,\beta=0.2,J=5\right)$}} & \multicolumn{5}{c}{{\scriptsize{}{}Panel F: cross-sectionally correlated errors}\textbf{\scriptsize{}{}
$\left(\rho=0,\beta=0.2,J=10\right)$}}\tabularnewline
\hline 
{\scriptsize{}{}$Pr(\xi<0\vert{\cal M}_{1})$ } & {\scriptsize{}100} & {\scriptsize{}100} & {\scriptsize{}100} & {\scriptsize{}100} & {\scriptsize{}100} & {\scriptsize{}0} & {\scriptsize{}100} & {\scriptsize{}100} & {\scriptsize{}100} & {\scriptsize{}100}\tabularnewline
{\scriptsize{}{}$Pr(\hat{k}>3)/\ Pr(\hat{k}<3)$} & {\scriptsize{}0/1} & {\scriptsize{}0/0} & {\scriptsize{}0.4/0} & {\scriptsize{}0/0} & {\scriptsize{}0.1/0} & {\scriptsize{}0/0.6} & {\scriptsize{}0/0} & {\scriptsize{}0.4/0} & {\scriptsize{}0/0} & {\scriptsize{}0.4/0}\tabularnewline
\hline 
 & \multicolumn{5}{c|}{{\scriptsize{}{}Panel G: cross-sectionally correlated errors}\textbf{\scriptsize{}{}
$\left(\rho=0,\beta=0.5,J=5\right)$}} & \multicolumn{5}{c}{{\scriptsize{}{}Panel H: serially and cross-sectionally correlated
errors}\textbf{\scriptsize{}{} $\left(\rho=0.2,\beta=0.2,J=5\right)$}}\tabularnewline
\hline 
{\scriptsize{}{}$Pr(\xi<0\vert{\cal M}_{1})$ } & {\scriptsize{}11.4} & {\scriptsize{}100} & {\scriptsize{}100} & {\scriptsize{}100} & {\scriptsize{}100} & {\scriptsize{}100} & {\scriptsize{}100} & {\scriptsize{}100} & {\scriptsize{}100} & {\scriptsize{}100}\tabularnewline
{\scriptsize{}{}$Pr(\hat{k}>3)/\ Pr(\hat{k}<3)$} & {\scriptsize{}0/0.6} & {\scriptsize{}0/0} & {\scriptsize{}0.2/0} & {\scriptsize{}0/0} & {\scriptsize{}0.2/0} & {\scriptsize{}0/0.6} & {\scriptsize{}0/0} & {\scriptsize{}0/0} & {\scriptsize{}0/0} & {\scriptsize{}0.4/0}\tabularnewline
\hline 
 & \multicolumn{5}{c|}{{\scriptsize{}{}Panel I: serially and cross-sectionally correlated
errors}\textbf{\scriptsize{}{} $\left(\rho=0.5,\beta=0.2,J=5\right)$}} & \multicolumn{5}{c}{{\scriptsize{}{}Panel L: serially and cross-sectionally correlated
errors}\textbf{\scriptsize{}{} $\left(\rho=0.2,\beta=0.5,J=5\right)$}}\tabularnewline
\hline 
{\scriptsize{}{}$Pr(\xi<0\vert{\cal M}_{1})$ } & {\scriptsize{}100} & {\scriptsize{}100} & {\scriptsize{}100} & {\scriptsize{}100} & {\scriptsize{}100} & {\scriptsize{}8.6} & {\scriptsize{}100} & {\scriptsize{}100} & {\scriptsize{}100} & {\scriptsize{}100}\tabularnewline
{\scriptsize{}{}$Pr(\hat{k}>3)/\ Pr(\hat{k}<3)$} & {\scriptsize{}0/0.6} & {\scriptsize{}0.2/0} & {\scriptsize{}0.6/0} & {\scriptsize{}0/0} & {\scriptsize{}0.3/0} & {\scriptsize{}0/0.68} & {\scriptsize{}0/0} & {\scriptsize{}1.2/0} & {\scriptsize{}0/0} & {\scriptsize{}0.3/0}\tabularnewline
\hline 
\end{tabular}\medskip{}
\par\end{centering}
The table contains the probabilities of correct model selection estimated
from the simulated unbalanced dataset when $r=0$ (first row in each
panel). For $r=3,$ the table reports the percentages $x\%/y\%$ of
the replications that result in overestimation / underestimation of
the number of factors (second row in each panel). The $\left(100-x-y\right)\%$
of the replications is the probability of correct estimation of the
number of unobservable factors. The simulations are based on the DGP
in Equation (\ref{DGP_AH}) for unbalanced panels with\textbf{ $\min\left(T_{i}\right)=240$
}. Panels A-L contain the results assuming different combinations
of parameters $\rho,\beta$ and $J$. 
\end{sidewaystable}

\newgeometry{top=1cm, bottom=1cm, left=2cm, right=2cm}\thispagestyle{empty}
\begin{sidewaystable}
\textbf{\protect\caption{\textbf{\label{MCtab_BalCase_OneOBSLatentFactor}Monte Carlo replications
of DGP \eqref{DGP_AH-1} on balanced panels with one observable factor
and $r=0$ or $3$ latent factors}}
}
\begin{centering}
{\scriptsize{}}%
\begin{tabular}{c|ccccc|ccccc|ccccc|ccccc}
\hline 
{\scriptsize{}{}$T$ } & \multicolumn{5}{c|}{{\scriptsize{}{}150}} & \multicolumn{5}{c|}{{\scriptsize{}{}500}} & \multicolumn{5}{c|}{{\scriptsize{}{}150}} & \multicolumn{5}{c}{{\scriptsize{}{}500}}\tabularnewline
\hline 
{\scriptsize{}{}$n$ } & {\scriptsize{}{}150 } & {\scriptsize{}{}500 } & {\scriptsize{}{}1,500 } & {\scriptsize{}{}3,000 } & {\scriptsize{}{}6,000 } & {\scriptsize{}{}150 } & {\scriptsize{}{}500 } & {\scriptsize{}{}1,500 } & {\scriptsize{}{}3,000 } & {\scriptsize{}{}6,000 } & {\scriptsize{}{}150 } & {\scriptsize{}{}500 } & {\scriptsize{}{}1,500 } & {\scriptsize{}{}3,000 } & {\scriptsize{}{}6,000 } & {\scriptsize{}{}150 } & {\scriptsize{}{}500 } & {\scriptsize{}{}1,500 } & {\scriptsize{}{}3,000 } & {\scriptsize{}{}6,000 }\tabularnewline
\hline 
 & \multicolumn{10}{c|}{{\scriptsize{}{}Panel A: i.i.d. errors}\textbf{\scriptsize{}{} $\left(\rho=0,\beta=0,J=0\right)$}} & \multicolumn{10}{c}{{\scriptsize{}{}Panel B: serially correlated errors}\textbf{\scriptsize{}{}
$\left(\rho=0.3,\beta=0,J=0\right)$}}\tabularnewline
\hline 
{\scriptsize{}{}$Pr(\xi<0\vert{\cal M}_{1})$ } & {\scriptsize{}100} & {\scriptsize{}100} & {\scriptsize{}100} & {\scriptsize{}100} & {\scriptsize{}100} & {\scriptsize{}100} & {\scriptsize{}100} & {\scriptsize{}100} & {\scriptsize{}100} & {\scriptsize{}100} & {\scriptsize{}100} & {\scriptsize{}100} & {\scriptsize{}100} & {\scriptsize{}100} & {\scriptsize{}100} & {\scriptsize{}100} & {\scriptsize{}100} & {\scriptsize{}100} & {\scriptsize{}100} & {\scriptsize{}100}\tabularnewline
{\scriptsize{}{}$Pr(\hat{k}>3)/\ Pr(\hat{k}<3)$} & {\scriptsize{}0/0} & {\scriptsize{}0/0} & {\scriptsize{}0/0} & {\scriptsize{}0/0} & {\scriptsize{}0/0} & {\scriptsize{}0/0} & {\scriptsize{}0/0} & {\scriptsize{}0/0} & {\scriptsize{}0/0} & {\scriptsize{}0/0} & {\scriptsize{}0/0} & {\scriptsize{}0/0} & {\scriptsize{}0/0} & {\scriptsize{}0/0} & {\scriptsize{}0/0} & {\scriptsize{}0/0} & {\scriptsize{}0/0} & {\scriptsize{}0/0} & {\scriptsize{}0/0} & {\scriptsize{}0/0}\tabularnewline
\hline 
 & \multicolumn{10}{c|}{{\scriptsize{}{}Panel C: serially correlated errors}\textbf{\scriptsize{}{}
$\left(\rho=0.5,\beta=0,J=0\right)$}} & \multicolumn{10}{c}{{\scriptsize{}{}Panel D: serially correlated errors}\textbf{\scriptsize{}{}
$\left(\rho=0.7,\beta=0,J=0\right)$}}\tabularnewline
\hline 
{\scriptsize{}{}$Pr(\xi<0\vert{\cal M}_{1})$ } & {\scriptsize{}100} & {\scriptsize{}100} & {\scriptsize{}100} & {\scriptsize{}100} & {\scriptsize{}100} & {\scriptsize{}100} & {\scriptsize{}100} & {\scriptsize{}100} & {\scriptsize{}100} & {\scriptsize{}100} & {\scriptsize{}100} & {\scriptsize{}100} & {\scriptsize{}100} & {\scriptsize{}100} & {\scriptsize{}100} & {\scriptsize{}100} & {\scriptsize{}100} & {\scriptsize{}100} & {\scriptsize{}100} & {\scriptsize{}100}\tabularnewline
{\scriptsize{}{}$Pr(\hat{k}>3)/\ Pr(\hat{k}<3)$} & {\scriptsize{}0/0} & {\scriptsize{}0/0} & {\scriptsize{}0/0} & {\scriptsize{}0/0} & {\scriptsize{}0/0} & {\scriptsize{}0/0} & {\scriptsize{}0/0} & {\scriptsize{}0/0} & {\scriptsize{}0/0} & {\scriptsize{}0/0} & {\scriptsize{}0/0} & {\scriptsize{}0/0} & {\scriptsize{}0/0} & {\scriptsize{}0/0} & {\scriptsize{}0/0} & {\scriptsize{}0/0} & {\scriptsize{}0/0} & {\scriptsize{}0/0} & {\scriptsize{}0/0} & {\scriptsize{}0/0}\tabularnewline
\hline 
 & \multicolumn{10}{c|}{{\scriptsize{}{}Panel E: cross-sectionally correlated errors}\textbf{\scriptsize{}{}
$\left(\rho=0,\beta=0.2,J=5\right)$}} & \multicolumn{10}{c}{{\scriptsize{}{}Panel F: cross-sectionally correlated errors}\textbf{\scriptsize{}{}
$\left(\rho=0,\beta=0.2,J=10\right)$}}\tabularnewline
\hline 
{\scriptsize{}{}$Pr(\xi<0\vert{\cal M}_{1})$ } & {\scriptsize{}100} & {\scriptsize{}100} & {\scriptsize{}100} & {\scriptsize{}100} & {\scriptsize{}100} & {\scriptsize{}100} & {\scriptsize{}100} & {\scriptsize{}100} & {\scriptsize{}100} & {\scriptsize{}100} & {\scriptsize{}1.5} & {\scriptsize{}100} & {\scriptsize{}100} & {\scriptsize{}100} & {\scriptsize{}100} & {\scriptsize{}0} & {\scriptsize{}99} & {\scriptsize{}100} & {\scriptsize{}100} & {\scriptsize{}100}\tabularnewline
{\scriptsize{}{}$Pr(\hat{k}>3)/\ Pr(\hat{k}<3)$} & {\scriptsize{}0/0} & {\scriptsize{}0/0} & {\scriptsize{}0/0} & {\scriptsize{}0/0} & {\scriptsize{}0/0} & {\scriptsize{}0/0} & {\scriptsize{}0/0} & {\scriptsize{}0/0} & {\scriptsize{}0/0} & {\scriptsize{}0/0} & {\scriptsize{}0/0} & {\scriptsize{}0/0} & {\scriptsize{}0/0} & {\scriptsize{}0/0} & {\scriptsize{}0/0} & {\scriptsize{}0/0} & {\scriptsize{}0/0} & {\scriptsize{}0/0} & {\scriptsize{}0/0} & {\scriptsize{}0/0}\tabularnewline
\hline 
 & \multicolumn{10}{c|}{{\scriptsize{}{}Panel G: cross-sectionally correlated errors}\textbf{\scriptsize{}{}
$\left(\rho=0,\beta=0.5,J=5\right)$}} & \multicolumn{10}{c}{{\scriptsize{}{}Panel H: serially and cross-sectionally correlated
errors}\textbf{\scriptsize{}{} $\left(\rho=0.2,\beta=0.2,J=5\right)$}}\tabularnewline
\hline 
{\scriptsize{}{}$Pr(\xi<0\vert{\cal M}_{1})$ } & {\scriptsize{}81.5} & {\scriptsize{}100} & {\scriptsize{}100} & {\scriptsize{}100} & {\scriptsize{}100} & {\scriptsize{}0} & {\scriptsize{}100} & {\scriptsize{}100} & {\scriptsize{}100} & {\scriptsize{}100} & {\scriptsize{}100} & {\scriptsize{}100} & {\scriptsize{}100} & {\scriptsize{}100} & {\scriptsize{}100} & {\scriptsize{}100} & {\scriptsize{}100} & {\scriptsize{}100} & {\scriptsize{}100} & {\scriptsize{}100}\tabularnewline
{\scriptsize{}{}$Pr(\hat{k}>3)/\ Pr(\hat{k}<3)$} & {\scriptsize{}0/0} & {\scriptsize{}0/0} & {\scriptsize{}0/0} & {\scriptsize{}0/0} & {\scriptsize{}0/0} & {\scriptsize{}0/0} & {\scriptsize{}0/0} & {\scriptsize{}0/0} & {\scriptsize{}0/0} & {\scriptsize{}0/0} & {\scriptsize{}0/0} & {\scriptsize{}0/0} & {\scriptsize{}0/0} & {\scriptsize{}0/0} & {\scriptsize{}0/0} & {\scriptsize{}0/0} & {\scriptsize{}0/0} & {\scriptsize{}0/0} & {\scriptsize{}0/0} & {\scriptsize{}0/0}\tabularnewline
\hline 
 & \multicolumn{10}{c|}{{\scriptsize{}{}Panel I: serially and cross-sectionally correlated
errors}\textbf{\scriptsize{}{} $\left(\rho=0.5,\beta=0.2,J=5\right)$}} & \multicolumn{10}{c}{{\scriptsize{}{}Panel L: serially and cross-sectionally correlated
errors}\textbf{\scriptsize{}{} $\left(\rho=0.2,\beta=0.5,J=5\right)$}}\tabularnewline
\hline 
{\scriptsize{}{}$Pr(\xi<0\vert{\cal M}_{1})$ } & {\scriptsize{}100} & {\scriptsize{}100} & {\scriptsize{}100} & {\scriptsize{}100} & {\scriptsize{}100} & {\scriptsize{}100} & {\scriptsize{}100} & {\scriptsize{}100} & {\scriptsize{}100} & {\scriptsize{}100} & {\scriptsize{}77} & {\scriptsize{}100} & {\scriptsize{}100} & {\scriptsize{}100} & {\scriptsize{}100} & {\scriptsize{}0} & {\scriptsize{}100} & {\scriptsize{}100} & {\scriptsize{}100} & {\scriptsize{}100}\tabularnewline
{\scriptsize{}{}$Pr(\hat{k}>3)/\ Pr(\hat{k}<3)$} & {\scriptsize{}0/0} & {\scriptsize{}0/0} & {\scriptsize{}0/0} & {\scriptsize{}0/0} & {\scriptsize{}0/0} & {\scriptsize{}0/0} & {\scriptsize{}0/0} & {\scriptsize{}0/0} & {\scriptsize{}0/0} & {\scriptsize{}0/0} & {\scriptsize{}0/0} & {\scriptsize{}0/0} & {\scriptsize{}0/0} & {\scriptsize{}0/0} & {\scriptsize{}0/0} & {\scriptsize{}0/0} & {\scriptsize{}0/0} & {\scriptsize{}0/0} & {\scriptsize{}0/0} & {\scriptsize{}0/0}\tabularnewline
\hline 
\end{tabular}\medskip{}
\par\end{centering}
The table contains the probabilities of correct model selection estimated
from the simulated balanced dataset when $r=0$ (first row in each
panel). For $r=3,$ the table reports the percentages $x\%/y\%$ of
the replications that result in overestimation / underestimation of
the number of factors (second row of each panel). The $\left(100-x-y\right)\%$
of the replications is the probability of correct estimation of the
number of unobservable factors. The simulations are based on the DGP
in Equation (\ref{DGP_AH-1}). Panels A-L contain the results assuming
different combinations of parameters $\rho,\beta$ and $J$. 
\end{sidewaystable}

\newgeometry{top=1cm, bottom=1cm, left=2cm, right=2cm}\thispagestyle{empty}
\begin{sidewaystable}
\textbf{\protect\caption{\textbf{\label{MCtab_UnBalCase_OneObsLatentFactor60}Monte Carlo replications
of DGP \eqref{DGP_AH-1} on unbalanced panels with $\min\left(T_{i}\right)=60$,
one observable factor and $r=0$ or $3$ latent factors only}}
}
\begin{centering}
{\scriptsize{}}%
\begin{tabular}{c|ccccc|ccccc|ccccc|ccccc}
\hline 
{\scriptsize{}{}$T$ } & \multicolumn{5}{c|}{{\scriptsize{}{}150}} & \multicolumn{5}{c|}{{\scriptsize{}{}500}} & \multicolumn{5}{c|}{{\scriptsize{}{}150}} & \multicolumn{5}{c}{{\scriptsize{}{}500}}\tabularnewline
\hline 
{\scriptsize{}{}$n$ } & {\scriptsize{}{}150 } & {\scriptsize{}{}500 } & {\scriptsize{}{}1,500 } & {\scriptsize{}{}3,000 } & {\scriptsize{}{}6,000 } & {\scriptsize{}{}150 } & {\scriptsize{}{}500 } & {\scriptsize{}{}1,500 } & {\scriptsize{}{}3,000 } & {\scriptsize{}{}6,000 } & {\scriptsize{}{}150 } & {\scriptsize{}{}500 } & {\scriptsize{}{}1,500 } & {\scriptsize{}{}3,000 } & {\scriptsize{}{}6,000 } & {\scriptsize{}{}150 } & {\scriptsize{}{}500 } & {\scriptsize{}{}1,500 } & {\scriptsize{}{}3,000 } & {\scriptsize{}{}6,000 }\tabularnewline
\hline 
 & \multicolumn{10}{c|}{{\scriptsize{}{}Panel A: i.i.d. errors}\textbf{\scriptsize{}{} $\left(\rho=0,\beta=0,J=0\right)$}} & \multicolumn{10}{c}{{\scriptsize{}{}Panel B: serially correlated errors}\textbf{\scriptsize{}{}
$\left(\rho=0.3,\beta=0,J=0\right)$}}\tabularnewline
\hline 
{\scriptsize{}{}$Pr(\xi<0\vert{\cal M}_{1})$ } & {\scriptsize{}100} & {\scriptsize{}100} & {\scriptsize{}100} & {\scriptsize{}100} & {\scriptsize{}100} & {\scriptsize{}100} & {\scriptsize{}100} & {\scriptsize{}100} & {\scriptsize{}100} & {\scriptsize{}100} & {\scriptsize{}100} & {\scriptsize{}100} & {\scriptsize{}100} & {\scriptsize{}100} & {\scriptsize{}100} & {\scriptsize{}100} & {\scriptsize{}100} & {\scriptsize{}100} & {\scriptsize{}100} & {\scriptsize{}100}\tabularnewline
{\scriptsize{}{}$Pr(\hat{k}>3)/\ Pr(\hat{k}<3)$} & {\scriptsize{}0/58} & {\scriptsize{}0/4.8} & {\scriptsize{}0/0} & {\scriptsize{}0/0} & {\scriptsize{}0/0} & {\scriptsize{}0/40} & {\scriptsize{}0.6/0} & {\scriptsize{}1.2/0} & {\scriptsize{}0/0} & {\scriptsize{}0.2/0} & {\scriptsize{}0/60} & {\scriptsize{}0/3} & {\scriptsize{}0/0} & {\scriptsize{}0/0} & {\scriptsize{}0/0} & {\scriptsize{}0/18.6} & {\scriptsize{}0.6/0} & {\scriptsize{}1.9/0} & {\scriptsize{}0/0} & {\scriptsize{}0.1/0}\tabularnewline
\hline 
 & \multicolumn{10}{c|}{{\scriptsize{}{}Panel C: serially correlated errors}\textbf{\scriptsize{}{}
$\left(\rho=0.5,\beta=0,J=0\right)$}} & \multicolumn{10}{c}{{\scriptsize{}{}Panel D: serially correlated errors}\textbf{\scriptsize{}{}
$\left(\rho=0.7,\beta=0,J=0\right)$}}\tabularnewline
\hline 
{\scriptsize{}{}$Pr(\xi<0\vert{\cal M}_{1})$ } & {\scriptsize{}100} & {\scriptsize{}100} & {\scriptsize{}100} & {\scriptsize{}100} & {\scriptsize{}100} & {\scriptsize{}100} & {\scriptsize{}100} & {\scriptsize{}100} & {\scriptsize{}100} & {\scriptsize{}100} & {\scriptsize{}100} & {\scriptsize{}100} & {\scriptsize{}84} & {\scriptsize{}0} & {\scriptsize{}0} & {\scriptsize{}100} & {\scriptsize{}100} & {\scriptsize{}100} & {\scriptsize{}100} & {\scriptsize{}100}\tabularnewline
{\scriptsize{}{}$Pr(\hat{k}>3)/\ Pr(\hat{k}<3)$} & {\scriptsize{}0/52} & {\scriptsize{}0/4.8} & {\scriptsize{}0/0} & {\scriptsize{}0/0} & {\scriptsize{}0/0} & {\scriptsize{}0/40} & {\scriptsize{}0.3/0} & {\scriptsize{}2.0/0} & {\scriptsize{}0/0} & {\scriptsize{}0/0} & {\scriptsize{}0/78} & {\scriptsize{}0/3.8} & {\scriptsize{}0/0} & {\scriptsize{}0/0} & {\scriptsize{}0/0} & {\scriptsize{}0/19.6} & {\scriptsize{}0.6/0} & {\scriptsize{}1.9/0} & {\scriptsize{}0/0} & {\scriptsize{}0.3/0}\tabularnewline
\hline 
 & \multicolumn{10}{c|}{{\scriptsize{}{}Panel E: cross-sectionally correlated errors}\textbf{\scriptsize{}{}
$\left(\rho=0,\beta=0.2,J=5\right)$}} & \multicolumn{10}{c}{{\scriptsize{}{}Panel F: cross-sectionally correlated errors}\textbf{\scriptsize{}{}
$\left(\rho=0,\beta=0.2,J=10\right)$}}\tabularnewline
\hline 
{\scriptsize{}{}$Pr(\xi<0\vert{\cal M}_{1})$ } & {\scriptsize{}100} & {\scriptsize{}100} & {\scriptsize{}100} & {\scriptsize{}100} & {\scriptsize{}100} & {\scriptsize{}100} & {\scriptsize{}100} & {\scriptsize{}100} & {\scriptsize{}100} & {\scriptsize{}100} & {\scriptsize{}70} & {\scriptsize{}100} & {\scriptsize{}100} & {\scriptsize{}100} & {\scriptsize{}100} & {\scriptsize{}100} & {\scriptsize{}100} & {\scriptsize{}100} & {\scriptsize{}100} & {\scriptsize{}100}\tabularnewline
{\scriptsize{}{}$Pr(\hat{k}>3)/\ Pr(\hat{k}<3)$} & {\scriptsize{}0/54} & {\scriptsize{}0/4.3} & {\scriptsize{}0/0} & {\scriptsize{}0/0} & {\scriptsize{}0/0} & {\scriptsize{}0/19} & {\scriptsize{}0.5/0} & {\scriptsize{}2.2/0} & {\scriptsize{}0/0} & {\scriptsize{}0.2/0} & {\scriptsize{}0/56} & {\scriptsize{}0/5.1} & {\scriptsize{}0/0} & {\scriptsize{}0/0} & {\scriptsize{}0/0} & {\scriptsize{}0/18.4} & {\scriptsize{}0.4/0} & {\scriptsize{}2.1/0} & {\scriptsize{}0/0} & {\scriptsize{}0.6/0}\tabularnewline
\hline 
 & \multicolumn{10}{c|}{{\scriptsize{}{}Panel G: cross-sectionally correlated errors}\textbf{\scriptsize{}{}
$\left(\rho=0,\beta=0.5,J=5\right)$}} & \multicolumn{10}{c}{{\scriptsize{}{}Panel H: serially and cross-sectionally correlated
errors}\textbf{\scriptsize{}{} $\left(\rho=0.2,\beta=0.2,J=5\right)$}}\tabularnewline
\hline 
{\scriptsize{}{}$Pr(\xi<0\vert{\cal M}_{1})$ } & {\scriptsize{}100} & {\scriptsize{}100} & {\scriptsize{}100} & {\scriptsize{}100} & {\scriptsize{}100} & {\scriptsize{}100} & {\scriptsize{}100} & {\scriptsize{}100} & {\scriptsize{}100} & {\scriptsize{}100} & {\scriptsize{}100} & {\scriptsize{}100} & {\scriptsize{}100} & {\scriptsize{}100} & {\scriptsize{}100} & {\scriptsize{}100} & {\scriptsize{}100} & {\scriptsize{}100} & {\scriptsize{}100} & {\scriptsize{}100}\tabularnewline
{\scriptsize{}{}$Pr(\hat{k}>3)/\ Pr(\hat{k}<3)$} & {\scriptsize{}0/68} & {\scriptsize{}0/1} & {\scriptsize{}0/0} & {\scriptsize{}0/0} & {\scriptsize{}0/0} & {\scriptsize{}0/18.5} & {\scriptsize{}0.2/0} & {\scriptsize{}1.4/0} & {\scriptsize{}0/0} & {\scriptsize{}0.1/0} & {\scriptsize{}0/58} & {\scriptsize{}0/4} & {\scriptsize{}0/0} & {\scriptsize{}0/0} & {\scriptsize{}0/0} & {\scriptsize{}0/18.3} & {\scriptsize{}0.4/0} & {\scriptsize{}1.9/0} & {\scriptsize{}0/0} & {\scriptsize{}0.1/0}\tabularnewline
\hline 
 & \multicolumn{10}{c|}{{\scriptsize{}{}Panel I: serially and cross-sectionally correlated
errors}\textbf{\scriptsize{}{} $\left(\rho=0.5,\beta=0.2,J=5\right)$}} & \multicolumn{10}{c}{{\scriptsize{}{}Panel L: serially and cross-sectionally correlated
errors}\textbf{\scriptsize{}{} $\left(\rho=0.2,\beta=0.5,J=5\right)$}}\tabularnewline
\hline 
{\scriptsize{}{}$Pr(\xi<0\vert{\cal M}_{1})$ } & {\scriptsize{}100} & {\scriptsize{}100} & {\scriptsize{}100} & {\scriptsize{}100} & {\scriptsize{}100} & {\scriptsize{}100} & {\scriptsize{}100} & {\scriptsize{}100} & {\scriptsize{}100} & {\scriptsize{}100} & {\scriptsize{}100} & {\scriptsize{}100} & {\scriptsize{}100} & {\scriptsize{}100} & {\scriptsize{}100} & {\scriptsize{}100} & {\scriptsize{}100} & {\scriptsize{}100} & {\scriptsize{}100} & {\scriptsize{}100}\tabularnewline
{\scriptsize{}{}$Pr(\hat{k}>3)/\ Pr(\hat{k}<3)$} & {\scriptsize{}0/54} & {\scriptsize{}0/4.6} & {\scriptsize{}0/0} & {\scriptsize{}0/0} & {\scriptsize{}0/0} & {\scriptsize{}0/18.5} & {\scriptsize{}0.4/0} & {\scriptsize{}1.5/0} & {\scriptsize{}0/0} & {\scriptsize{}0.1/0} & {\scriptsize{}0/54} & {\scriptsize{}0/4.5} & {\scriptsize{}0/0} & {\scriptsize{}0/0} & {\scriptsize{}0/0} & {\scriptsize{}0/18.2} & {\scriptsize{}0.1/0} & {\scriptsize{}2.8/0} & {\scriptsize{}0/0} & {\scriptsize{}0.2/0}\tabularnewline
\hline 
\end{tabular}\medskip{}
\par\end{centering}
The table contains the probabilities of correct model selection estimated
from the simulated unbalanced dataset when $r=0$ (first row in each
panel). For $r=3,$ the table reports the percentages $x\%/y\%$ of
the replications that result in overestimation / underestimation of
the number of factors (second row in each panel). The $\left(100-x-y\right)\%$
of the replications is the probability of correct estimation of the
number of unobservable factors. The simulations are based on the DGP
in Equation (\ref{DGP_AH-1}) for unbalanced panels with\textbf{ $\min\left(T_{i}\right)=60$
}. Panels A-L contain the results assuming different combinations
of parameters $\rho,\beta$ and $J$. 
\end{sidewaystable}

\newgeometry{top=1cm, bottom=1cm, left=2cm, right=2cm}\thispagestyle{empty}
\begin{sidewaystable}
\textbf{\protect\caption{\textbf{\label{MCtab_UnBalCase_OneObsLatentFactorONLY120}Monte Carlo
replications of DGP \eqref{DGP_AH-1} on unbalanced panels with $\min\left(T_{i}\right)=120$,
one observable factor and $r=0$ or $3$ latent factors only}}
}
\begin{centering}
{\scriptsize{}}%
\begin{tabular}{c|ccccc|ccccc|ccccc|ccccc}
\hline 
{\scriptsize{}{}$T$ } & \multicolumn{5}{c|}{{\scriptsize{}{}150}} & \multicolumn{5}{c|}{{\scriptsize{}{}500}} & \multicolumn{5}{c|}{{\scriptsize{}{}150}} & \multicolumn{5}{c}{{\scriptsize{}{}500}}\tabularnewline
\hline 
{\scriptsize{}{}$n$ } & {\scriptsize{}{}150 } & {\scriptsize{}{}500 } & {\scriptsize{}{}1,500 } & {\scriptsize{}{}3,000 } & {\scriptsize{}{}6,000 } & {\scriptsize{}{}150 } & {\scriptsize{}{}500 } & {\scriptsize{}{}1,500 } & {\scriptsize{}{}3,000 } & {\scriptsize{}{}6,000 } & {\scriptsize{}{}150 } & {\scriptsize{}{}500 } & {\scriptsize{}{}1,500 } & {\scriptsize{}{}3,000 } & {\scriptsize{}{}6,000 } & {\scriptsize{}{}150 } & {\scriptsize{}{}500 } & {\scriptsize{}{}1,500 } & {\scriptsize{}{}3,000 } & {\scriptsize{}{}6,000 }\tabularnewline
\hline 
 & \multicolumn{10}{c|}{{\scriptsize{}{}Panel A: i.i.d. errors}\textbf{\scriptsize{}{} $\left(\rho=0,\beta=0,J=0\right)$}} & \multicolumn{10}{c}{{\scriptsize{}{}Panel B: serially correlated errors}\textbf{\scriptsize{}{}
$\left(\rho=0.3,\beta=0,J=0\right)$}}\tabularnewline
\hline 
{\scriptsize{}{}$Pr(\xi<0\vert{\cal M}_{1})$ } & {\scriptsize{}100} & {\scriptsize{}100} & {\scriptsize{}100} & {\scriptsize{}100} & {\scriptsize{}100} & {\scriptsize{}100} & {\scriptsize{}100} & {\scriptsize{}100} & {\scriptsize{}100} & {\scriptsize{}100} & {\scriptsize{}100} & {\scriptsize{}100} & {\scriptsize{}100} & {\scriptsize{}100} & {\scriptsize{}100} & {\scriptsize{}100} & {\scriptsize{}100} & {\scriptsize{}100} & {\scriptsize{}100} & {\scriptsize{}100}\tabularnewline
{\scriptsize{}{}$Pr(\hat{k}>3)/\ Pr(\hat{k}<3)$} & {\scriptsize{}0/30} & {\scriptsize{}0/0} & {\scriptsize{}0/0} & {\scriptsize{}0/0} & {\scriptsize{}0/0} & {\scriptsize{}0/13.2} & {\scriptsize{}0/0} & {\scriptsize{}0.2/0} & {\scriptsize{}0/0} & {\scriptsize{}0/0} & {\scriptsize{}0/57.4} & {\scriptsize{}0/0} & {\scriptsize{}0/0} & {\scriptsize{}0/0} & {\scriptsize{}0/0} & {\scriptsize{}0/13.3} & {\scriptsize{}1/0} & {\scriptsize{}0.1/0} & {\scriptsize{}0/0} & {\scriptsize{}0/0}\tabularnewline
\hline 
 & \multicolumn{10}{c|}{{\scriptsize{}{}Panel C: serially correlated errors}\textbf{\scriptsize{}{}
$\left(\rho=0.5,\beta=0,J=0\right)$}} & \multicolumn{10}{c}{{\scriptsize{}{}Panel D: serially correlated errors}\textbf{\scriptsize{}{}
$\left(\rho=0.7,\beta=0,J=0\right)$}}\tabularnewline
\hline 
{\scriptsize{}{}$Pr(\xi<0\vert{\cal M}_{1})$ } & {\scriptsize{}100} & {\scriptsize{}100} & {\scriptsize{}100} & {\scriptsize{}100} & {\scriptsize{}100} & {\scriptsize{}100} & {\scriptsize{}100} & {\scriptsize{}100} & {\scriptsize{}100} & {\scriptsize{}100} & {\scriptsize{}100} & {\scriptsize{}100} & {\scriptsize{}100} & {\scriptsize{}99.8} & {\scriptsize{}51} & {\scriptsize{}100} & {\scriptsize{}100} & {\scriptsize{}100} & {\scriptsize{}100} & {\scriptsize{}100}\tabularnewline
{\scriptsize{}{}$Pr(\hat{k}>3)/\ Pr(\hat{k}<3)$} & {\scriptsize{}0/36} & {\scriptsize{}0/0} & {\scriptsize{}0/0} & {\scriptsize{}0/0} & {\scriptsize{}0/0} & {\scriptsize{}0/13} & {\scriptsize{}0/0} & {\scriptsize{}0/0} & {\scriptsize{}0/0} & {\scriptsize{}0/0} & {\scriptsize{}0/58} & {\scriptsize{}0/0} & {\scriptsize{}0/0} & {\scriptsize{}0/0} & {\scriptsize{}0/0} & {\scriptsize{}0/12.5} & {\scriptsize{}1/0} & {\scriptsize{}0.3/0} & {\scriptsize{}0/0} & {\scriptsize{}0.2/0}\tabularnewline
\hline 
 & \multicolumn{10}{c|}{{\scriptsize{}{}Panel E: cross-sectionally correlated errors}\textbf{\scriptsize{}{}
$\left(\rho=0,\beta=0.2,J=5\right)$}} & \multicolumn{10}{c}{{\scriptsize{}{}Panel F: cross-sectionally correlated errors}\textbf{\scriptsize{}{}
$\left(\rho=0,\beta=0.2,J=10\right)$}}\tabularnewline
\hline 
{\scriptsize{}{}$Pr(\xi<0\vert{\cal M}_{1})$ } & {\scriptsize{}100} & {\scriptsize{}100} & {\scriptsize{}100} & {\scriptsize{}100} & {\scriptsize{}100} & {\scriptsize{}100} & {\scriptsize{}100} & {\scriptsize{}100} & {\scriptsize{}100} & {\scriptsize{}100} & {\scriptsize{}4} & {\scriptsize{}100} & {\scriptsize{}100} & {\scriptsize{}100} & {\scriptsize{}100} & {\scriptsize{}100} & {\scriptsize{}100} & {\scriptsize{}100} & {\scriptsize{}100} & {\scriptsize{}100}\tabularnewline
{\scriptsize{}{}$Pr(\hat{k}>3)/\ Pr(\hat{k}<3)$} & {\scriptsize{}0/36} & {\scriptsize{}0/0} & {\scriptsize{}0/0} & {\scriptsize{}0/0} & {\scriptsize{}0/0} & {\scriptsize{}0/13.8} & {\scriptsize{}0/0} & {\scriptsize{}0/0} & {\scriptsize{}0/0} & {\scriptsize{}0/0} & {\scriptsize{}0/28} & {\scriptsize{}0/0} & {\scriptsize{}0/0} & {\scriptsize{}0/0} & {\scriptsize{}0/0} & {\scriptsize{}0/11.7} & {\scriptsize{}1/0} & {\scriptsize{}0.2/0} & {\scriptsize{}0/0} & {\scriptsize{}0.1/0}\tabularnewline
\hline 
 & \multicolumn{10}{c|}{{\scriptsize{}{}Panel G: cross-sectionally correlated errors}\textbf{\scriptsize{}{}
$\left(\rho=0,\beta=0.5,J=5\right)$}} & \multicolumn{10}{c}{{\scriptsize{}{}Panel H: serially and cross-sectionally correlated
errors}\textbf{\scriptsize{}{} $\left(\rho=0.2,\beta=0.2,J=5\right)$}}\tabularnewline
\hline 
{\scriptsize{}{}$Pr(\xi<0\vert{\cal M}_{1})$ } & {\scriptsize{}96} & {\scriptsize{}100} & {\scriptsize{}100} & {\scriptsize{}100} & {\scriptsize{}100} & {\scriptsize{}95.5} & {\scriptsize{}100} & {\scriptsize{}100} & {\scriptsize{}100} & {\scriptsize{}100} & {\scriptsize{}100} & {\scriptsize{}100} & {\scriptsize{}100} & {\scriptsize{}100} & {\scriptsize{}100} & {\scriptsize{}100} & {\scriptsize{}100} & {\scriptsize{}100} & {\scriptsize{}100} & {\scriptsize{}100}\tabularnewline
{\scriptsize{}{}$Pr(\hat{k}>3)/\ Pr(\hat{k}<3)$} & {\scriptsize{}0/38} & {\scriptsize{}0/0} & {\scriptsize{}0/0} & {\scriptsize{}0/0} & {\scriptsize{}0/0} & {\scriptsize{}0/15.3} & {\scriptsize{}1/0} & {\scriptsize{}0.3/0} & {\scriptsize{}0/0} & {\scriptsize{}0/0} & {\scriptsize{}0/34} & {\scriptsize{}0/0} & {\scriptsize{}0/0} & {\scriptsize{}0/0} & {\scriptsize{}0/0} & {\scriptsize{}0/13.6} & {\scriptsize{}0/0} & {\scriptsize{}0.3/0} & {\scriptsize{}0/0} & {\scriptsize{}0/0}\tabularnewline
\hline 
 & \multicolumn{10}{c|}{{\scriptsize{}{}Panel I: serially and cross-sectionally correlated
errors}\textbf{\scriptsize{}{} $\left(\rho=0.5,\beta=0.2,J=5\right)$}} & \multicolumn{10}{c}{{\scriptsize{}{}Panel L: serially and cross-sectionally correlated
errors}\textbf{\scriptsize{}{} $\left(\rho=0.2,\beta=0.5,J=5\right)$}}\tabularnewline
\hline 
{\scriptsize{}{}$Pr(\xi<0\vert{\cal M}_{1})$ } & {\scriptsize{}100} & {\scriptsize{}100} & {\scriptsize{}100} & {\scriptsize{}100} & {\scriptsize{}100} & {\scriptsize{}100} & {\scriptsize{}100} & {\scriptsize{}100} & {\scriptsize{}100} & {\scriptsize{}100} & {\scriptsize{}88} & {\scriptsize{}100} & {\scriptsize{}100} & {\scriptsize{}100} & {\scriptsize{}100} & {\scriptsize{}91.6} & {\scriptsize{}100} & {\scriptsize{}100} & {\scriptsize{}100} & {\scriptsize{}100}\tabularnewline
{\scriptsize{}{}$Pr(\hat{k}>3)/\ Pr(\hat{k}<3)$} & {\scriptsize{}0/46} & {\scriptsize{}0/0} & {\scriptsize{}0/0} & {\scriptsize{}0/0} & {\scriptsize{}0/0} & {\scriptsize{}0/10} & {\scriptsize{}0/0} & {\scriptsize{}0/0} & {\scriptsize{}0/0} & {\scriptsize{}0/0} & {\scriptsize{}0/36} & {\scriptsize{}0/0} & {\scriptsize{}0/0} & {\scriptsize{}0/0} & {\scriptsize{}0/0} & {\scriptsize{}0/13.1} & {\scriptsize{}0/0} & {\scriptsize{}0.3/0} & {\scriptsize{}0/0} & {\scriptsize{}0.4/0}\tabularnewline
\hline 
\end{tabular}\medskip{}
\par\end{centering}
The table contains the probabilities of correct model selection estimated
from the simulated unbalanced dataset when $r=0$ (first row in each
panel). For $r=3,$ the table reports the percentages $x\%/y\%$ of
the replications that result in overestimation / underestimation of
the number of factors (second row in each panel). The $\left(100-x-y\right)\%$
of the replications is the probability of correct estimation of the
number of unobservable factors. The simulations are based on the DGP
in Equation (\ref{DGP_AH-1}) for unbalanced panels with\textbf{ $\min\left(T_{i}\right)=120$}.
Panels A-L contain the results assuming different combinations of
parameters $\rho,\beta$ and $J$. 
\end{sidewaystable}

\newgeometry{top=1cm, bottom=1cm, left=2cm, right=2cm}\thispagestyle{empty}
\begin{sidewaystable}
\textbf{\protect\caption{\textbf{\label{MCtab_UnBalCase_OneObsLatentFactor240}Monte Carlo
replications of DGP \eqref{DGP_AH-1} on unbalanced panels with $\min\left(T_{i}\right)=240$,
one observable factor and $r=0$ or $3$ latent factors only}}
}
\begin{centering}
{\scriptsize{}}%
\begin{tabular}{c|ccccc|ccccc}
\hline 
{\scriptsize{}{}$T$ } & \multicolumn{5}{c|}{{\scriptsize{}{}500}} & \multicolumn{5}{c}{{\scriptsize{}{}500}}\tabularnewline
\hline 
{\scriptsize{}{}$n$ } & {\scriptsize{}{}150 } & {\scriptsize{}{}500 } & {\scriptsize{}{}1,500 } & {\scriptsize{}{}3,000 } & {\scriptsize{}{}6,000 } & {\scriptsize{}{}150 } & {\scriptsize{}{}500 } & {\scriptsize{}{}1,500 } & {\scriptsize{}{}3,000 } & {\scriptsize{}{}6,000 }\tabularnewline
\hline 
 & \multicolumn{5}{c|}{{\scriptsize{}{}Panel A: i.i.d. errors}\textbf{\scriptsize{}{} $\left(\rho=0,\beta=0,J=0\right)$}} & \multicolumn{5}{c}{{\scriptsize{}{}Panel B: serially correlated errors}\textbf{\scriptsize{}{}
$\left(\rho=0.3,\beta=0,J=0\right)$}}\tabularnewline
\hline 
{\scriptsize{}{}$Pr(\xi<0\vert{\cal M}_{1})$ } & {\scriptsize{}100} & {\scriptsize{}100} & {\scriptsize{}100} & {\scriptsize{}100} & {\scriptsize{}100} & {\scriptsize{}100} & {\scriptsize{}100} & {\scriptsize{}100} & {\scriptsize{}100} & {\scriptsize{}100}\tabularnewline
{\scriptsize{}{}$Pr(\hat{k}>3)/\ Pr(\hat{k}<3)$} & {\scriptsize{}0/1.4} & {\scriptsize{}0/0} & {\scriptsize{}0/0} & {\scriptsize{}0/0} & {\scriptsize{}0/0} & {\scriptsize{}0/1.5} & {\scriptsize{}0/0} & {\scriptsize{}0/0} & {\scriptsize{}0/0} & {\scriptsize{}0/0}\tabularnewline
\hline 
 & \multicolumn{5}{c|}{{\scriptsize{}{}Panel C: serially correlated errors}\textbf{\scriptsize{}{}
$\left(\rho=0.5,\beta=0,J=0\right)$}} & \multicolumn{5}{c}{{\scriptsize{}{}Panel D: serially correlated errors}\textbf{\scriptsize{}{}
$\left(\rho=0.7,\beta=0,J=0\right)$}}\tabularnewline
\hline 
{\scriptsize{}{}$Pr(\xi<0\vert{\cal M}_{1})$ } & {\scriptsize{}100} & {\scriptsize{}100} & {\scriptsize{}100} & {\scriptsize{}100} & {\scriptsize{}100} & {\scriptsize{}100} & {\scriptsize{}100} & {\scriptsize{}100} & {\scriptsize{}100} & {\scriptsize{}100}\tabularnewline
{\scriptsize{}{}$Pr(\hat{k}>3)/\ Pr(\hat{k}<3)$} & {\scriptsize{}0/0.8} & {\scriptsize{}0.1/0} & {\scriptsize{}0/0} & {\scriptsize{}0/0} & {\scriptsize{}0/0} & {\scriptsize{}0/1.5} & {\scriptsize{}0.1/0} & {\scriptsize{}0/0} & {\scriptsize{}0/0} & {\scriptsize{}0/0}\tabularnewline
\hline 
 & \multicolumn{5}{c|}{{\scriptsize{}{}Panel E: cross-sectionally correlated errors}\textbf{\scriptsize{}{}
$\left(\rho=0,\beta=0.2,J=5\right)$}} & \multicolumn{5}{c}{{\scriptsize{}{}Panel F: cross-sectionally correlated errors}\textbf{\scriptsize{}{}
$\left(\rho=0,\beta=0.2,J=10\right)$}}\tabularnewline
\hline 
{\scriptsize{}{}$Pr(\xi<0\vert{\cal M}_{1})$ } & {\scriptsize{}100} & {\scriptsize{}100} & {\scriptsize{}100} & {\scriptsize{}100} & {\scriptsize{}100} & {\scriptsize{}0} & {\scriptsize{}100} & {\scriptsize{}100} & {\scriptsize{}100} & {\scriptsize{}100}\tabularnewline
{\scriptsize{}{}$Pr(\hat{k}>3)/\ Pr(\hat{k}<3)$} & {\scriptsize{}0/1.1} & {\scriptsize{}0.1/0} & {\scriptsize{}0.1/0} & {\scriptsize{}0/0} & {\scriptsize{}0/0} & {\scriptsize{}0/0.6} & {\scriptsize{}0.1/0} & {\scriptsize{}0/0} & {\scriptsize{}0/0} & {\scriptsize{}0/0}\tabularnewline
\hline 
 & \multicolumn{5}{c|}{{\scriptsize{}{}Panel G: cross-sectionally correlated errors}\textbf{\scriptsize{}{}
$\left(\rho=0,\beta=0.5,J=5\right)$}} & \multicolumn{5}{c}{{\scriptsize{}{}Panel H: serially and cross-sectionally correlated
errors}\textbf{\scriptsize{}{} $\left(\rho=0.2,\beta=0.2,J=5\right)$}}\tabularnewline
\hline 
{\scriptsize{}{}$Pr(\xi<0\vert{\cal M}_{1})$ } & {\scriptsize{}100} & {\scriptsize{}100} & {\scriptsize{}100} & {\scriptsize{}100} & {\scriptsize{}100} & {\scriptsize{}100} & {\scriptsize{}100} & {\scriptsize{}100} & {\scriptsize{}100} & {\scriptsize{}100}\tabularnewline
{\scriptsize{}{}$Pr(\hat{k}>3)/\ Pr(\hat{k}<3)$} & {\scriptsize{}0/1.4} & {\scriptsize{}0.1/0} & {\scriptsize{}0/0} & {\scriptsize{}0/0} & {\scriptsize{}0/0} & {\scriptsize{}0/1.2} & {\scriptsize{}0.1/0} & {\scriptsize{}0/0} & {\scriptsize{}0/0} & {\scriptsize{}0/0}\tabularnewline
\hline 
 & \multicolumn{5}{c|}{{\scriptsize{}{}Panel I: serially and cross-sectionally correlated
errors}\textbf{\scriptsize{}{} $\left(\rho=0.5,\beta=0.2,J=5\right)$}} & \multicolumn{5}{c}{{\scriptsize{}{}Panel L: serially and cross-sectionally correlated
errors}\textbf{\scriptsize{}{} $\left(\rho=0.2,\beta=0.5,J=5\right)$}}\tabularnewline
\hline 
{\scriptsize{}{}$Pr(\xi<0\vert{\cal M}_{1})$ } & {\scriptsize{}100} & {\scriptsize{}100} & {\scriptsize{}100} & {\scriptsize{}100} & {\scriptsize{}100} & {\scriptsize{}10} & {\scriptsize{}100} & {\scriptsize{}100} & {\scriptsize{}100} & {\scriptsize{}100}\tabularnewline
{\scriptsize{}{}$Pr(\hat{k}>3)/\ Pr(\hat{k}<3)$} & {\scriptsize{}0/0.7} & {\scriptsize{}0/0} & {\scriptsize{}0/0} & {\scriptsize{}0/0} & {\scriptsize{}0/0} & {\scriptsize{}0/1.1} & {\scriptsize{}0.1/0} & {\scriptsize{}0/0} & {\scriptsize{}0/0} & {\scriptsize{}0/0}\tabularnewline
\hline 
\end{tabular}\medskip{}
\par\end{centering}
The table contains the probabilities of correct model selection estimated
from the simulated unbalanced dataset when $r=0$ (first row in each
panel). For $r=3,$ the table reports the percentages $x\%/y\%$ of
the replications that result in overestimation / underestimation of
the number of factors (second row in each panel). The $\left(100-x-y\right)\%$
of the replications is the probability of correct estimation of the
number of unobservable factors. The simulations are based on the DGP
in Equation (\ref{DGP_AH-1}) for unbalanced panels with\textbf{ $\min\left(T_{i}\right)=240$
}. Panels A-L contain the results assuming different combinations
of parameters $\rho,\beta$ and $J$. 
\end{sidewaystable}

\end{document}